\documentclass[twocolumn]{svjour3}

\smartqed  

\usepackage{graphicx}
\usepackage{cite}
\usepackage{color}
\usepackage{adjustbox}
\usepackage{amssymb}
\usepackage{amsmath}
\usepackage{url}

\graphicspath{{figs/}}

\begin{document}

\title{%
	Space-time adaptive ADER-DG finite element method 
	with LST-DG predictor and \textit{a posteriori} sub-cell WENO finite-volume limiting for
	simulation of non-stationary compressible multicomponent reactive flows
}

\titlerunning{%
	ADER-DG method for non-stationary compressible multicomponent reactive flows
}

\author{Popov I.S.}

\institute{
	Popov I.S. \at
	Department of Theoretical Physics, 
	Dostoevsky Omsk State University, Omsk, Russia \\
	\email{diphosgen@mail.ru, popovis@omsu.ru}
}

\date{Received: date / Accepted: date}

\maketitle

\begin{abstract}
The space-time adaptive ADER finite element DG method with a posteriori correction technique of solutions on subcells by the finite-volume ADER-WENO limiter was used to simulate non-stationary compressible multicomponent reactive flows. The multicomponent composition of the reacting medium and the reactions occurring in it were described by expanding the original system of Euler equations by a system of non-stationary convection-reaction equations. The use of this method to simulate high stiff problems associated with reactions occurring in a multicomponent medium requires the use of the adaptive change in the time step. The solution of the classical problem related to the formation and propagation of a ZND detonation wave is carried out. It was shown that the space-time adaptive ADER finite element DG method with a posteriori correction technique of solutions on subcells by the finite-volume ADER-WENO limiter can be used to simulate flows without using of splitting in directions and fractional step methods.

\end{abstract}

\section*{Introduction}
\label{intro}

Modern problems of fluids, gases and plasma dynamics are accompanied by the problems of simulating quite complex multi-scale flows~\cite{Fortov_ESM, Fortov_HEDP, Zeldovich_Raizer_2002, Drake_2018}. Typical situations of flows arising within the framework of such problems contain forming, propagating and changing discontinuous solution components, such as shock waves and contact discontinuities; large scale and small scale flow turbulence; complex structure of boundary layers; coherent structures and a complicated set of nonlinear feedbacks. These peculiarities of the solution place strict conditions on the possibilities and accuracy of numerical schemes for simulating hydrodynamic flows.

Non-stationary solutions of classical problems of hydrodynamics and gas dynamics are characterized by the formation of discontinuous solution components, such as shock waves and contact discontinuities, even from smooth initial conditions~\cite{Kulikovskii, Rozhdestvenskii_Janenko}. This leads to the imposition of rather contradictory requirements~\cite{Handbook_NMHP_BF_2016, Handbook_NMHP_AM_2017} on the numerical schemes for modeling compressible flows. On the one hand, for the correct resolution of discontinuous components of the solution, the numerical scheme must have no more than the first order of accuracy, which is determined by the famous Godunov's theorem, and have a sufficiently low numerical dispersion, or the numerical scheme must be nonlinear, which usually leads to the concept of hybrid numerical schemes. On the other hand, for a correct description of the propagation of small scale perturbations and non-stationary processes of the formation of complex solution structures, the numerical scheme must have a high order of accuracy and have a rather weak numerical dissipation.

Non-stationary compressible flows of multicomponent reactive flows are characterized by some significant additional difficulties that are not encountered in classical gas-dynamic problems~\cite{Oran_Boris_2005, Lunev_2017, Nagnibeda_2009, Anderson_1989}. The kinetics of reactions in a multicomponent gas medium is described by the equations of chemical kinetics, such as the law of mass action. The reaction rates and the local energy yield of the reactions strongly and significantly nonlinearly depend on local density, temperature and composition fields. This imposes significant restrictions on the resolution accuracy of discontinuous components and sharp solution gradients. The strong dependence of the reaction parameters on the local temperature fields and concentrations of the medium components and the strong dependence of the fields of hydrodynamic quantities on the reaction fields (especially on the energy yield) lead to the formation of a complex feedback structure. This imposes significant restrictions on the accuracy of reproduction in the numerical solution of small perturbations, which is especially important in the study of non-stationary processes~\cite{Oran_Boris_2005}.

Additional reactions terms in the convective transport equations of individual components of a multicomponent medium and in the equation(s) of the energy of the current have an extremely high stiffness~\cite{Oran_Boris_2005}. These peculiarities, inherent in non-stationary compressible flows of multicomponent reactive flows, often lead to the use of numerical schemes for splitting into physical processes or methods of schemes with fractional time steps. However, the use of splitting methods usually leads to a significant reduction in the formal order of accuracy of the numerical scheme (usually, down to the first order). In certain situations, in particular, those associated with modeling the formation of detonation waves, the use of splitting methods can lead to a fundamentally incorrect solution (the detailed description see in~\cite{frac_steps_detwave_sim_2000}), which is associated with a significant difference in the local rates of hydrodynamic processes and reactions.

The described difficulties of numerical simulation of non-stationary compressible flows of multicomponent reactive flows require the development of high-precision numerical schemes that, on the one hand, would allow to resolve discontinuous components of the solution quite accurately and accurately reproduce small perturbations against the background of the general flow, on the other hand, would allow not to use direct methods of splitting by physical processes.

Among modern methods of computational fluid dynamics, in this case, high-precision DG methods (see review in~\cite{chem_kin_hrs_rev}) and finite-volume methods with high-precision solution reconstruction~\cite{chem_kin_hrs_weno} are considered as perspective. In the case of DG methods, this is primarily due to the possibility of obtaining an arbitrarily high order of accuracy, especially when using the well-known ADER paradigm~\cite{ader_init_1, ader_init_2}, as well as the high stability of these methods with respect to the occurrence of stiff terms in the system of equations~\cite{ader_stiff_1, ader_stiff_2}, using a local space-time DG predictor (LST-DG) instead of the problem-dependent Cauchy-Kowalewski procedure in the ADER paradigm. However, one can't just take and refuse using the splitting methods in high-precision finite volume methods, especially in terms of computational efficiency~\cite{chem_kin_hrs_weno}.

Computational methods such as the space-time adaptive ADER finite element DG method with \textit{a posteriori} correction technique of solutions on sub-cells by the finite-volume ADER-WENO limiter using adaptive mesh refinement~\cite{ader_dg_ideal_flows} demonstrate unprecedented accuracy of numerical solution and resolution of discontinuities, and may be considered as a new generation of shock capturing schemes for computational fluid dynamics. This computational scheme has arisen as a result of the development of special schemes~\cite{mood_par} using a paradigm MOOD, using high-precision ADER-DG methods~\cite{ader_dg_dev_1, ader_dg_dev_2} and adaptive correction with a sufficiently accurate and stable ADER-WENO-FV limiter~\cite{ader_weno_lstdg}. This scheme was extended for application in simulation of dissipative GD and MHD flows in~\cite{ader_dg_diss_flows}; to the case of unstructured mesh, with application of direct ALE schemes in~\cite{ader_dg_ale}; it was applied for solution of GRGD and GRMHD problems in~\cite{ader_dg_grmhd, ader_dg_gr_prd}. In the article~\cite{ader_dg_simple_mod} in the Appendix A it was described the main actions that must be taken to transform a given DG code into a DG code that is stabilized with an a posteriori finite volume sub-cell limiter. In work~\cite{ader_dg_PNPM}, a generalization of the ADER-DG method with a posteriori correction technique of solutions on sub-cells by the finite-volume ADER-WENO limiter using adaptive mesh refinement to the $P_{N} P_{M}$-schemes~\cite{PNPM_DG} was constructed. Peculiarities of mathematical formulation and efficiency possibility of implementation are discussed in~\cite{ader_dg_eff_impl, fron_phys} and~\cite{exahype, ader_dg_hpc_impl_1, ader_dg_hpc_impl_2, ader_dg_hpc_impl_3, ader_dg_hpc_impl_4}. The modern state of development of the space-time adaptive ADER finite element DG method with a posteriori correction technique of solutions on sub-cells by the finite-volume limiter using AMR for use on unstructured meshes and using the ALE approach is presented in the works~\cite{ader_dg_mod_1, ader_dg_mod_2}. The article~\cite{Vilar_subcell_lim_2019} presents a new development of these methods, associated with the use of a finite-volume limiter on the scale of individual subcells rather than the entire subgrid of the cell. An important feature of the space-time adaptive ADER finite element DG method with a posteriori correction technique of solutions on sub-cells by the finite-volume ADER-WENO limiter, as a new generation of high-resolution shock capturing numerical schemes, is the property of high-precision discontinuity resolution in the numerical solution --- shock waves and contact discontinuities are resolved within a single finite element cell, while contact discontinuities do not spread over time; at the same time with high-precision resolution of small (acoustic) components of the solution.

This work is devoted to the use of space-time adaptive ADER finite element DG method with a posteriori correction technique of solutions on subcells by the finite-volume ADER-WENO limiter scheme for simulation of non-stationary compressible multicomponent reactive flows. The multicomponent composition of the reacting medium and the reactions occurring in it were described by expanding the original system of Euler equations by a system of non-stationary convection-reaction equations.

\section{General description of the numerical method}
\label{sec:1}

\subsection{Mathematical formulation of the problem}
\label{subsec:1:1}

In this work, a classic mathematical problem formulation has been used to describe non-stationary gas-dynamic flow of multicomponent reactive media. The general properties of the gas-dynamic flow were described using a system of non-stationary one-di\-men\-sio\-nal Euler equations. The multicomponent composition of the reacting medium and the reactions occurring in it were described by expanding the original system of Euler equations by a system of non-stationary one-di\-men\-sio\-nal convection-reaction equations. The transfer velocities of the components in the convection-reaction equations were chosen to be the same and equal to the gas-dynamic velocity of the medium (one-speed approximation of the transfer). The density of the components was chosen by introducing into consideration the mass concentrations of the components, while the density of the medium was also determined from the Euler equations. In this case, the source term associated with the energy yield from the reactions occurring in the medium was included in the system of Euler equations on the right side of the energy equation. Thus, the system of non-stationary one-di\-men\-sio\-nal Euler equations and the system of non-stationary one-di\-men\-sio\-nal convection-reaction equations formed a quasilinear system of hyperbolic equations for describing non-sta\-tio\-na\-ry compressible reacting flows, which takes the following form:
\begin{equation}\label{system_of_equations}
\frac{\partial\mathbf{U}}{\partial t} +
\frac{\partial\mathbf{F}(\mathbf{U})}{\partial x} =
\mathbf{S}(\mathbf{U});
\end{equation}
where $\mathbf{U}$ is the vector of conserved values, 
$\mathbf{F}$ is the flux terms and $\mathbf{S}$ is the source term:
\begin{eqnarray}
\mathbf{U} \hspace{-0.5mm} = \hspace{-1mm}\left[
\begin{array}{c}
\rho\\
\rho u\\
\varepsilon\\
\rho \mathbf{c}
\end{array}
\right]\hspace{-1.5mm};\hspace{5mm}
\mathbf{F} \hspace{-0.5mm} = \hspace{-1mm}\left[
\begin{array}{c}
\rho u\\
\rho u^{2} + p\\
(\varepsilon + p) u\\
\rho \mathbf{c} u
\end{array}
\right]\hspace{-1.5mm};\hspace{5mm}
\mathbf{S} \hspace{-0.5mm} = \hspace{-1mm}\left[
\begin{array}{c}
0\\
0\\
S_{e}\\
\mathbf{S}_{r}
\end{array}
\right]\hspace{-1.5mm};
\end{eqnarray}
where $\rho$ is the mass density; $u$ is the velocity; $p$ is the pressure; $\varepsilon$ is the total energy density including the thermal $e$ and the kinetic contributions $\varepsilon = e + \frac{1}{2} \rho v^{2}$; $\mathbf{c}^{T} = [c_{1}, \ldots, c_{R}]$ is a vector of mass concentrations $c_{k}$ of the component of the reacting medium, which determines the mass fraction of the $k$-th component in the mixture: the density of the $k$-th component can be obtained by the formula $\rho_{k} = \rho c_{k}$, and thus the relation $\sum_{k} c_{k} = 1$; $R$ is the amount of components in the mixture; $\mathrm{S}_{e}$ is the source term that determine energy yield associated with reactions occurring in the medium; $\mathbf{S}_{r}$ is the source terms that determines the rates of reactions. The multicomponent and reaction properties of the flow are considered in the form of non-stationary one-dimensional convection-reaction equations
\begin{equation}\label{convection_reaction_equations}
\frac{\partial \left(\rho\mathbf{c}\right)}{\partial t} +
\frac{\partial\left(\rho\mathbf{c} u\right)}{\partial x} =
\mathbf{S}_{r},
\end{equation}
where, as can be seen from the system of equations (\ref{system_of_equations}), the transfer velocity of the mixture components and the total mass density are determined from the Euler equations. The reaction rates included in the term $\mathbf{S}_{r}$ were chosen in the form of the law of mass action, where the reaction rate constants could have an arbitrary dependence on the temperature $T$ of the medium. The equation of state of the multicomponent mixture was chosen in the form of the perfect gas equation. The caloric equation of state was specified in the form $p = (\gamma - 1) e$, where $\gamma$ was calculated from the ratio of the weighted average, by mass concentrations $c_{k}$, specific heat capacities of components. Empirical values of the constants, $R$ and $\gamma$, which are used in the local approximation of the wide-range equations of state by the equation of state of a perfect gas~\cite{Zeldovich_Raizer_2002, Lunev_2017, Anderson_1989}, can be used instead of the classical values.

Thus, in the present work, a computational scheme was used for the simplest mathematical model of unsteady compressible reacting flows, however, it is quite in demand in solving applied problems~\cite{Oran_Boris_2005, Lunev_2017, Nagnibeda_2009, Anderson_1989}.

\subsection{Formulation of the numerical method}
\label{subsec:1:2}

The numerical scheme used in this work is based on the space-time adaptive ADER finite-element DG method, which has a high accuracy in resolution the smooth components of the solution. Nonphysical anomalies of the numerical solution that arise in the areas of discontinuities and strong gradients of the solution, associated with the linearity of the basic numerical scheme and Godunov's theorem, are corrected a posteriori by a sufficiently stable high-precision ADER-WENO finite-volume scheme. An excellent detailed description of this computational scheme is given in the basic works~\cite{ader_dg_ideal_flows, ader_dg_dev_1, ader_dg_dev_2, ader_weno_lstdg, ader_dg_diss_flows, ader_dg_ale, ader_dg_grmhd, ader_dg_gr_prd, ader_dg_PNPM, PNPM_DG, ader_dg_eff_impl, fron_phys, exahype, ader_dg_hpc_impl_1, ader_dg_hpc_impl_2, ader_dg_hpc_impl_3, ader_dg_hpc_impl_4} of the developers of this method. Details of the internal structure of the ADER-DG are presented in the works~\cite{ader_dg_dev_1, ader_dg_dev_2, PNPM_DG}. Details of the internal structure of the ADER-DG and ADER-WENO finite volume methods, in which the LST-DG prediction method is used, are presented in the work~\cite{ader_weno_lstdg}. Peculiarities of mathematical formulation and efficiency possibility of implementation are discussed in~\cite{ader_dg_eff_impl, fron_phys} and~\cite{exahype, ader_dg_hpc_impl_1, ader_dg_hpc_impl_2, ader_dg_hpc_impl_3, ader_dg_hpc_impl_4}. The modern state of development of the space-time adaptive ADER finite element DG method with a posteriori correction technique of solutions on sub-cells by the finite-volume limiter using AMR for use on unstructured meshes and using the ALE approach is presented in the works~\cite{ader_dg_mod_1, ader_dg_mod_2}. For this reason, the description of the computational method in this paragraph will be given only briefly enough to understand the general structure of the method used in this particular case.

The space-time adaptive ADER-DG finite element method with LST-DG predictor and a posteriori sub-cell WENO finite-volume limiting involves a sequence of steps~\cite{ader_dg_ideal_flows, ader_dg_dev_1, ader_dg_dev_2, ader_weno_lstdg, ader_dg_diss_flows, ader_dg_ale, ader_dg_grmhd, ader_dg_gr_prd, ader_dg_PNPM, PNPM_DG, ader_dg_eff_impl, fron_phys, exahype, ader_dg_hpc_impl_1, ader_dg_hpc_impl_2, ader_dg_hpc_impl_3, ader_dg_hpc_impl_4}:
\begin{itemize}
	\item a LST-DG predictor, using which a local discrete space-time solution in the small is obtained;
	\item a pure ADER discontinuous Galerkin $\mathbb{P}_{N}\mathbb{P}_{N}$ scheme, using which a preliminary high accuracy solution is obtained;
	\item a determination of the admissibility of the obtained high accuracy solution and identification of ``troubled'' cells;
	\item a recalculation of the solution in ``troubled'' cells by a stable ADER-WENO finite-volume limiter.
\end{itemize}

The one-di\-men\-sio\-nal computational domain $\Omega = [a, b]$ was represented by the mesh $\Omega = \cup_{i} \Omega_{i}$, where $\Omega_{i} = [x_{i-\frac12}, x_{i+\frac12}]$. The piecewise polynomials DG-re\-pre\-sen\-ta\-tion $\mathbf{U}_{h}(x, t^{n})$ are based on the Legendre interpolation polynomials of degree $N$, specified on the space cell $\Omega_{i}$ (mapped into the space reference element $\omega = [0; 1]$):
\begin{equation}\label{DG_representation}
\mathbf{U}_{h}(x, t^{n}) = 
\sum\limits_{p} \hat{\mathbf{u}}_{p}^{n} \varphi_{p}\left(\xi(x)\right),\;\;\;\;\;\;
\mathbf{r} \in \Omega_{i} \leftrightarrow \mathbf{\xi} \in \omega,
\end{equation}
where $\hat{\mathbf{u}}_{p}^{n}$ are the DG-representation coefficients; $\varphi_{p} = \varphi_{p}(\xi)$ are the basis functions; and index $p \in [0, N]$. The interpolation nodes of the Legendre polynomials $\varphi_{p}(\xi)$ were chosen at the nodal points of the Gauss-Legendre quadrature formula.

The solution on a new $(n+1)$-th time step is obtained using the fully discrete one-step ADER-DG sche\-me~\cite{ader_dg_ideal_flows, ader_dg_dev_1, ader_dg_dev_2, ader_weno_lstdg, ader_dg_diss_flows, ader_dg_ale, ader_dg_grmhd, ader_dg_gr_prd, ader_dg_PNPM, PNPM_DG, ader_dg_eff_impl, fron_phys, exahype, ader_dg_hpc_impl_1, ader_dg_hpc_impl_2, ader_dg_hpc_impl_3, ader_dg_hpc_impl_4}, that was chosen in the classical form of projection onto the elements $\varphi_{h}(\xi)$ of the set of basis functions $\left\{\varphi_{h}(\xi)\right\}$:
\begin{equation}
\begin{split}\label{ADER_DG_scheme}
\int\limits_{t^{n}}^{t^{n+1}}\hspace{-2mm}\int\limits_{\Omega_{i}}&
\varphi_{h}(\xi(x)) \frac{\partial \mathbf{U}_{h}(x, t)}{\partial t}\ dx dt\hspace{1mm} \\
+\int\limits_{t^{n}}^{t^{n+1}}\hspace{-2mm}&\left[
\varphi_{h}\left(\xi(x_{i+\frac12})\right) \mathbf{\mathfrak{G}}\Big(\mathbf{q}_{h}^{R}(x_{i-\frac12}, t), \mathbf{q}_{h}(x_{i+\frac12}, t)\Big) \right.\\
- &\left.\hspace{1.2mm}\varphi_{h}\left(\xi(x_{i-\frac12})\right) \mathbf{\mathfrak{G}}\Big(\mathbf{q}_{h}(x_{i-\frac12}, t), \mathbf{q}_{h}^{L}(x_{i+\frac12}, t)\Big)\right] dt \\
- \int\limits_{t^{n}}^{t^{n+1}}\hspace{-2mm}&\int\limits_{\Omega_{i}}
\left[\frac{d}{dx} \varphi_{h}(\xi(x))\right] \mathbf{F}\Big(\mathbf{q}_{h}(x, t)\Big) dx dt\\
= \int\limits_{t^{n}}^{t^{n+1}}\hspace{-2mm}&\int\limits_{\Omega_{i}}
\varphi_{h}(\xi(x)) \mathbf{S}\Big(\mathbf{q}_{h}(x, t)\Big) dx dt,\\
\end{split}
\end{equation}
where $\mathbf{q}_{h} = \mathbf{q}_{h}(x, t)$ is the discrete space-time solution obtained by using the LST-DG-predictor; $\mathbf{\mathfrak{G}} = \mathbf{\mathfrak{G}}(\mathbf{q}_{h}^{-}, \mathbf{q}_{h}^{+})$ is the Riemann solver; $\mathbf{q}_{h}^{L}$ and $\mathbf{q}_{h}^{R}$ is the discrete space-time solution in left and right cells; left and right states $(\mathbf{q}_{h}^{-}, \mathbf{q}_{h}^{+})$ of the Riemann problem $\mathbf{\mathfrak{G}}$ are chosen as a solution at cell $\Omega_{i}$ interfaces. In this work, the Rusanov flux~\cite{Rusanov_solver} (as noted in the work~\cite{ader_dg_ideal_flows}, the Rusanov flux sometimes referred to as the local Lax-Friedrichs flux) and HLLE Riemann solver~\cite{HLLE_solver_1, HLLE_solver_2} were used as a Riemann solver~\cite{Toro_solvers_2009}; HLLE Riemann solver is characterized by high stability when used in high rarefaction ranges~\cite{HLLE_solver_2, Toro_solvers_2009}. In this paper, the author would also like to note a very interesting shock-stable modification of the HLLC Riemann solver with reduced numerical dissipation~\cite{HLLC_solver_shock_stable_2020}, which may arouse interest in using it with the ADER-DG schemes.

The computational scheme was based on space-time adaptive ADER finite-element discontinuous Galerkin method, which is linear in the sense of Godunov's theorem. Therefore, when using a DG-representation with a degree $N \geqslant 1$, with a formal order of accuracy not less than $2$, nonphysical oscillations arise in the numerical solution. Nonphysical anomalies of the solution were found in the solution obtained at a new time step, using the physical admissibility criteria (PAC) and the relaxed discrete maximum principle (DMP) in the sense of polynomials (numerical admissibility criteria; NAC). The physical admissibility criteria in this work were chosen in a form similar to the work~\cite{ader_dg_ideal_flows}: density, pressure (more precisely, internal energy) must be positive; the concentrations of all components must be non-negative; the normalization condition $\sum_{k} c_{k} = 1$ for the mass concentrations of the medium components must be satisfied (this automatically includes a check for the condition $c_{k} \leqslant 1$); the values calculated on the new time step must not be \texttt{nan}, which can occur due to a serious loss of accuracy. The numerical admissibility criteria, in the form of DMP, without significant changes are taken from works~\cite{ader_dg_ideal_flows, ader_dg_diss_flows}:
\begin{equation}
\begin{split}\label{dmp_formula}
\min\limits_{x' \in V_{i}}&\left(\mathbf{U}(x', t^{n})\right) - \mathbf{\delta}
\leqslant \mathbf{U}(x, t^{n+1}) \\
&\leqslant \max\limits_{x' \in V_{i}}\left(\mathbf{U}(x', t^{n})\right) + \mathbf{\delta},\quad \forall x \in \Omega_{i},
\end{split}
\end{equation}
where $V_{i} = \Omega_{i-1}\, \cup \Omega_{i}\, \cup \Omega_{i+1}$, $\mathbf{\delta}$ is chosen to be a solution-dependent tolerance given by
\begin{equation}
\begin{split}
\mathbf{\delta} = \max\Big[\delta_{0}, \epsilon\cdot\Big(&\max\limits_{x' \in V_{i}}\left(\mathbf{U}(x', t^{n})\right)\\ 
&- \min\limits_{x' \in V_{i}}\left(\mathbf{U}(x', t^{n})\right)\Big)\Big],
\end{split}
\end{equation}
where $\delta_{0} = 10^{-4}$ and $\epsilon = 10^{-3}$, in accordance with the recommendations of the works~\cite{ader_dg_ideal_flows, ader_dg_diss_flows}. The inequality (\ref{dmp_formula}) is checked separately for each value of the vector of conserved values. The $\min$ and $\max$ functions return vectors of conserved values. Violation of inequality for at least one value leads to violation of the criterion.


Simultaneous verification by the PAC and the NAC for a finite-element cell $\Omega_{i}$ determines that the numerical solution obtained by the space-time adaptive ADER finite-element DG method is admissible, while the cell is marked with indicator $\beta = 0$; otherwise, the cell is marked as ``troubled'', with indicator $\beta = 1$~\cite{ader_dg_ideal_flows}. In such ``troubled'' cells, a sub-grid with $N_{s}$ spatial sub-cells $\Omega_{i, k}$ was generated, where the piecewise-constant alternative data representation $\mathbf{v}_{h} = \hat{\mathit{P}} \mathbf{U}_{h}$ was constructed using the reversible conservative $L_{2}$-in\-ter\-po\-la\-ti\-on procedure with operator $\hat{\mathit{P}}$: 
\begin{equation}
\begin{split}\label{P_operator}
\mathbf{v}_{h, k} = \left[\hat{\mathit{P}} \mathbf{U}_{h}\right]_{k} = \frac{1}{|\Omega_{i, k}|}\int\limits_{\Omega_{i, k}} \mathbf{U}_{h}(x, t^{n}) dx,
\end{split}
\end{equation}
The verification of the PAC and the NAC, in algorithmic and software implementations, was carried out not on the DG-representation $\mathbf{U}_{h}$ of the solution, but immediately on the piecewise-constant alternative data representation $\mathbf{v}_{h}$; for which a small tolerance $\mathbf{\delta}$ was included in the inequalities of the DMP (see details in~\cite{ader_dg_ideal_flows, ader_dg_diss_flows}). 

The solution in ``troubled'' cells was recalculated using the stable high-precision ADER-WENO-FV scheme, that was chosen in the classical form~\cite{ader_dg_ideal_flows, ader_dg_dev_1, ader_dg_dev_2, ader_weno_lstdg, ader_dg_diss_flows, ader_dg_ale, ader_dg_grmhd, ader_dg_gr_prd, ader_dg_PNPM, PNPM_DG, ader_dg_eff_impl, fron_phys, exahype, ader_dg_hpc_impl_1, ader_dg_hpc_impl_2, ader_dg_hpc_impl_3, ader_dg_hpc_impl_4}:
\begin{equation}
\begin{split}\label{ADER_WENO_FV_scheme}
\Big(\mathbf{v}^{n+1}_{h, k}& - \mathbf{v}^{n}_{h, k}\Big)\cdot\tau^{n} \hspace{1mm} \\
+ &\int\limits_{t^{n}}^{t^{n+1}}\hspace{-2mm}\left[
\mathbf{\mathfrak{G}}\Big(\mathbf{q}_{h}^{R}(x_{i-\frac12}, t), \mathbf{q}_{h}(x_{i+\frac12}, t)\Big) \right.\\
&\hspace{2mm}-\left.\mathbf{\mathfrak{G}}\Big(\mathbf{q}_{h}(x_{i-\frac12}, t), \mathbf{q}_{h}^{L}(x_{i+\frac12}, t)\Big)\right] dt \\
= & \int\limits_{t^{n}}^{t^{n+1}}\hspace{-2mm}\int\limits_{\Omega_{i}}
\mathbf{S}\Big(\mathbf{q}_{h}(x, t)\Big) dx dt,\\
\end{split}
\end{equation}
where the introduced notation corresponds to the notation for the formula (\ref{ADER_DG_scheme}); $\tau^{n} = t^{n+1} - t^{n}$ is the time step. The main difference is that the discrete space-time solution $\mathbf{q}_{h}(x, t)$ is obtained by the LST-DG-predictor based on the WENO-reconstruction $\mathbf{w}_{h}(x, t^{n})$ as an initial condition, instead of DG-representation $\mathbf{U}^{n}_{h}$. The resulting finite-volume solution, in the form of the pi\-e\-ce\-wi\-se-constant alternative data representation $\mathbf{v}^{n+1}_{h}$, is converted into a DG-representation $\mathbf{U}^{n+1}_{h}$ using the inverse transformation specified by the suitable high order accurate reconstruction operator $\hat{\mathit{R}}$, which determines the relation $\mathbf{U}^{n+1}_{h} = \hat{\mathit{R}} \mathbf{v}^{n+1}_{h}$. The suitable high order accurate reconstruction operator $\hat{\mathit{R}}$ is implemented using the least squares method implemented using a pseudo-inverse matrix.

It is important to note that the operators $\hat{\mathit{P}}$ and $\hat{\mathit{R}}$ satisfy the reversibility condition $\hat{\mathit{R}} \circ \hat{\mathit{P}} = 1$, however, this condition is satisfied only in one direction: direct transformation is reversible $\mathbf{U}_{h} \rightarrow \mathbf{v}_{h} \rightarrow \mathbf{U}_{h}$, but the reverse transformation is irreversible, in the general case, $\mathbf{v}_{h} \rightarrow \mathbf{U}_{h} \rightarrow \mathbf{v}^{*}_{h} \neq \mathbf{v}_{h}$; so there is the relation $\hat{\mathit{P}} \circ \hat{\mathit{R}} \neq 1$. This is due to the fact that the amount of sub-cells $N_{s}$ of sub-grid in cell is greater than the amount of coefficients $N+1$ of the DG-representation, and the system of equations for obtaining the DG-representation $\mathbf{U}_{h}$ from the pi\-e\-ce\-wi\-se-constant representation $\mathbf{v}_{h}$ is overdetermined (thus, it is solved by the least squares method, using a pseudo-inverse matrix), and the system of equations for obtaining pi\-e\-ce\-wi\-se-constant representation $\mathbf{v}_{h}$ from a DG-representation $\mathbf{U}_{h}$ is un\-der\-de\-ter\-min\-ed. Therefore, in algorithmic and software implementations, each cell is associated with the status (such as \texttt{is\_dg\_the\_primary\_representation}) that de\-ter\-mi\-nes which of the DG-representations of the solution at the time step $t^{n}$ is primary --- whether the DG-representation $\mathbf{U}^{n}_{h}$ is obtained as a result of using the space-time adaptive ADER finite-element DG scheme and is its solution, or the DG-representation $\mathbf{U}^{n}_{h}$ is the result of a transformation from a pi\-e\-ce\-wi\-se-constant representation $\mathbf{v}_{h}$, using an operator $\hat{\mathit{R}}$. This is an important aspect of the implementation, since the next time step $t^{n+1}$ of ADER-DG scheme may result in non-admissible DG-representation $\mathbf{U}^{n+1}_{h}$ (``troubled'' cell) of the solution, so it will be necessary to recalculate the solution by the ADER-WENO-FV scheme, which will require a solution $\mathbf{v}^{n}_{h}$ at this time step $t^{n}$. Depending on which solution $\mathbf{v}^{n}_{h}$ is primary at this time step $t^{n}$, it will be used as a pi\-e\-ce\-wi\-se-constant representation for the ADER-WENO-FV scheme. Thus, if the cell remains ``troubled'' for several time steps, all these time steps will be performed only by the ADER-WENO-FV scheme, and there will be no problem with the irreversibility of the transformation operators $\hat{\mathit{P}} \circ \hat{\mathit{R}} \neq 1$.

The WENO-reconstruction in high-precision ADER-WENO-FV scheme was carried out according to the pi\-e\-ce\-wi\-se-constant alternative data representation, for each sub-cell, based on the expansion in basis functions (similar to the representation (\ref{DG_representation}), however, in this case, mapped into the space reference element $\omega = [0; 1]$ is performed separately for each sub-cell $\Omega_{i, k}$; so each ``troubled'' cell $\Omega_{i}$ contains $N_{s}$ separate representations of the WENO-reconstructions):
\begin{equation}\label{WENO_reconstruction}
\mathbf{w}_{h}(x, t^{n}) = 
\sum\limits_{p} \hat{\mathbf{w}}_{p}^{n} \varphi_{p}\left(\xi(x)\right),\;\;\;\;\;\;
\mathbf{r} \in \Omega_{i} \leftrightarrow \mathbf{\xi} \in \omega,
\end{equation}
where $\hat{\mathbf{w}}_{p}^{n}$ are the WENO-reconstruction coefficients. A detailed description of the reconstruction procedure is presented in the work~\cite{ader_weno_lstdg}. The piecewise constant representation used to perform the reconstruction is taken from the piecewise constant representation of the given problem cell and two neighboring cells --- ``ghost'' subgrids are built in neighboring cells. Technically, in algorithmic and software implementations, the piecewise-constant representation of two neighboring cells is already computed, whether they are ``troubled'' or not; the piecewise-constant representation calculation is still performed for the verification of the PAC and the NAC of the cells; at the same time, NAC still requires the calculation of the piecewise-constant representation in neighboring cells. The resulting WENO-reconstruction $\mathbf{w}_{h}(x, t^{n})$ is used as an initial condition in the LST-DG-predictor to obtain a discrete space-time solution $\mathbf{q}_{h}(x, t)$ to be used in the high-precision ADER-WENO-FV scheme (\ref{ADER_WENO_FV_scheme}) for each sub-cell $\Omega_{i, k}$.

ADER paradigm and the solution of a Generalized Riemann Problem (GRP) in this computational scheme are based on the use of a local discrete space-time solution $\mathbf{q}_{h}(x, t)$. The space-time adaptive ADER-DG sche\-me (\ref{ADER_DG_scheme}) and ADER-WENO-FV scheme (\ref{ADER_WENO_FV_scheme}) use the discrete space-time solution $\mathbf{q}_{h}(x, t)$ in the space-time cell $\Omega_{i} \times [t^{n}, t^{n+1}]$ for ADER-DG scheme and in the space-time cell $\Omega_{i, k} \times [t^{n}, t^{n+1}]$ for ADER-WENO-FV scheme (map\-ped into the space-time reference element $[0; 1]\times[0; 1]$). The discrete space-time solution $\mathbf{q}_{h}(x, t)$ is chosen in the form~\cite{ader_dg_ideal_flows, ader_dg_dev_1, ader_dg_dev_2, ader_weno_lstdg, ader_dg_diss_flows, ader_dg_ale, ader_dg_grmhd, ader_dg_gr_prd, ader_dg_PNPM, PNPM_DG, ader_dg_eff_impl, fron_phys, exahype, ader_dg_hpc_impl_1, ader_dg_hpc_impl_2, ader_dg_hpc_impl_3, ader_dg_hpc_impl_4}:
\begin{equation}
\begin{split}
&\mathbf{q}(x, t) = 
\sum\limits_{\mathfrak{p}} \hat{\mathbf{q}}_{\mathfrak{p}}
\Phi_{\mathfrak{p}}\big(\xi(x), \tau(t)\big),\\
&(\mathbf{\xi}, \tau) \in [0; 1]\times[0; 1],
\end{split}
\end{equation}
where basis functions $\Phi_{\mathfrak{p}}(\xi, \tau)$ are chosen as tensor-products of Lagrange interpolation polynomials $\varphi_{n}(\xi)$ with degree $N$, which were also used in the DG-re\-pre\-sen\-ta\-ti\-on (\ref{DG_representation}) and WENO-reconstruction (\ref{WENO_reconstruction}):
\begin{equation}
\Phi_{\mathfrak{p}}\big(\xi(x), \tau(t)\big) = \varphi_{p_{0}}\big(\tau(t)\big)\ \varphi_{p_{1}}\big(\xi(x)\big), 
\end{equation}
where tensor index $\mathfrak{p} = (p_{0}, p_{1})$, and $0 \leqslant p_{0},\ p_{1} \leqslant N$. In this case, the following method of solving the GRP is used: first evolve the data locally in the small inside each element and then interact the evolved data at the element interfaces via a classical Riemann solver~\cite{PNPM_DG, ADER_GRP_LST_DG_1, ADER_GRP_LST_DG_2, ADER_GRP_LST_DG_3, ADER_GRP_LST_DG_4, ADER_GRP_LST_DG_5} (references in work~\cite{ader_dg_ideal_flows}).

The LST-DG predictor for obtaining the discrete space-time solution $\mathbf{q}_{h}(\xi(x), \tau(t))$ in the space-time control volume $\Omega \times [t^{n}, t^{n+1}]$ is chosen~\cite{ader_dg_ideal_flows, ader_dg_dev_1, ader_dg_dev_2, ader_weno_lstdg, ader_dg_diss_flows, ader_dg_ale, ader_dg_grmhd, ader_dg_gr_prd, ader_dg_PNPM, PNPM_DG, ader_dg_eff_impl, fron_phys, exahype, ader_dg_hpc_impl_1, ader_dg_hpc_impl_2, ader_dg_hpc_impl_3, ader_dg_hpc_impl_4} as
\begin{equation}\label{lstdg_full_form}
\begin{split}
\Bigg[&\int\limits_{0}^{1} \Phi_{\mathfrak{p}}(\mathbf{\xi},1)
\Phi_{\mathfrak{q}}(\mathbf{\xi},1) d\mathbf{\xi}\\
&-\int\limits_{0}^{1}\int\limits_{0}^{1} \frac{\partial\Phi_{\mathfrak{p}}(\mathbf{\xi},\tau)}{\partial\tau}
\Phi_{\mathfrak{q}}(\mathbf{\xi},\tau) d\mathbf{\xi} d\tau \Bigg]\cdot \hat{\mathbf{q}}_{\mathfrak{q}}\\
+& \int\limits_{0}^{1}\int\limits_{0}^{1} \Phi_{\mathfrak{p}}(\mathbf{\xi},\tau)
\frac{\partial\Phi_{\mathfrak{p}}(\mathbf{\xi},\tau)}{\partial\xi} d\mathbf{\xi} d\tau\cdot
\tilde{\mathbf{F}}(\hat{\mathbf{q}}_{\mathfrak{q}}) \\
&\hspace{5mm}=\int\limits_{0}^{1} \Phi_{\mathfrak{p}}(\mathbf{\xi},0)
\Phi_{k}(\mathbf{\xi}) d\mathbf{\xi}\cdot \hat{\mathbf{u}}_{k}\\
&\hspace{5mm}+ \int\limits_{0}^{1}\int\limits_{0}^{1} \Phi_{\mathfrak{p}}(\mathbf{\xi},\tau)
\Phi_{\mathfrak{q}}(\mathbf{\xi},\tau) d\mathbf{\xi} d\tau\cdot \tilde{\mathbf{S}}(\hat{\mathbf{q}}_{\mathfrak{q}}),
\end{split}
\end{equation}
where summation is assumed over the tensor index $\mathfrak{q} = (q_{0}, q_{1})$ and scalar index $k$ (Einstein notation); $\hat{\mathbf{u}}_{k}$ is the representation coefficients of the initial conditions (at this time step $t^{n}$): for the case of the ADER-DG scheme $\hat{\mathbf{u}}_{k}$ is the DG-representation coefficients --- $\hat{\mathbf{u}}_{k} = \hat{\mathbf{U}}^{n}_{k}$; for the case of the ADER-WENO-FV scheme $\hat{\mathbf{u}}_{k}$ is the WENO-reconstruction coefficients --- $\hat{\mathbf{u}}_{k} = \hat{\mathbf{w}}^{n}_{k}$. Functions $\tilde{\mathbf{F}}$ and $\tilde{\mathbf{S}}$ are rescaled flux $\mathbf{F}$ and source $\mathbf{S}$ terms due to mapping $\Omega \times [t^{n}, t^{n+1}] \leftrightarrow [0; 1]\times[0; 1]$; it is important to note here that in the cases of the ADER-DG scheme and the ADER-WENO-FV scheme, the spatial step of the cells differs --- the cell step $h$ and the sub-cell step $h_{s}$ differ by $N_{s}$ times; so the rescaling factors will be different. The LST-DG predictor is used to obtain the discrete space-time solution $\mathbf{q}_{h}(\xi(x), \tau(t))$ in each cell $\Omega_{i}$ in the case of a ADER-DG scheme, as well as in each subcell $\Omega_{i, k}$ of each ``troubled'' cell $\Omega_{i}$ (and also, in certain cases, in nearby sub-cells to the ``troubled'' cells, to calculate the left flux terms in the left sub-cell and right flux terms in the right sub-cell, if neighboring cells are not ``troubled'' and WENO-reconstruction is not required in them). LST-DG predictor (\ref{lstdg_full_form}) takes a compact notation~\cite{Zanotti_lectures_2016, Jackson_2017}
\begin{equation}\label{lstdg_compact_form}
\mathbb{K}_{\mathfrak{p}\mathfrak{q}}^{\tau} \hat{\mathbf{q}}_{\mathfrak{q}} +
\mathbb{K}_{\mathfrak{p}\mathfrak{q}}^{\xi} \tilde{\mathbf{F}}(\hat{\mathbf{q}}_{\mathfrak{q}}) =
\mathbb{F}_{\mathfrak{p},k} \hat{\mathbf{u}}_{k} +
\mathbb{M}_{\mathfrak{p}\mathfrak{q}} \tilde{\mathbf{S}}(\hat{\mathbf{q}}_{\mathfrak{q}}),
\end{equation}
where $\mathbb{K}_{\mathfrak{p}\mathfrak{q}}^{\tau}$, $\mathbb{K}_{\mathfrak{p}\mathfrak{q}}^{\xi}$, $\mathbb{F}_{\mathfrak{p}\mathfrak{h}}$ and $\mathbb{M}_{\mathfrak{p}\mathfrak{q}}$ are the matrices of integrals of $\varphi_{p}(\xi)$ and $d\varphi_{p}(\xi)/d\xi$, which have a convenient index structure due to the orthogonality of the set of basis functions, and can be pre-computed in the code. The process of obtaining a solution $\hat{\mathbf{q}}_{\mathfrak{p}}$ to system of nonlinear algebraic equations (\ref{lstdg_compact_form}) with stiff terms can be organized using one of the iterative scheme~\cite{ader_dg_ideal_flows, ader_dg_dev_1, ader_dg_dev_2}
\begin{equation}\label{stiff_iter}
\begin{split}
\hat{\mathbf{q}}^{(i+1)}_{\mathfrak{p}} - &
\left[\left(\mathbb{K}^{\tau}\right)^{-1}\right]_{\mathfrak{p}\mathfrak{q}}
\left[\mathbb{M}_{\mathfrak{q}\mathfrak{r}} \tilde{\mathbf{S}}\left(\hat{\mathbf{q}}^{(i+1)}_{\mathfrak{r}}\right)\right] =\\
&\left[\left(\mathbb{K}^{\tau}\right)^{-1}\right]_{\mathfrak{p}\mathfrak{q}} \left[
	\mathbb{F}_{\mathfrak{q},k} \hat{\mathbf{u}}_{k} -
	\mathbb{K}_{\mathfrak{q}\mathfrak{r}}^{\xi} \tilde{\mathbf{F}}\left(\hat{\mathbf{q}}^{(i)}_{\mathfrak{r}}\right)
\right],
\end{split}
\end{equation}
where matrix products on the right side have certain properties: the matrix $[(\mathbb{K}^{\tau})^{-1}]_{\mathfrak{p}\mathfrak{q}} \mathbb{K}_{\mathfrak{q}\mathfrak{r}}^{\xi}$ has zero eigenvalues~\cite{Zanotti_lectures_2016, Jackson_2017}, therefore, in the case of a problem without source terms, a simple iterative procedure is converged to a fixed point (Banach fixed point theorem); the matrix $[(\mathbb{K}^{\tau})^{-1}]_{\mathfrak{p}\mathfrak{q}} \mathbb{F}_{\mathfrak{q},k}$ also has an interesting property:
\begin{equation}
\left\{\left[\left(\mathbb{K}^{\tau}\right)^{-1}\right]_{\mathfrak{p}\mathfrak{q}} \mathbb{F}_{\mathfrak{q},k}\right\}_{p_{0}, p_{1}, k} = \mathbb{I}_{p_{0}} \delta_{p_{1}, k},
\end{equation}
where $\delta_{p_{1}, k}$ is the Kronecker delta symbol and $\mathbb{I}_{p_{0}} = 1$ is just a one; which is a consequence of the correspondence principle: in the case of null fluxes terms $\mathbf{F} \equiv 0$ and null sources terms $\mathbf{S} \equiv 0$, the solution $\mathbf{q}(x, t)$ should not depend on time $t$ and for the representation coefficients the solution $\hat{\mathbf{q}}_{p_{0}, k} = \hat{\mathbf{u}}_{p}$ can be obtained. In the present work, the system of equations was solved by an iterative method, the left side of the system of equations was solved by an internal iterative procedure. In the case of a weakly stiff terms of the problem, the internal iterative procedure was a simple fixed point method. In the case of a strong stiff terms of the problem, the internal iterative procedure was implemented by Newton's method.

Despite the fact that the methods used in this paper are locally implicit, the resulting ADER-DG and ADER-WENO-FV schemes are explicit (local implicitness of the LST-DG predictor step). Therefore, the Courant-Friedrichs-Lewy (CFL) condition is imposed on the time discretization step $\tau^{n} = t^{n+1} - t^{n}$, which takes the form~\cite{ader_dg_ideal_flows, ADER_DG_time_step_1, ADER_DG_time_step_2}:
\begin{equation}
\begin{split}
&\tau^{n}_{DG} = \frac{1}{2N + 1}\frac{h}{|\lambda_{max}|},\\
&\tau^{n}_{FV} = \frac{h_{s}}{|\lambda_{max}|},
\end{split}
\end{equation}
where $|\lambda_{max}|$ is the maximum signal velocity, $h$ is the spatial characteristic mesh size and $h_{s} = h/N_{s}$ is the spatial characteristic sub-grid size. If $N_{s} = 2N + 1$ then the discretization time steps $\tau^{n}_{DG} = \tau^{n}_{FV}$~\cite{ader_dg_ideal_flows}, therefore, in this work, the condition $N_{s} = 2N + 1$ was chosen.

The software implementation was made using the C++ programming language. Multithreading execution was provided using the OpenMP standard. Parallelization by threads was performed primitively --- loops were parallelized using OpenMP pragmas, with the allocation of critical code sections and private and shared variables. Multiprocessing execution and inter-process communication was provided within the framework of the MPI standard. Parallelization by processes was also performed primitively --- the full spatial mesh was divided into ranges, in each of which calculations were performed by separate processes; at each step, data was transferred from the boundary cells of the selected ranges. Load balancing of processes and threads in the parallelized version of the program was not performed, which in some cases led to a slight idle time of computing resources --- the iterative processes of obtaining discrete space-time solutions by the LST-DG predictor for various initial states often converged for a different number of iterations, but this difference was not significant in terms of computation time.

\section{Applications of the numerical method}
\label{sec:2}

\subsection{Accuracy and convergence}
\label{subsec:2:1}

\paragraph{Formulation of the problem}

\begin{figure*}
\centering
\includegraphics[width=0.32\textwidth]{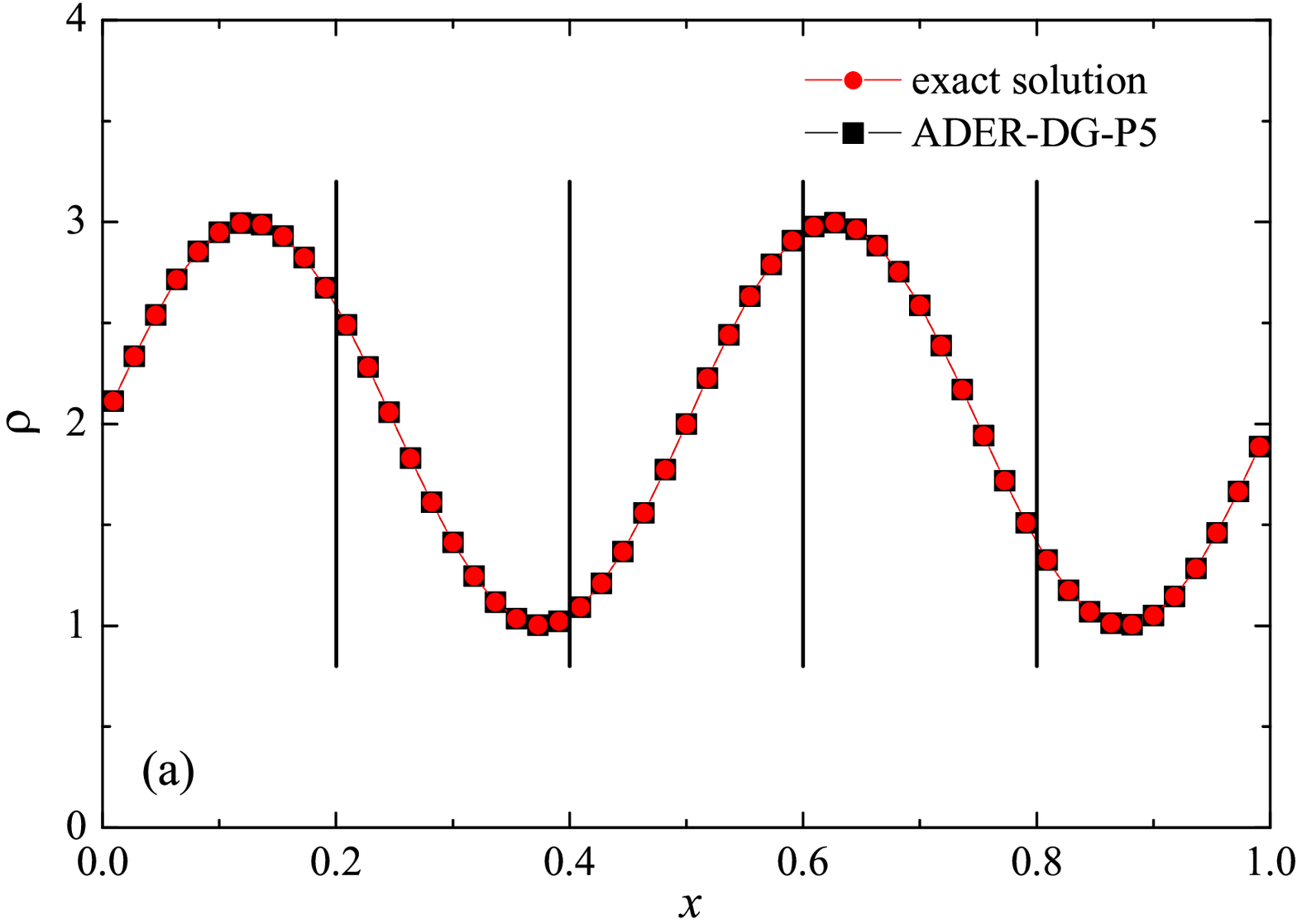}
\includegraphics[width=0.32\textwidth]{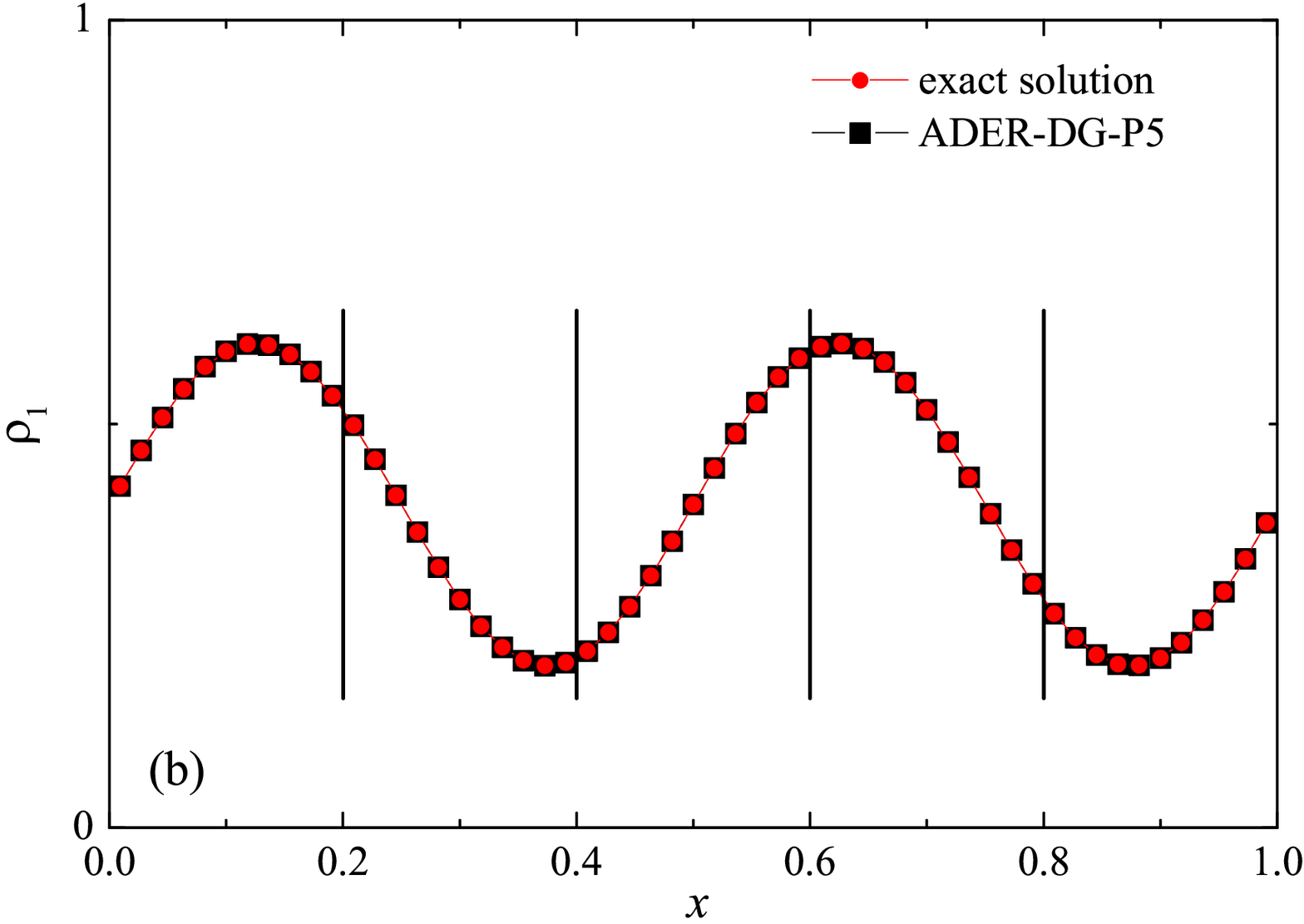}
\includegraphics[width=0.32\textwidth]{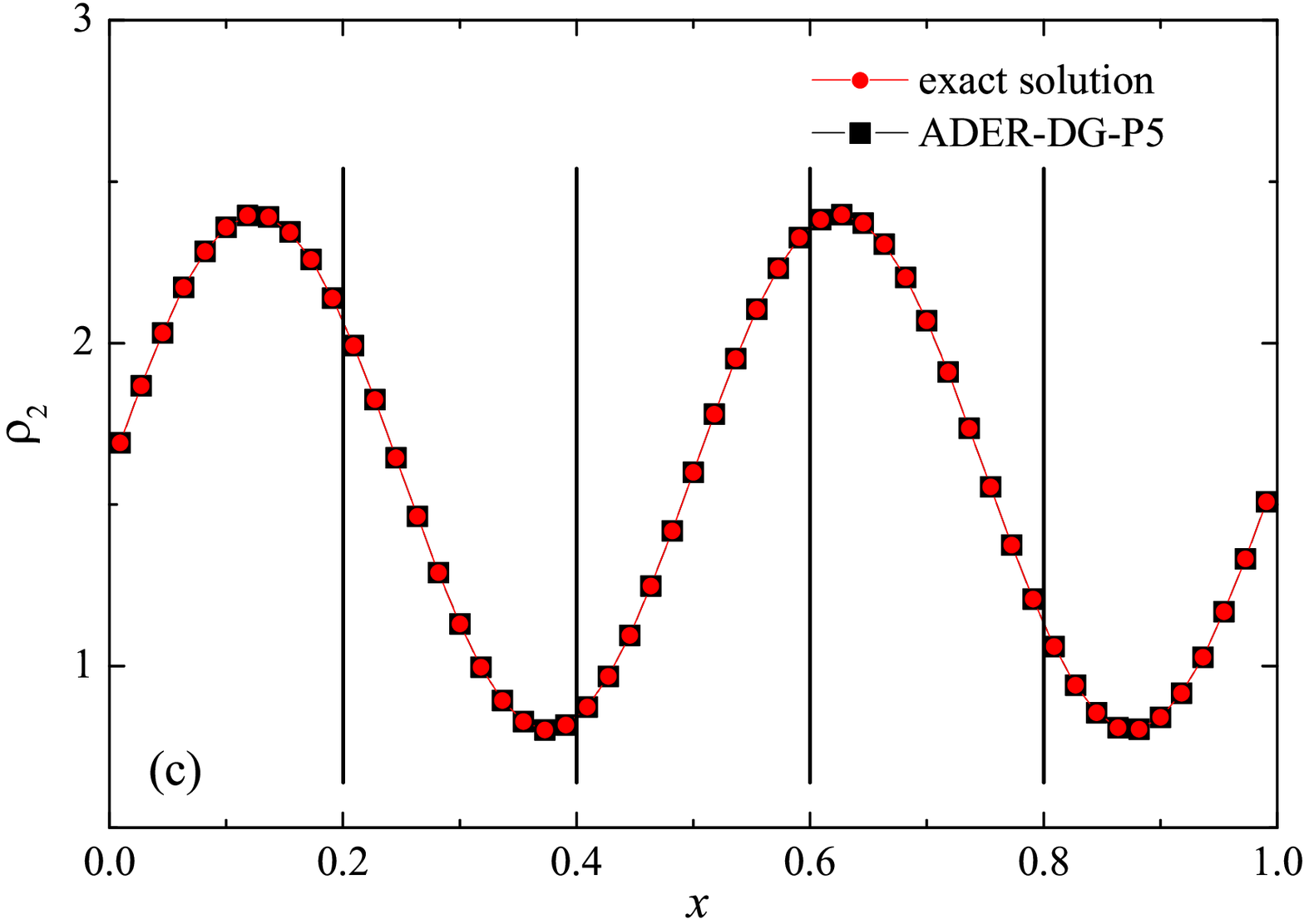}\\
\includegraphics[width=0.32\textwidth]{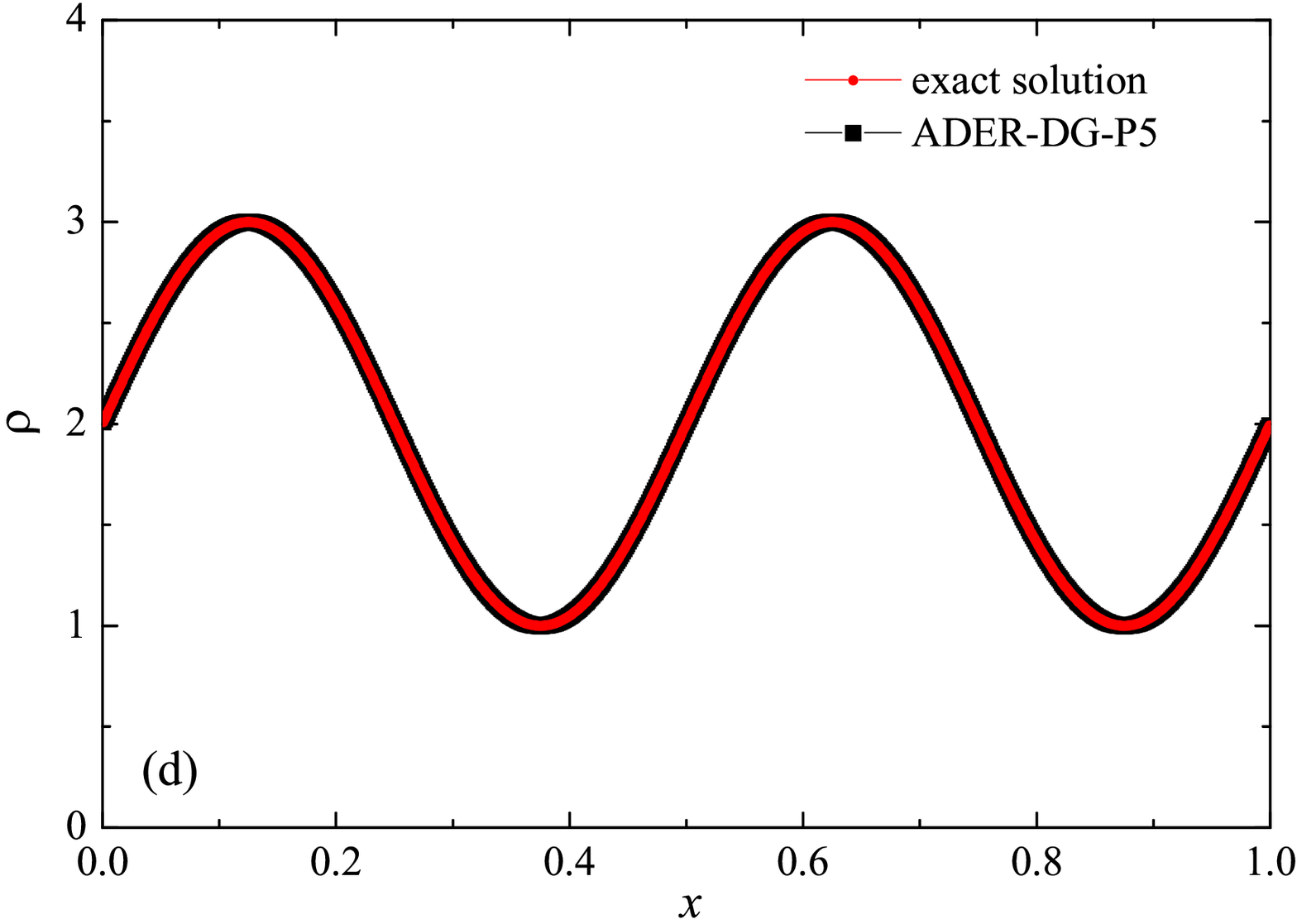}
\includegraphics[width=0.32\textwidth]{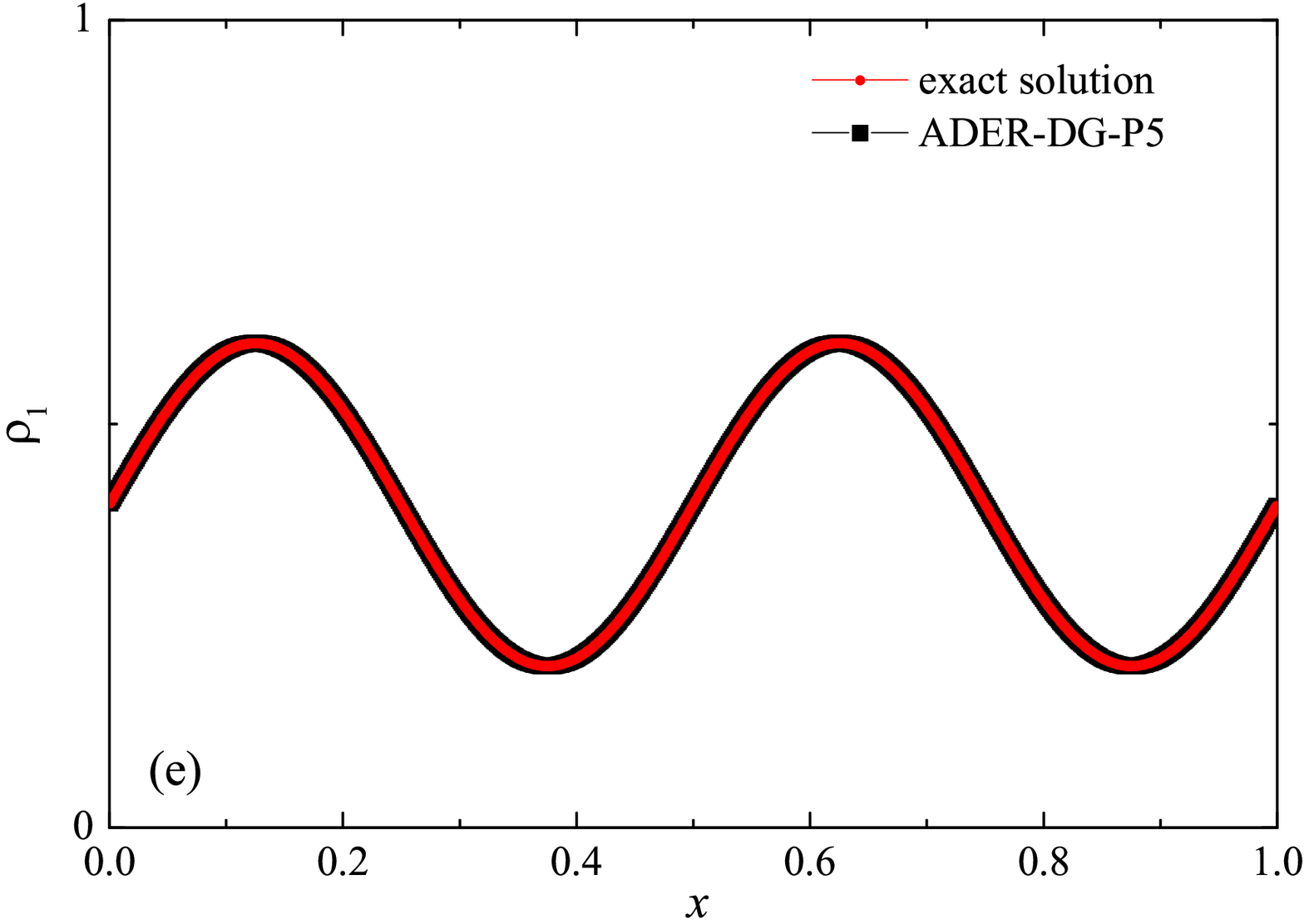}
\includegraphics[width=0.32\textwidth]{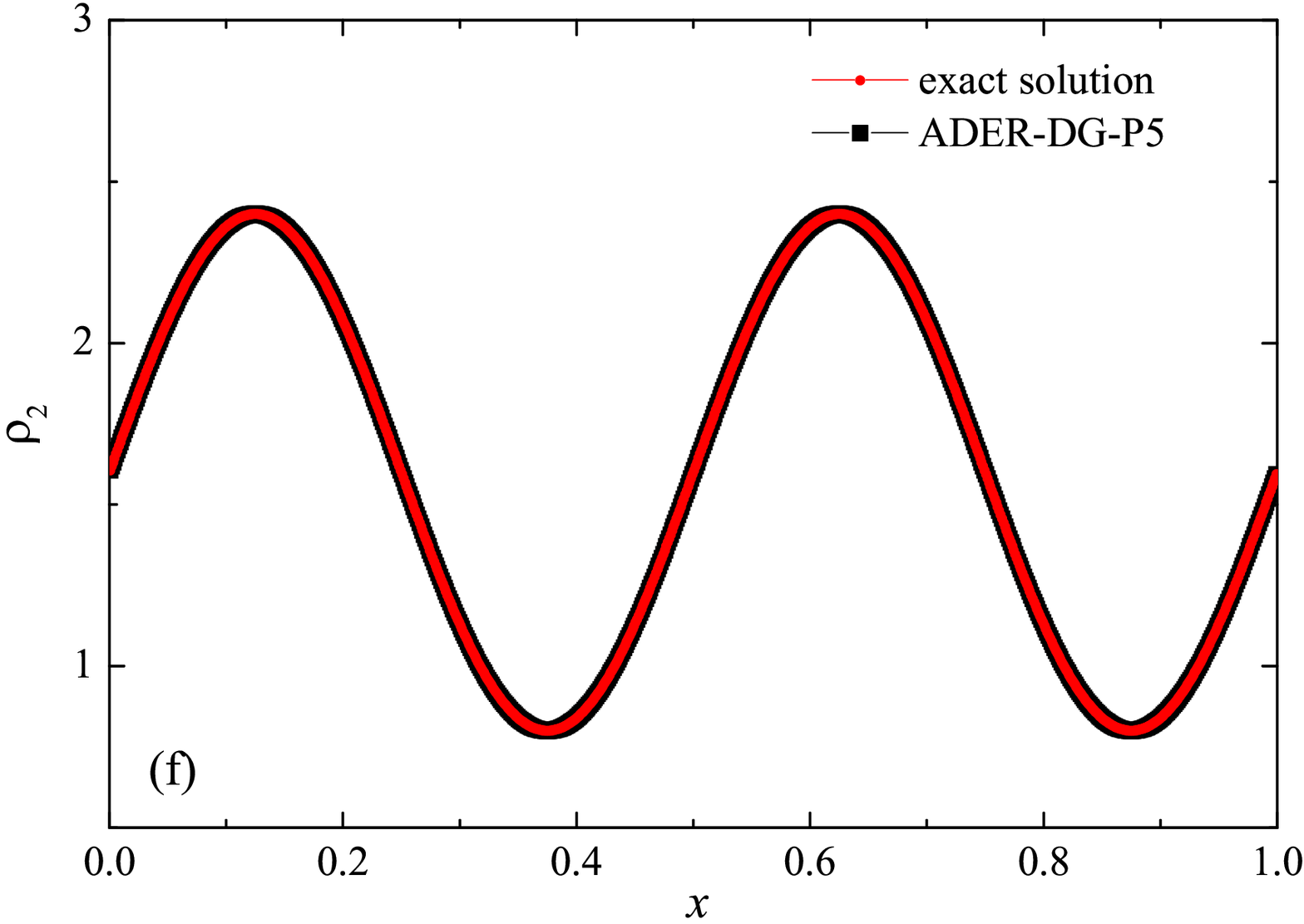}\\
\caption{%
Numerical solution of the classical problem of advection flow for a multicomponent medium 
(a detailed statement of the problem is presented in the text),
using the computational scheme $\mathrm{ADER}$-$\mathrm{DG}$-$\mathbb{P}_5$ with a posteriori 
limitation of the solution by a $\mathrm{ADER}$-$\mathrm{WENO}5$ finite volume limiter,
on a coordinate mesh with $5$ (top figures) and $800$ (bottom figures) finite element cells.
Each finite element contains $N_{s} = 11$ subcells. 
In the top figures, vertical lines indicate the coordinate boundaries of each finite element.
The graphs show the coordinate dependencies of subcells finite-volume representation of density $\rho$ (a, d) and 
densities $\rho_{k} = \rho c_{k}$ (b, c, e, f) of individual components $k = 1,\, 2$ of the multicomponent medium,
at the final time $t_{\rm final} = 1.0$.
The black square symbols represent the numerical solution; 
the red circle symbols represents the exact analytical solution of the problem.
}
\label{fig:adv_test_fig}
\end{figure*} 

\begin{table*}[h!]
\centering
\caption{%
$L^{1}$, $L^{2}$ and $L^{\infty}$ norms of errors and convergence rates orders
by density $\rho$, computational scheme $\mathrm{ADER}$-$\mathrm{DG}$-$\mathbb{P}_{N}$ with a posteriori 
limitation of the solution by a $\mathrm{ADER}$-$\mathrm{WENO} (\mathbb{P}_{N})$ finite volume limiter.
}
\label{tab:adv_test_fig}
\begin{tabular}{|c|l|ccc|ccc|c|}
\hline
& cells
& ~~~~~~$L^{1}$ error~~~
& ~~~$L^{2}$ error~~~
& ~~~$L^{\infty}$ error~~~~~~
& $L^{1}$ order
& $L^{2}$ order
& $L^{\infty}$ order
& ~~theor.~~\\
\hline
ADER-DG-$\mathbb{P}_{1}$	&	$5$	&	$4.4476\mathrm{E}\!-\!01$	&	$5.0488\mathrm{E}\!-\!01$	&	$6.7112\mathrm{E}\!-\!01$		&	$-$	&	$-$	&	$-$	&	$2$\\
&	$10$	&	$3.6731\mathrm{E}\!-\!01$	&	$4.0975\mathrm{E}\!-\!01$	&	$5.5141\mathrm{E}\!-\!01$		&	$0.276$	&	$0.301$	&	$0.283$	&	\\
&	$25$	&	$8.5188\mathrm{E}\!-\!02$	&	$1.0203\mathrm{E}\!-\!01$	&	$2.0990\mathrm{E}\!-\!01$		&	$1.595$	&	$1.517$	&	$1.054$	&	\\
&	$50$	&	$1.8983\mathrm{E}\!-\!02$	&	$2.6755\mathrm{E}\!-\!02$	&	$6.8839\mathrm{E}\!-\!02$		&	$2.166$	&	$1.931$	&	$1.608$	&	\\
&	$100$	&	$2.8155\mathrm{E}\!-\!03$	&	$4.9650\mathrm{E}\!-\!03$	&	$1.7083\mathrm{E}\!-\!02$		&	$2.753$	&	$2.430$	&	$2.011$	&	\\
&	$200$	&	$1.4200\mathrm{E}\!-\!04$	&	$1.6443\mathrm{E}\!-\!04$	&	$2.4239\mathrm{E}\!-\!04$		&	$4.309$	&	$4.916$	&	$6.139$	&	\\
&	$400$	&	$3.4883\mathrm{E}\!-\!05$	&	$4.0442\mathrm{E}\!-\!05$	&	$5.9769\mathrm{E}\!-\!05$		&	$2.025$	&	$2.024$	&	$2.020$	&	\\
&	$800$	&	$8.6832\mathrm{E}\!-\!06$	&	$1.0056\mathrm{E}\!-\!05$	&	$1.4869\mathrm{E}\!-\!05$		&	$2.006$	&	$2.008$	&	$2.007$	&	\\
\hline
ADER-DG-$\mathbb{P}_{2}$	&	$5$	&	$5.5317\mathrm{E}\!-\!02$	&	$6.0282\mathrm{E}\!-\!02$	&	$8.6083\mathrm{E}\!-\!02$		&	$-$	&	$-$	&	$-$	&	$3$\\
&	$10$	&	$1.3787\mathrm{E}\!-\!02$	&	$1.5171\mathrm{E}\!-\!02$	&	$1.9636\mathrm{E}\!-\!02$		&	$2.004$	&	$1.990$	&	$2.132$	&	\\
&	$25$	&	$5.7508\mathrm{E}\!-\!04$	&	$7.0939\mathrm{E}\!-\!04$	&	$1.2220\mathrm{E}\!-\!03$		&	$3.467$	&	$3.343$	&	$3.031$	&	\\
&	$50$	&	$5.1458\mathrm{E}\!-\!05$	&	$6.5010\mathrm{E}\!-\!05$	&	$1.4852\mathrm{E}\!-\!04$		&	$3.482$	&	$3.448$	&	$3.041$	&	\\
&	$100$	&	$4.6608\mathrm{E}\!-\!06$	&	$5.1810\mathrm{E}\!-\!06$	&	$7.3933\mathrm{E}\!-\!06$		&	$3.465$	&	$3.649$	&	$4.328$	&	\\
&	$200$	&	$5.8233\mathrm{E}\!-\!07$	&	$6.4685\mathrm{E}\!-\!07$	&	$9.1489\mathrm{E}\!-\!07$		&	$3.001$	&	$3.002$	&	$3.015$	&	\\
&	$400$	&	$7.2854\mathrm{E}\!-\!08$	&	$8.0903\mathrm{E}\!-\!08$	&	$1.1444\mathrm{E}\!-\!07$		&	$2.999$	&	$2.999$	&	$2.999$	&	\\
&	$800$	&	$9.1131\mathrm{E}\!-\!09$	&	$1.0119\mathrm{E}\!-\!08$	&	$1.4315\mathrm{E}\!-\!08$		&	$2.999$	&	$2.999$	&	$2.999$	&	\\
\hline
ADER-DG-$\mathbb{P}_{3}$	&	$5$	&	$5.7454\mathrm{E}\!-\!03$	&	$6.5334\mathrm{E}\!-\!03$	&	$8.7552\mathrm{E}\!-\!03$		&	$-$	&	$-$	&	$-$	&	$4$\\
&	$10$	&	$4.1255\mathrm{E}\!-\!04$	&	$4.8525\mathrm{E}\!-\!04$	&	$8.3217\mathrm{E}\!-\!04$		&	$3.800$	&	$3.751$	&	$3.395$	&	\\
&	$25$	&	$7.3836\mathrm{E}\!-\!06$	&	$8.4286\mathrm{E}\!-\!06$	&	$1.5863\mathrm{E}\!-\!05$		&	$4.391$	&	$4.423$	&	$4.322$	&	\\
&	$50$	&	$4.1056\mathrm{E}\!-\!07$	&	$4.7356\mathrm{E}\!-\!07$	&	$7.0085\mathrm{E}\!-\!07$		&	$4.169$	&	$4.154$	&	$4.500$	&	\\
&	$100$	&	$2.5826\mathrm{E}\!-\!08$	&	$2.9882\mathrm{E}\!-\!08$	&	$4.4280\mathrm{E}\!-\!08$		&	$3.991$	&	$3.986$	&	$3.984$	&	\\
&	$200$	&	$1.6177\mathrm{E}\!-\!09$	&	$1.8749\mathrm{E}\!-\!09$	&	$2.7768\mathrm{E}\!-\!09$		&	$3.997$	&	$3.994$	&	$3.995$	&	\\
&	$400$	&	$1.0126\mathrm{E}\!-\!10$	&	$1.1738\mathrm{E}\!-\!10$	&	$1.7383\mathrm{E}\!-\!10$		&	$3.998$	&	$3.998$	&	$3.998$	&	\\
&	$800$	&	$6.3580\mathrm{E}\!-\!12$	&	$7.3703\mathrm{E}\!-\!12$	&	$1.1028\mathrm{E}\!-\!11$		&	$3.993$	&	$3.993$	&	$3.978$	&	\\
\hline
ADER-DG-$\mathbb{P}_{4}$	&	$5$	&	$2.3897\mathrm{E}\!-\!04$	&	$2.7028\mathrm{E}\!-\!04$	&	$4.0639\mathrm{E}\!-\!04$		&	$-$	&	$-$	&	$-$	&	$5$\\
&	$10$	&	$1.3196\mathrm{E}\!-\!05$	&	$1.4964\mathrm{E}\!-\!05$	&	$2.1046\mathrm{E}\!-\!05$		&	$4.179$	&	$4.175$	&	$4.271$	&	\\
&	$25$	&	$1.5401\mathrm{E}\!-\!07$	&	$1.7074\mathrm{E}\!-\!07$	&	$2.4186\mathrm{E}\!-\!07$		&	$4.857$	&	$4.882$	&	$4.874$	&	\\
&	$50$	&	$4.9460\mathrm{E}\!-\!09$	&	$5.4955\mathrm{E}\!-\!09$	&	$7.7642\mathrm{E}\!-\!09$		&	$4.961$	&	$4.957$	&	$4.961$	&	\\
&	$100$	&	$1.5594\mathrm{E}\!-\!10$	&	$1.7351\mathrm{E}\!-\!10$	&	$2.7513\mathrm{E}\!-\!10$		&	$4.987$	&	$4.985$	&	$4.819$	&	\\
&	$200$	&	$4.9148\mathrm{E}\!-\!12$	&	$5.4550\mathrm{E}\!-\!12$	&	$7.8031\mathrm{E}\!-\!12$		&	$4.988$	&	$4.991$	&	$5.140$	&	\\
&	$400$	&	$1.4970\mathrm{E}\!-\!13$	&	$1.6845\mathrm{E}\!-\!13$	&	$3.2685\mathrm{E}\!-\!13$		&	$5.037$	&	$5.017$	&	$4.577$	&	\\
\hline
ADER-DG-$\mathbb{P}_{5}$	&	$5$	&	$2.4059\mathrm{E}\!-\!05$	&	$2.5733\mathrm{E}\!-\!05$	&	$3.5695\mathrm{E}\!-\!05$		&	$-$	&	$-$	&	$-$	&	$6$\\
&	$10$	&	$5.8109\mathrm{E}\!-\!07$	&	$6.3412\mathrm{E}\!-\!07$	&	$8.8554\mathrm{E}\!-\!07$		&	$5.372$	&	$5.343$	&	$5.333$	&	\\
&	$25$	&	$2.5136\mathrm{E}\!-\!09$	&	$2.8697\mathrm{E}\!-\!09$	&	$4.2690\mathrm{E}\!-\!09$		&	$5.940$	&	$5.891$	&	$5.822$	&	\\
&	$50$	&	$3.9997\mathrm{E}\!-\!11$	&	$4.6238\mathrm{E}\!-\!11$	&	$6.8677\mathrm{E}\!-\!11$		&	$5.974$	&	$5.956$	&	$5.958$	&	\\
&	$100$	&	$1.8509\mathrm{E}\!-\!12$	&	$3.7825\mathrm{E}\!-\!12$	&	$1.3581\mathrm{E}\!-\!11$		&	$4.434$	&	$3.612$	&	$2.338$	&	\\
&	$200$	&	$6.1555\mathrm{E}\!-\!14$	&	$7.2808\mathrm{E}\!-\!14$	&	$1.6587\mathrm{E}\!-\!13$		&	$4.910$	&	$5.699$	&	$6.355$	&	\\
\hline
\end{tabular}
\end{table*} 

The computer program developed in this work implements the space-time adaptive ADER-DG finite element method with LST-DG predictor and a posteriori sub-cell WENO finite-volume limiting for simulation of non-stationary compressible multicomponent reactive flows. The first problem of testing the developed program that implements the numerical method was checking the accuracy and convergence of the computational scheme. In the work~\cite{ader_dg_ideal_flows} where the ADER-DG finite element method, with LST-DG predictor and a posteriori sub-cell limiting by ADER-WENO-FV scheme, was originally proposed, it was shown that the convergence of the numerical solution for smooth solutions has an order of convergence $N+1$ for polynomials with a degree $N$, while the order of convergence was calculated for the computational scheme with AMR. 

In this work, testing of the accuracy and convergence of the computational scheme was carried out using a gas-dynamic test with an advective flow. The spatial domain of the flow was chosen as $x \in [0.0, 1.0]$. The initial values of velocity $u_{\rm init} = 1.0$ and pressure $p_{\rm init} = 1.0$ were chosen to be constant for the entire spatial domain of the flow. The initial values of mass concentrations $c_{1, \mathrm{init}} = 0.2$ and $c_{2, \mathrm{init}} = 0.8$ were also chosen to be constant. The initial coordinate dependence of the density $\rho(x, t = 0) = \rho_{\rm init}(x)$ was chosen in the following form:
\begin{equation}
	\rho_{\rm init}(x) = 2 + \sin\left(4 \pi x\right).
\end{equation}
Periodic boundary conditions were set at the boundaries of the spatial domain of the flow. The solution of this problem represents the transfer for the coordinate dependencies of the density $\rho(x, t) = \rho_{\rm init}(x - u t)$ and component densities $\rho_{k}(x, t) = c_{k, \mathrm{init}}\cdot\rho_{\rm init}(x - u t)$ with a constant flow velocity $u$. In this case, the coordinate dependencies of the flow velocity $u(x, t)$, pressure $p(x, t)$, and mass concentrations $c_{k}(x, t)$ of the components do not change with time $t$ and remain constant values determined at the initial moment of time. As a result of choosing the conditions of the testing problem, the exact solutions for the flow density $\rho(x, t)$ and the component densities $\rho_{k}(x, t)$ are recovered to the coordinate dependence form after a time interval equal to an integer value. Testing of accuracy and convergence was carried out for the first recovery of the solution, with the final time of the simulation $t_{\rm final} = 1.0$. The study of the convergence of the numerical solution was carried out using a sequence of $8$ mesh sizes, with the number $N_{\rm cells}$ of finite element cells: $5$, $10$, $25$, $50$, $100$, $200$, $400$ and $800$. Numerical schemes ADER-DG-$\mathbb{P}_{N}$ with the degrees $N = 1$, $2$, $3$, $4$ and $5$ of polynomials were investigated. The sub-cell limiter for the numerical scheme ADER-DG-$\mathbb{P}_{N}$ was the ADER-WENO ($\mathbb{P}_{N}$) finite-volume scheme. 

\paragraph{Results}

The solution to the advection problem is shown in Fig.~\ref{fig:adv_test_fig} for two demonstration cases: $N_{\rm cells} = 5$ and $N_{\rm cells} = 800$, with degrees $N = 5$ of polynomials. The figure shows the coordinate dependencies of subcells finite-volume representation of density $\rho$ and densities $\rho_{k} = \rho c_{k}$ of individual components. The subcells finite-volume representation was calculated for the numerical solution and for the analytical solution (analytical calculation of integrals). The solution to the problem is smooth, so it was expected that the subcell limiter would never be activated, and the property was indeed satisfied in all cases of computation. The presented results show that the numerical method has a high subgrid resolution --- five finite element cells are enough to correctly represent such solution.

The assessment of accuracy and convergence was carried out in the classical norms $L^{1}$, $L^{2}$ and $L^{\infty}$ of error. Orders of convergence were calculated as ratios of logarithms $\ln\left(\varepsilon_{0}/\varepsilon_{1}\right)/\ln\left(N_{\mathrm{cells}, 1}/N_{\mathrm{cells}, 0}\right)$, where the errors $\varepsilon_{0}$ and $\varepsilon_{1}$ values correspond to the sequential numbers of cells $N_{\mathrm{cells}, 0}$ and $N_{\mathrm{cells}, 1}$, respectively. The error has been calculated relative to the available analytical solution at the time $t_{\rm final}$. The error was calculated for the density $\rho(x, t)$ and component densities $\rho_{k}(x, t)$. It should be noted that exact analytical integration over the coordinate was carried out in the calculation of the norm values (numerically represented in the calculation of the differences for the finite volume representation, in subcells, of the numerical solution and of the exact solution). The results of calculating the errors in these norms are presented in the Table~\ref{tab:adv_test_fig}, together with the calculated values of the orders of convergence and the expected theoretical values of the orders of convergence, which are equal to the formal orders of approximation of a smooth solution. Only the results for the error by density $\rho$ are presented, the results for the error by component densities $\rho_{k}$ are similar and do not have significant differences. There is a clear trend that with an increase in the degree $N$ of polynomials, there is a decrease in the error for a fixed number $N_{\rm cells}$ of finite element cells. There are no violations of the monotonicity of the decrease in the error with an increase in the values of $N$ or $N_{\rm cells}$. In cases $N \geqslant 4$, solutions for meshes of not all sizes are presented, because for larger meshes, round-off errors in the representation of floating point numbers make a significant contribution to the test results. In the case $N = 1$, there is some ``over-convergence'' for intermediate mesh sizes, with orders of convergence $\simeq 4$-$6$, but with an increase in the number of cells and reaching an asymptotic dependence, the order of convergence tends to theoretical value $2$. For all other values $N$ studied, a similar behavior is observed when the order of convergence is greater than the theoretical result, which disappears in the asymptotics of large values $N_{\rm cells}$, where there is a correspondence with the theoretical result $N+1$. It should be noted that for values $N = 4$-$5$ in the range $N_{\rm cells} \geqslant 200$, round-off errors already begin to affect.

Thus, as a result of this carried out study of accuracy and convergence, it was revealed that the computational scheme ADER-DG-$\mathbb{P}_{N}$ shows the expected~\cite{ader_dg_ideal_flows} high orders $N+1$ of convergence and the computer program correctly implements this numerical method. This testing applies mainly only to the computational scheme ADER-DG-$\mathbb{P}_{N}$, because for a smooth solution, the limiter, which is the computational scheme $\mathrm{ADER}$-$\mathrm{WENO} (\mathbb{P}_{N})$, has never been used in obtaining this numerical solution.

\subsection{Classical gas dynamics problems}
\label{subsec:2:2}

\paragraph{Formulation of the problems}

The computer program developed in this work implements the space-time adaptive ADER-DG finite element method with LST-DG predictor and a posteriori sub-cell WENO finite-volume limiting for simulation of non-stationary compressible multicomponent reactive flows. This subsection presents the results of calculating classical test cases based on exactly solvable classical Riemann problems~\cite{Toro_solvers_2009}. The developed simulation method and its implementation are designed to simulate multicomponent reacting medium, therefore, in the selected test cases, they are extended to the case of multicomponent medium, with convective transfer. These problems have well-known exact analytical solutions, since in the case of passive convective transport of components (with the same values of the adiabatic exponent $\gamma_{k}$, otherwise, the problem associated with the equation of state of a multicomponent gas becomes much more complicated; then, in the general case, the splitting of the system of equations into two independent systems is violated), the Euler equations are split off from the equations of convective transport of components, which can then be solved~\cite{Kulikovskii, Rozhdestvenskii_Janenko} on the basis of the obtained solution of the Euler equations. The mass concentrations of individual flow components, as well as the densities of individual components (see formula (\ref{convection_reaction_equations})), for a non-reacting medium satisfy the simple equations of convective transport:
\begin{equation}
\begin{split}
\frac{\partial \left(\rho \mathbf{c}\right)}{\partial t} + &\frac{\partial \left(\rho \mathbf{c} u\right)}{\partial t} \\
& = \rho \left[\frac{\partial \mathbf{c}}{\partial t} + u \frac{\partial \mathbf{c}}{\partial x}\right] +
\mathbf{c}\left[\frac{\partial \rho}{\partial t} + \frac{\partial \left(\rho u\right)}{\partial x}\right]\\
& = \rho \left[\frac{\partial \mathbf{c}}{\partial t} + u \frac{\partial \mathbf{c}}{\partial x}\right] = \mathbf{S} = 0;\\
& \Rightarrow \frac{\partial \mathbf{c}}{\partial t} + u \frac{\partial \mathbf{c}}{\partial x} = 0;
\end{split}
\end{equation}
where the continuity equation was explicitly taken into account. Thus, the solution is constant along the characteristic curves $\frac{dx}{dt} = u(x, t)$, the slopes of which are piecewise separated in the case of the classical Riemann problem of the Euler equations, for which the exact analytical solution is known (including the flow velocity $u = u(x, t)$). In the test problems chosen in this work, the coordinate dependencies of mass concentrations $\mathbf{c}(x, t)$ remain pi\-e\-ce\-wi\-se-constant over time $t$ in the areas to the left and to the right of the contact discontinuity, if it exists in the solution.

\begin{table}[h!]
\caption{%
Data for four Riemann problem tests.
The parameter values $(\rho_{L}, u_{L}, p_{L})$ correspond to the state of the flow to the left of the discontinuity; 
the parameter values $(\rho_{R}, u_{R}, p_{R})$ correspond to the state of the flow to the right of the discontinuity.
}
\label{tab:tests_table}
\begin{tabular}{|c||c|c|c||c|c|c|}
\noalign{\smallskip}\hline\noalign{\smallskip}
test & $\rho_{L}$ & $u_{L}$ & $p_{L}$ & $\rho_{R}$ & $u_{R}$ & $p_{R}$ \\
\noalign{\smallskip}\hline\noalign{\smallskip}
1 & $1.000$ & $0.000$ & $1.000$ & $0.125$ & $0.000$ & $0.100$ \\
\noalign{\smallskip}\hline\noalign{\smallskip}
2 & $0.445$ & $0.698$ & $3.528$ & $0.500$ & $0.000$ & $0.571$ \\
\noalign{\smallskip}\hline\noalign{\smallskip}
3 & $1.000$ & $-1.000$ & $1.000$ & $1.000$ & $+1.000$ & $1.000$ \\
\noalign{\smallskip}\hline\noalign{\smallskip}
4 & $1.000$ & $+1.000$ & $1.000$ & $1.000$ & $-1.000$ & $1.000$ \\
\noalign{\smallskip}\hline
\end{tabular}
\end{table}

\begin{figure*}
\centering
\includegraphics[width=0.24\textwidth]{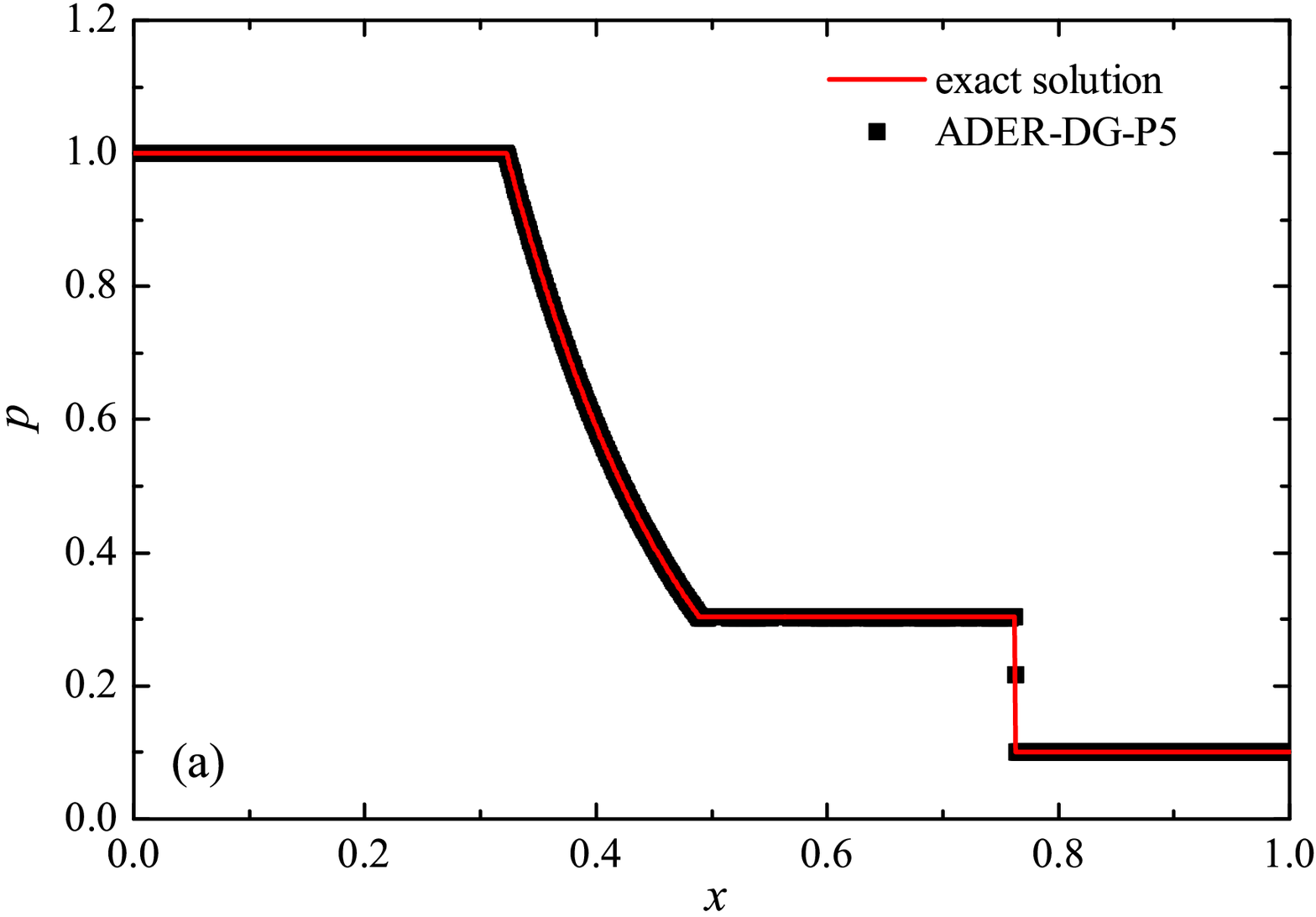}
\includegraphics[width=0.24\textwidth]{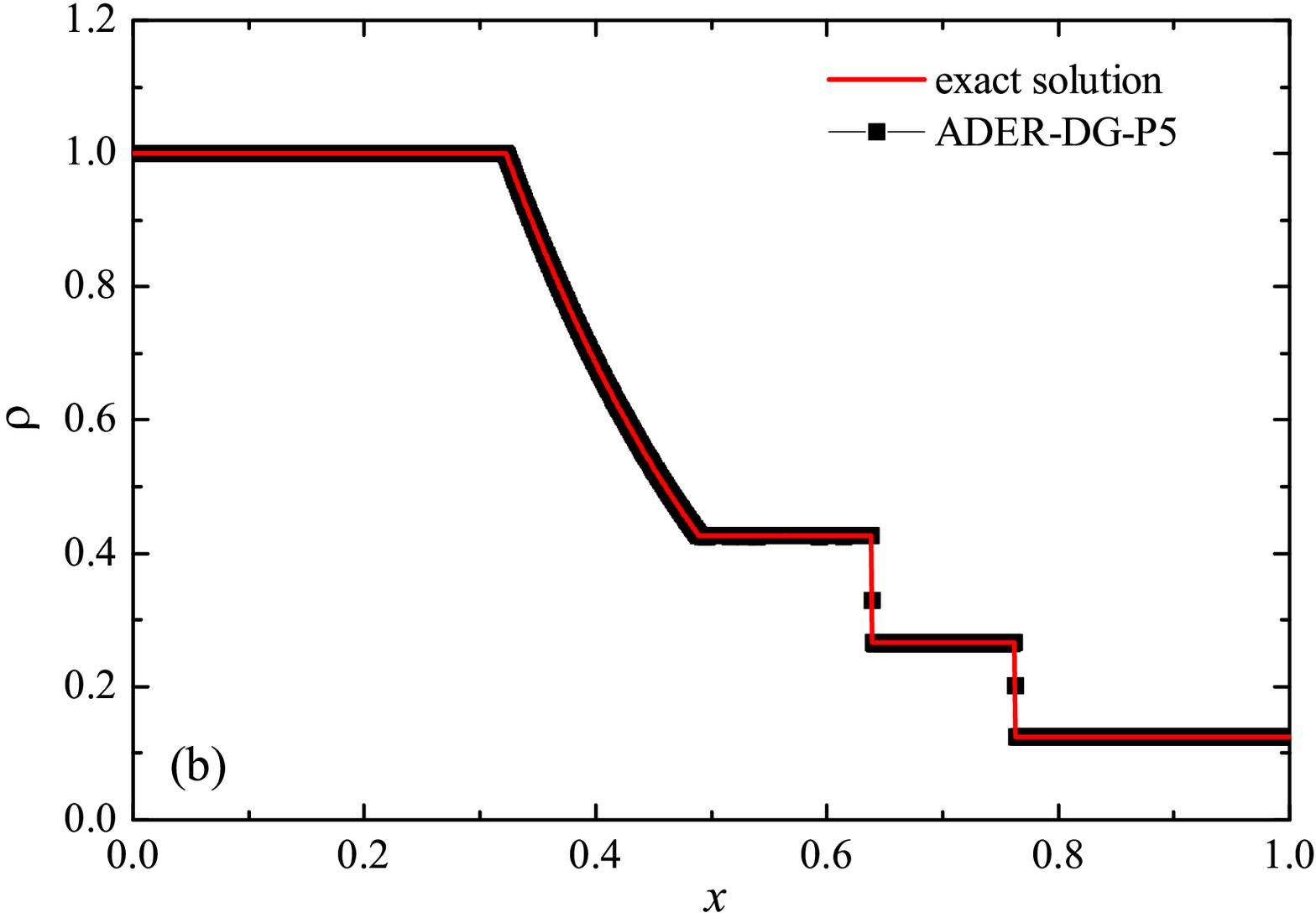}
\includegraphics[width=0.24\textwidth]{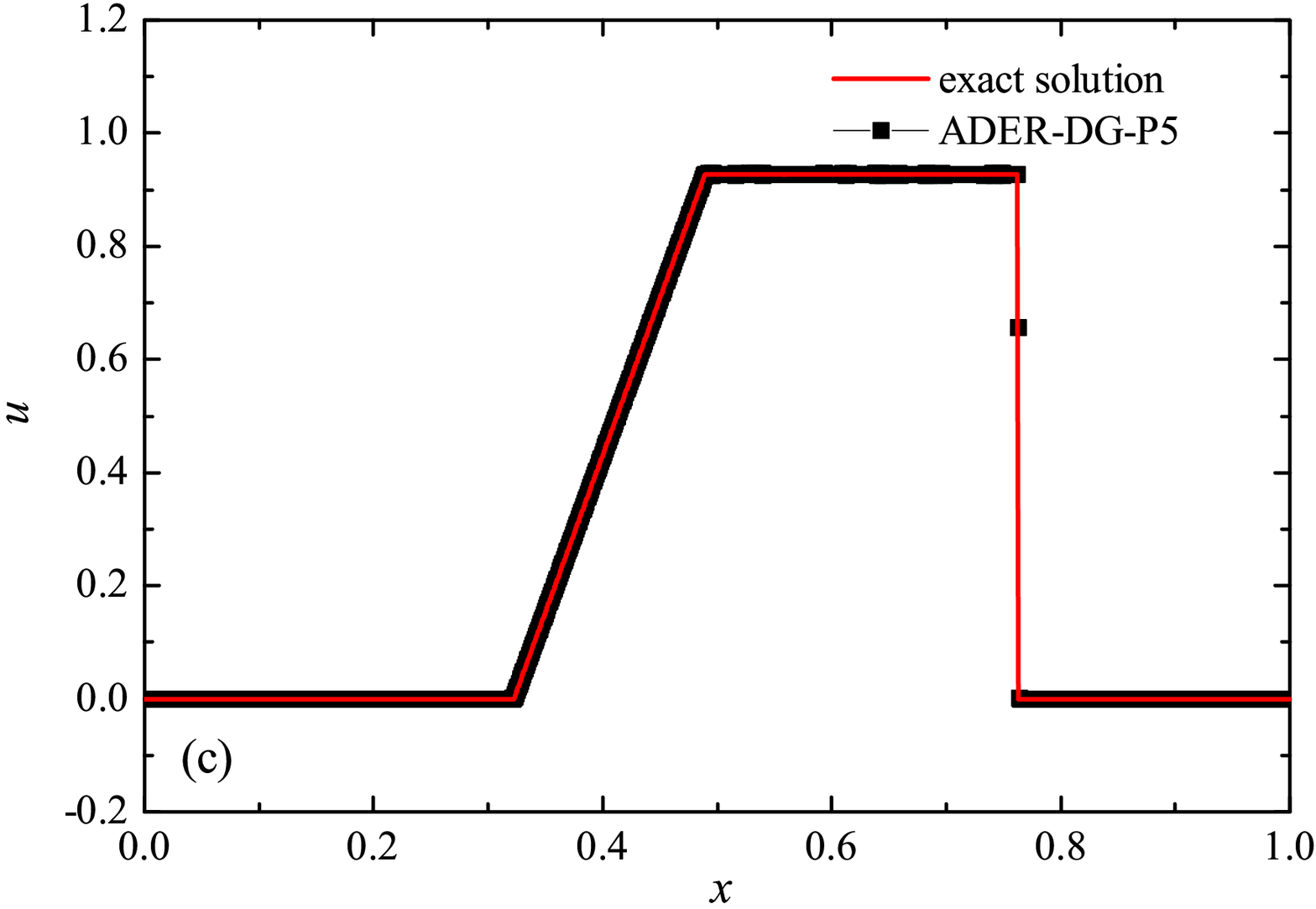}
\includegraphics[width=0.24\textwidth]{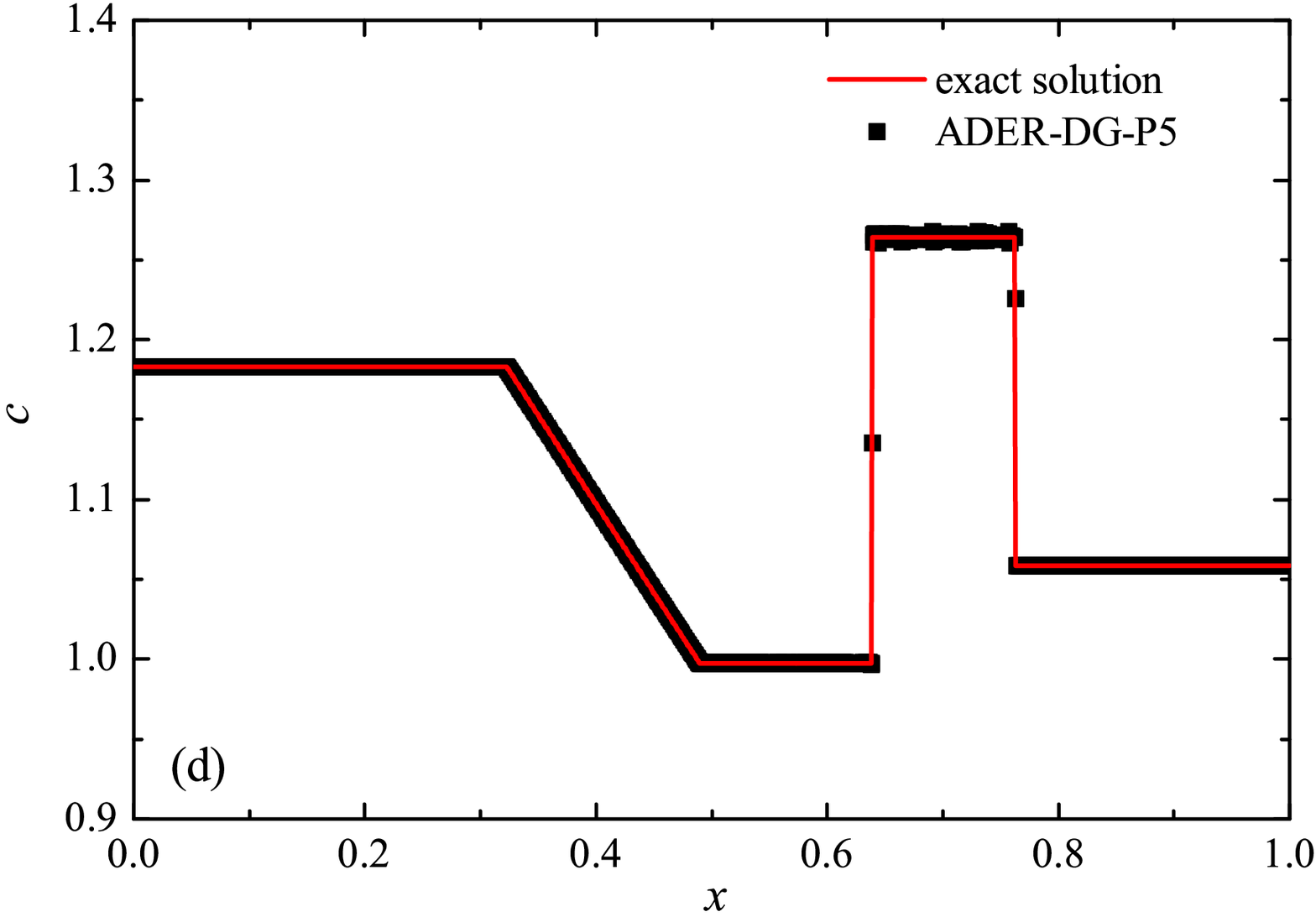}\\
\includegraphics[width=0.24\textwidth]{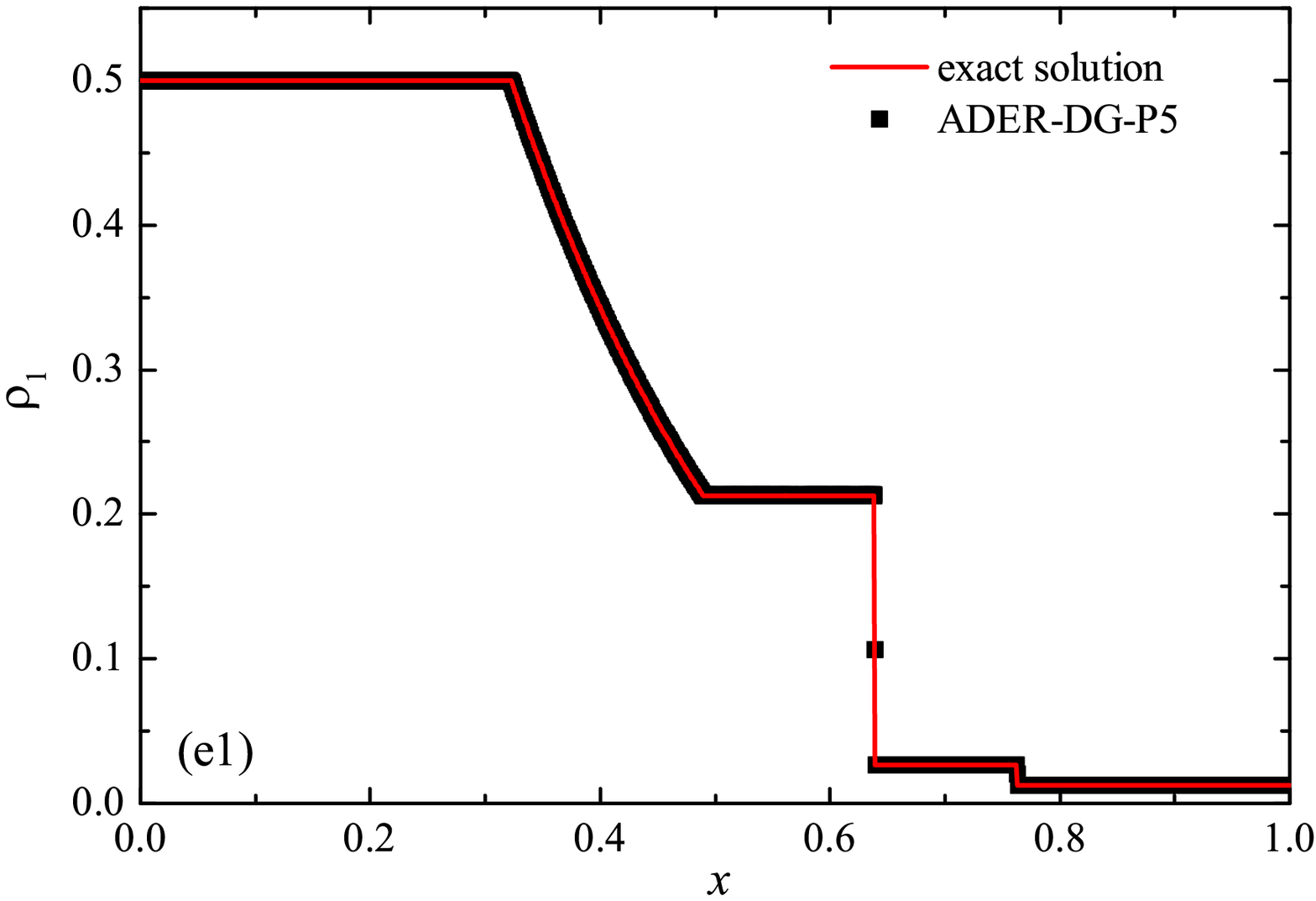}
\includegraphics[width=0.24\textwidth]{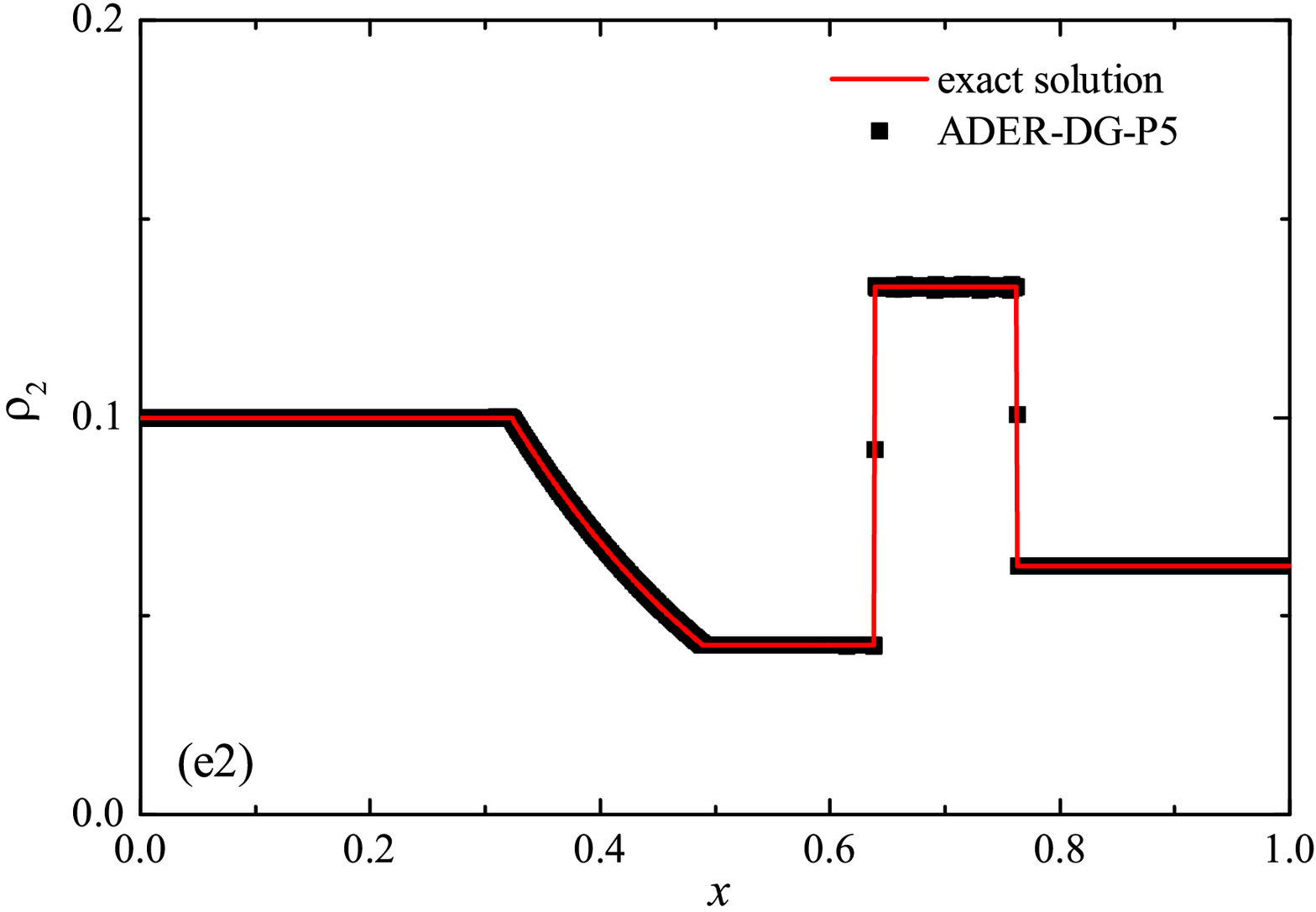}
\includegraphics[width=0.24\textwidth]{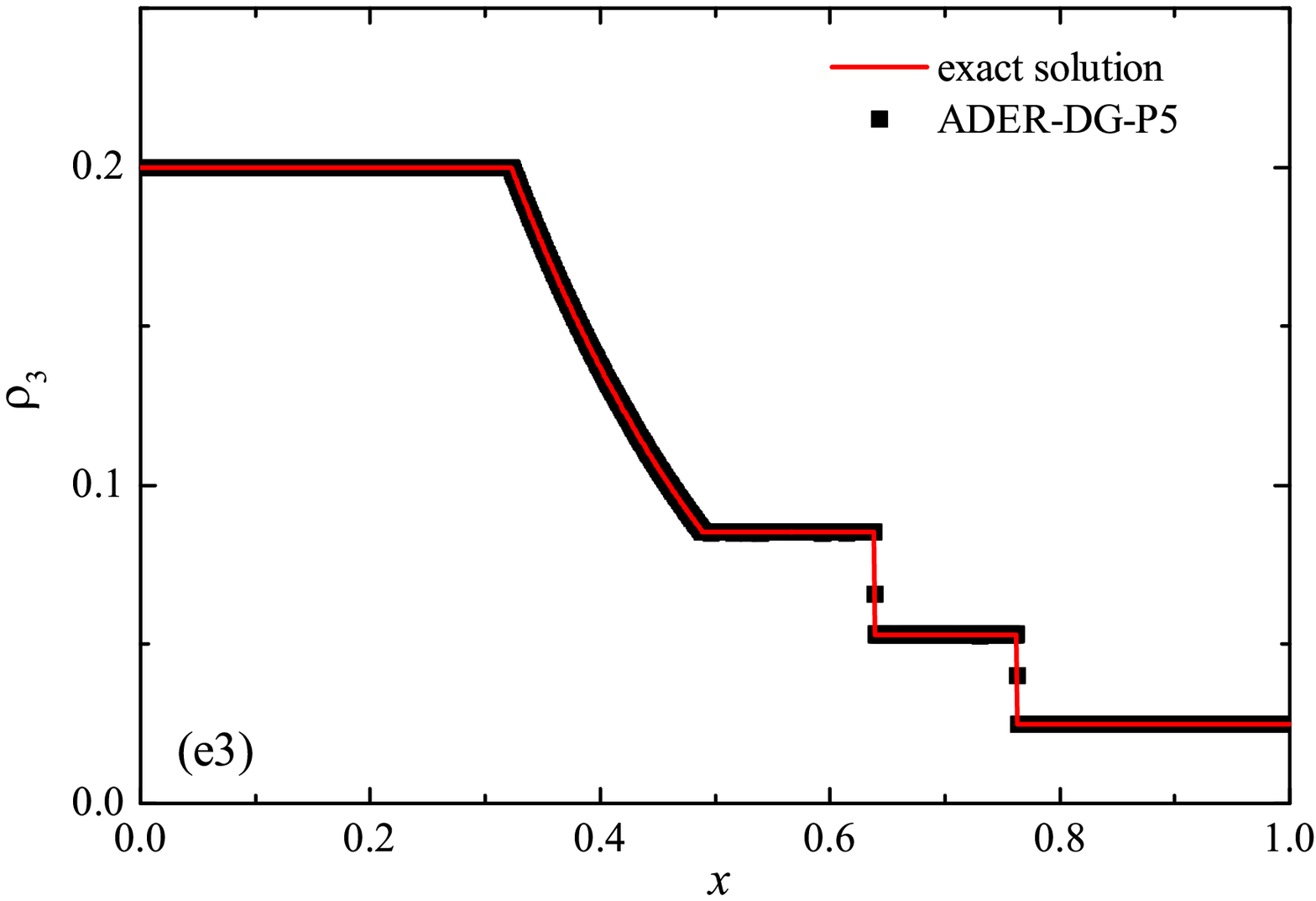}
\includegraphics[width=0.24\textwidth]{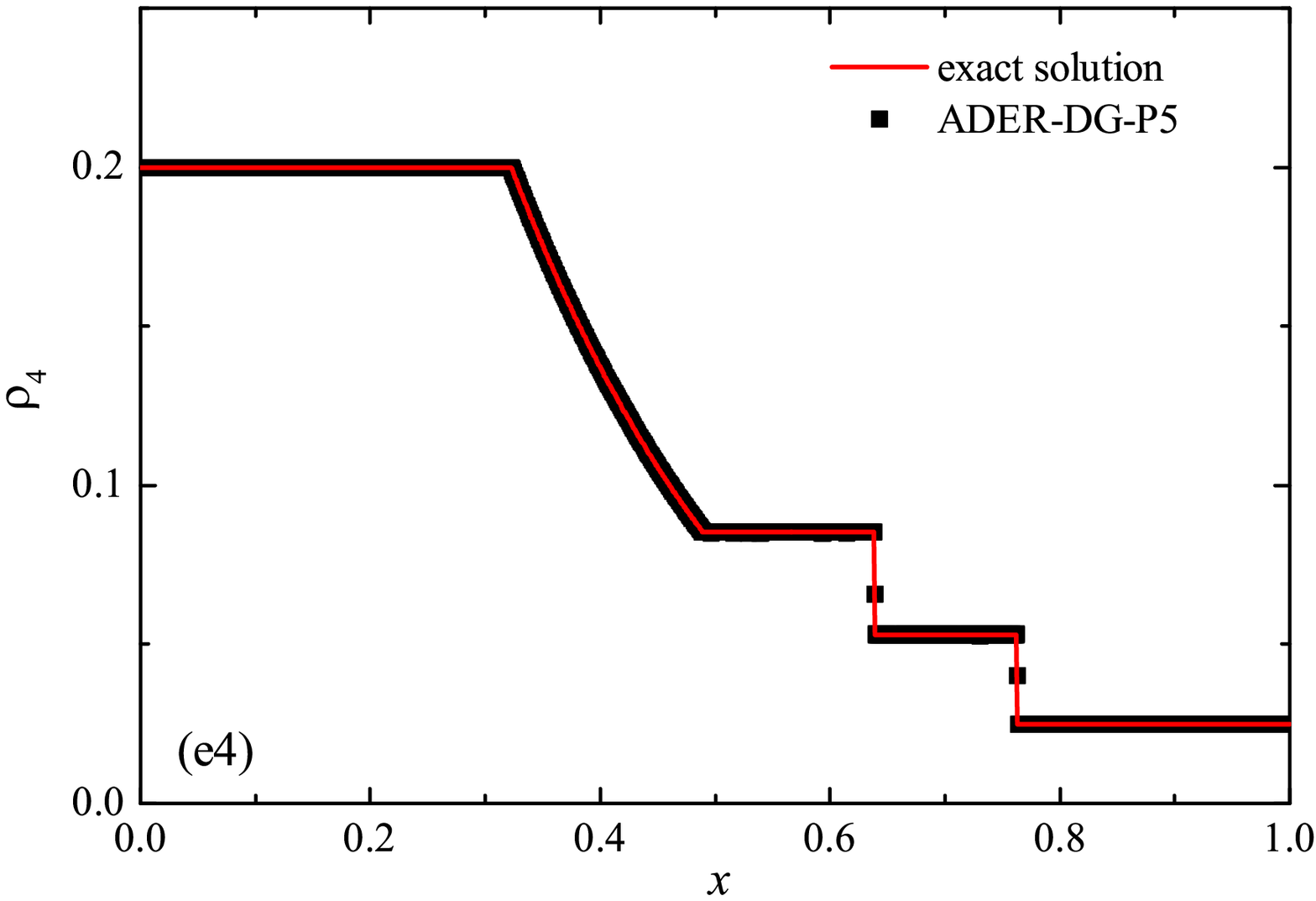}\\
\caption{%
	Numerical solution of the classical Sod problem for a multicomponent medium 
	(a detailed statement of the problem is presented in the text),
	using the computational scheme $\mathrm{ADER}$-$\mathbb{P}_5$ with a posteriori 
	limitation of the solution by a $\mathrm{ADER}$-$\mathrm{WENO}5$ finite volume limiter,
	on a coordinate mesh with $1800$ finite element cells.
	The graphs show the coordinate dependencies of pressure $p$ (a), density $\rho$ (b), flow velocity $u$ (c), 
	sound speed $c$ (d), and densities $\rho_{k} = \rho c_{k}$ (e1-e4) of individual components $k$ of the multicomponent medium,
	at the final time $t_{\rm final} = 0.15$. The black square symbols represent the numerical solution; 
	the red solid lines represents the exact analytical solution of the problem.
}
\label{fig:sod_test_fig}
\end{figure*}

\begin{figure*}
\centering
\includegraphics[width=0.24\textwidth]{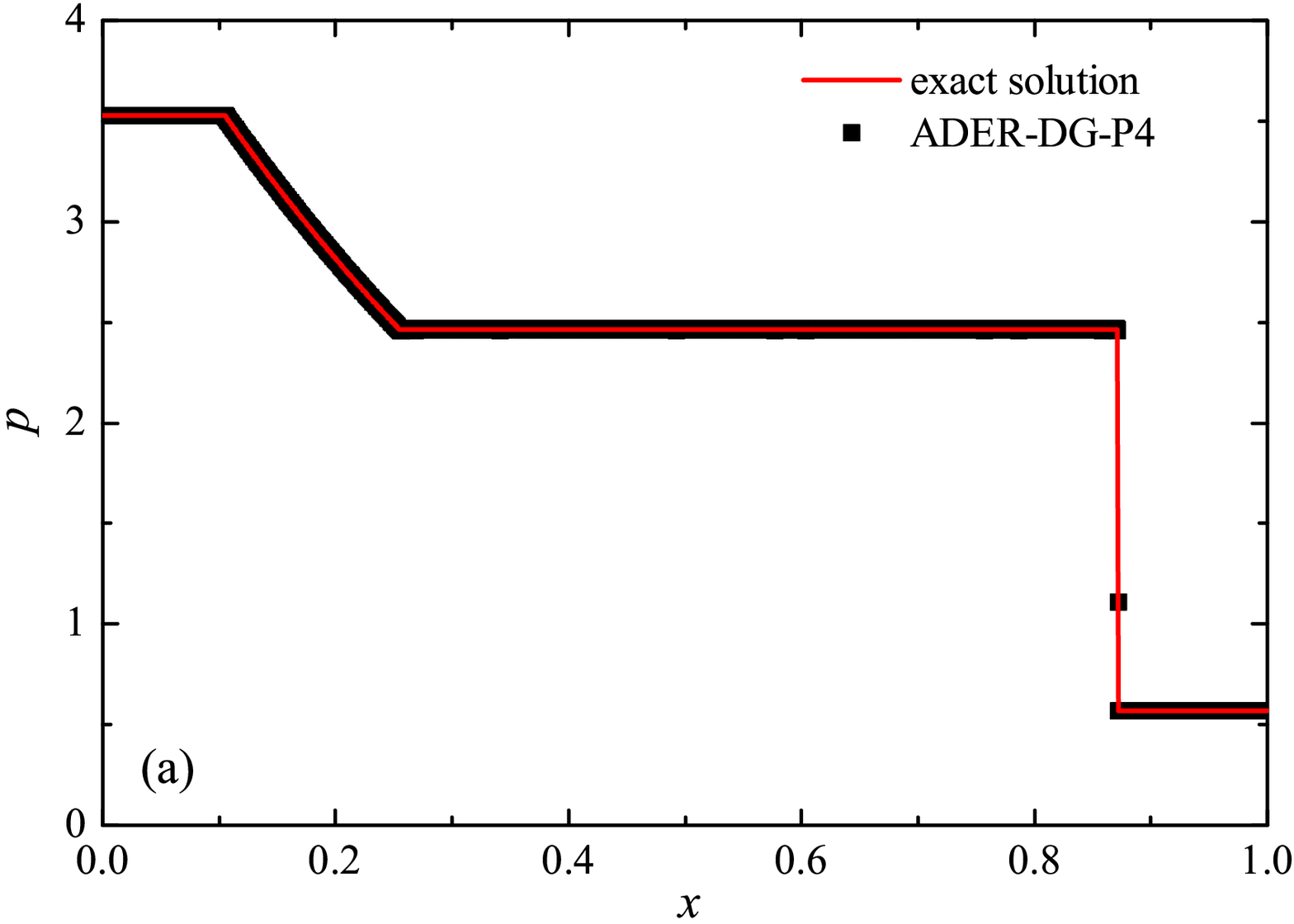}
\includegraphics[width=0.24\textwidth]{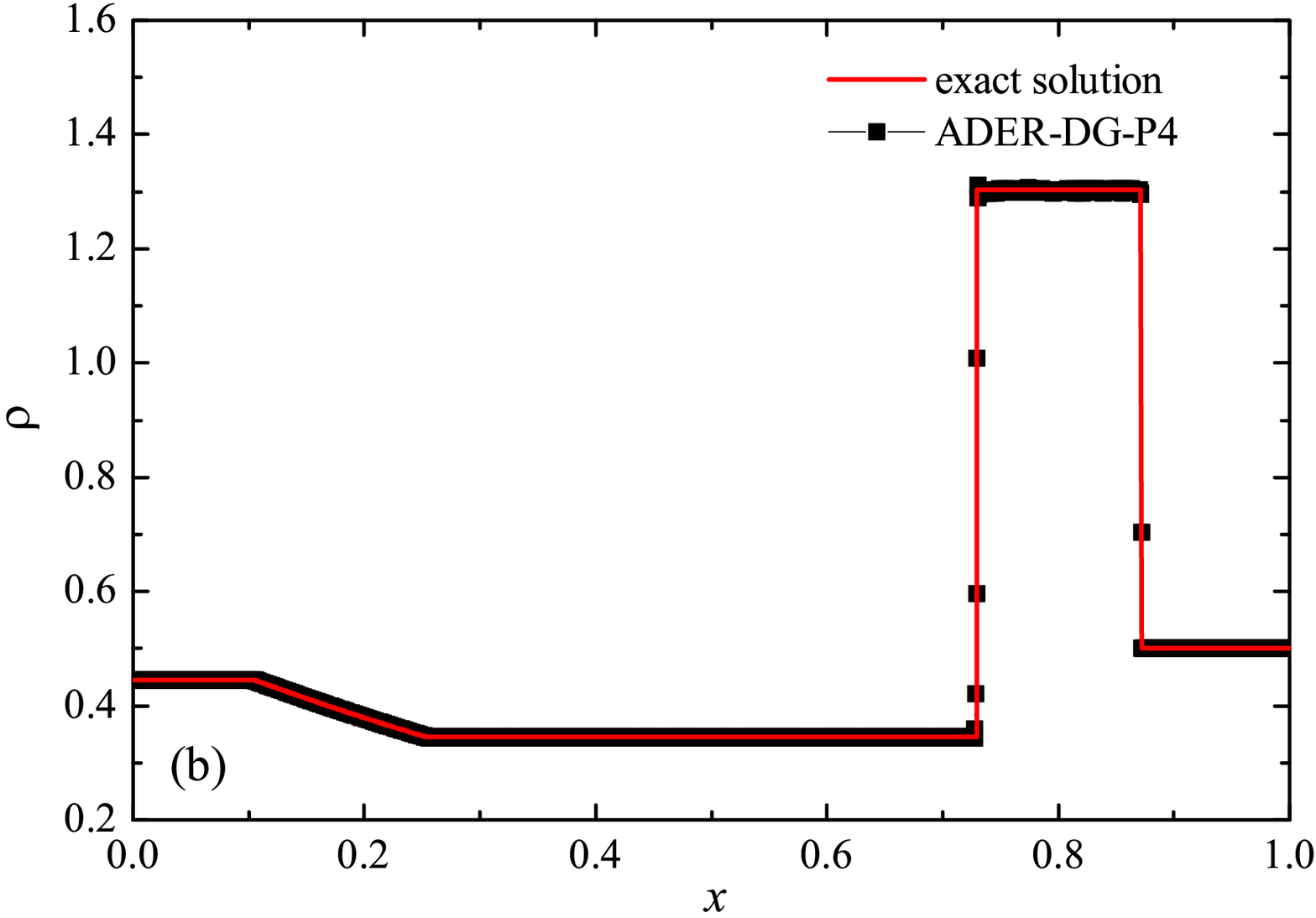}
\includegraphics[width=0.24\textwidth]{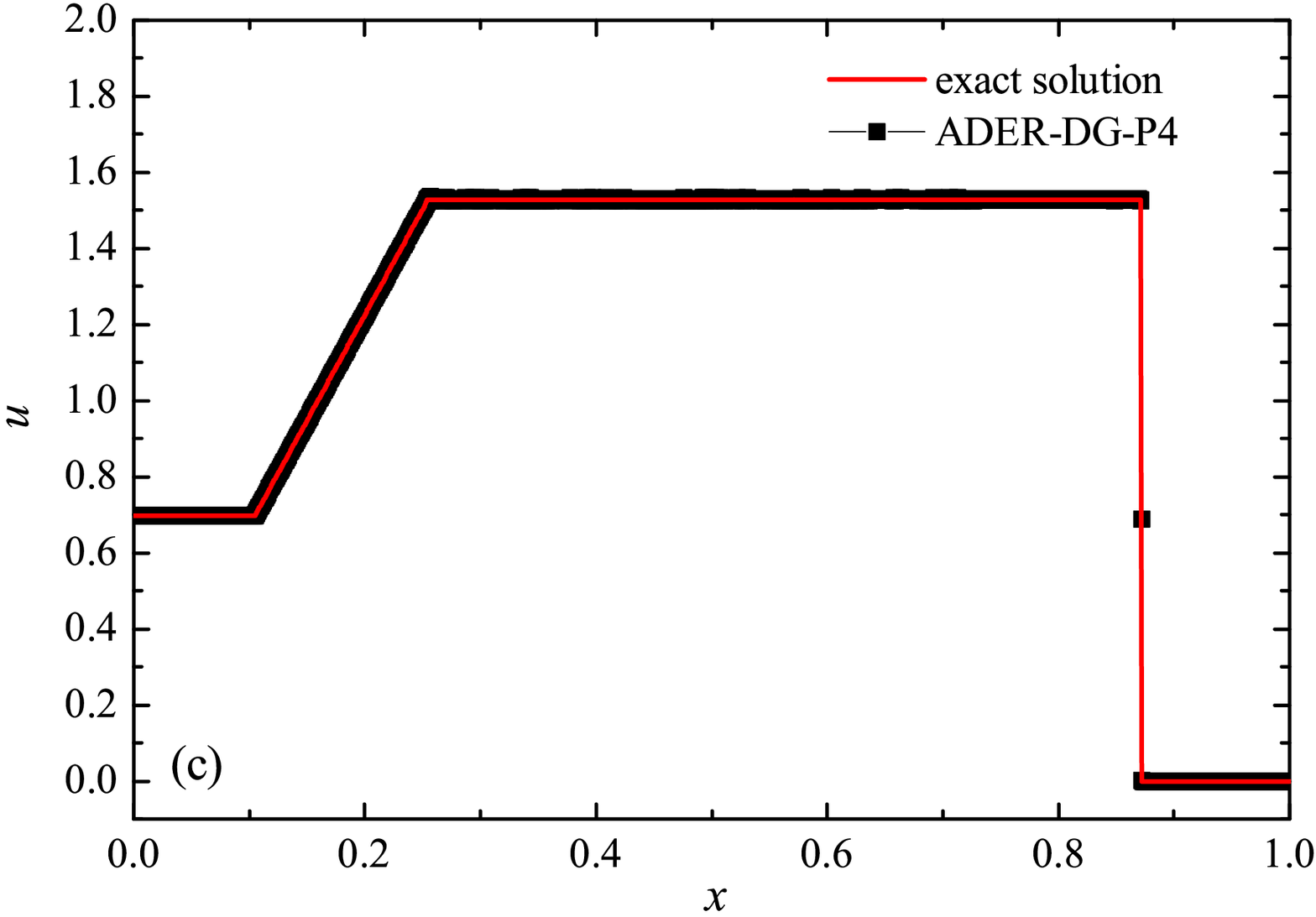}
\includegraphics[width=0.24\textwidth]{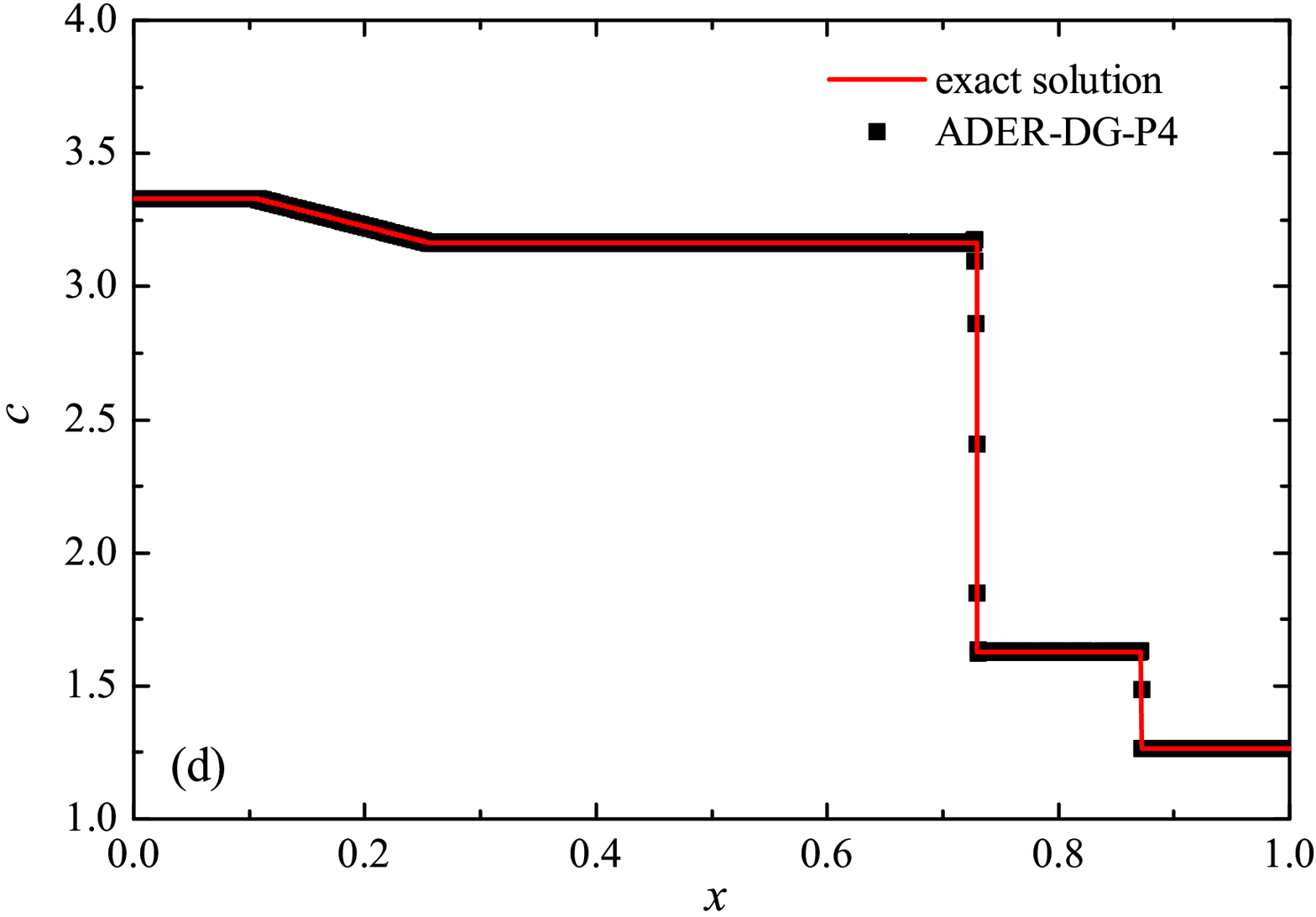}\\
\includegraphics[width=0.24\textwidth]{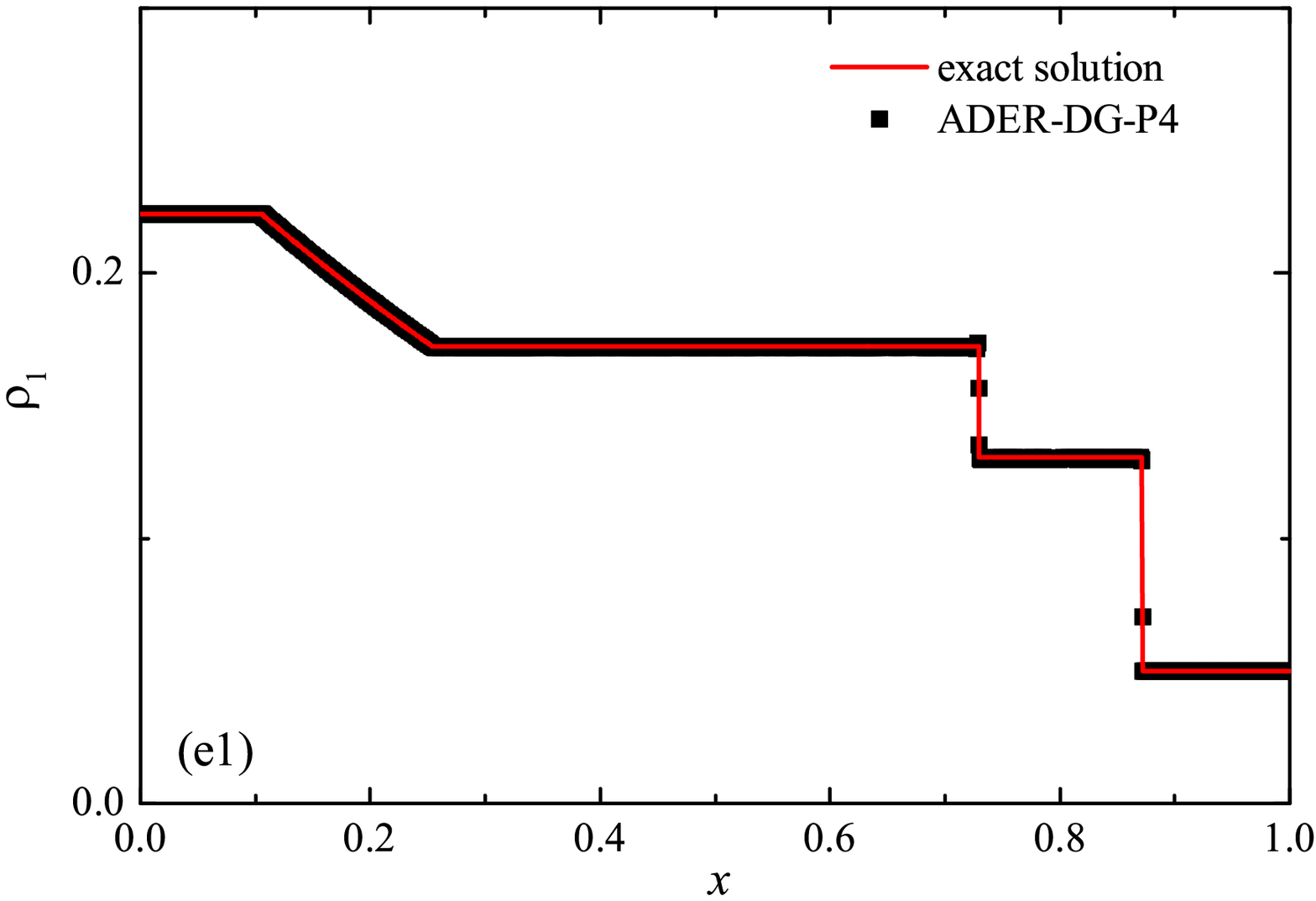}
\includegraphics[width=0.24\textwidth]{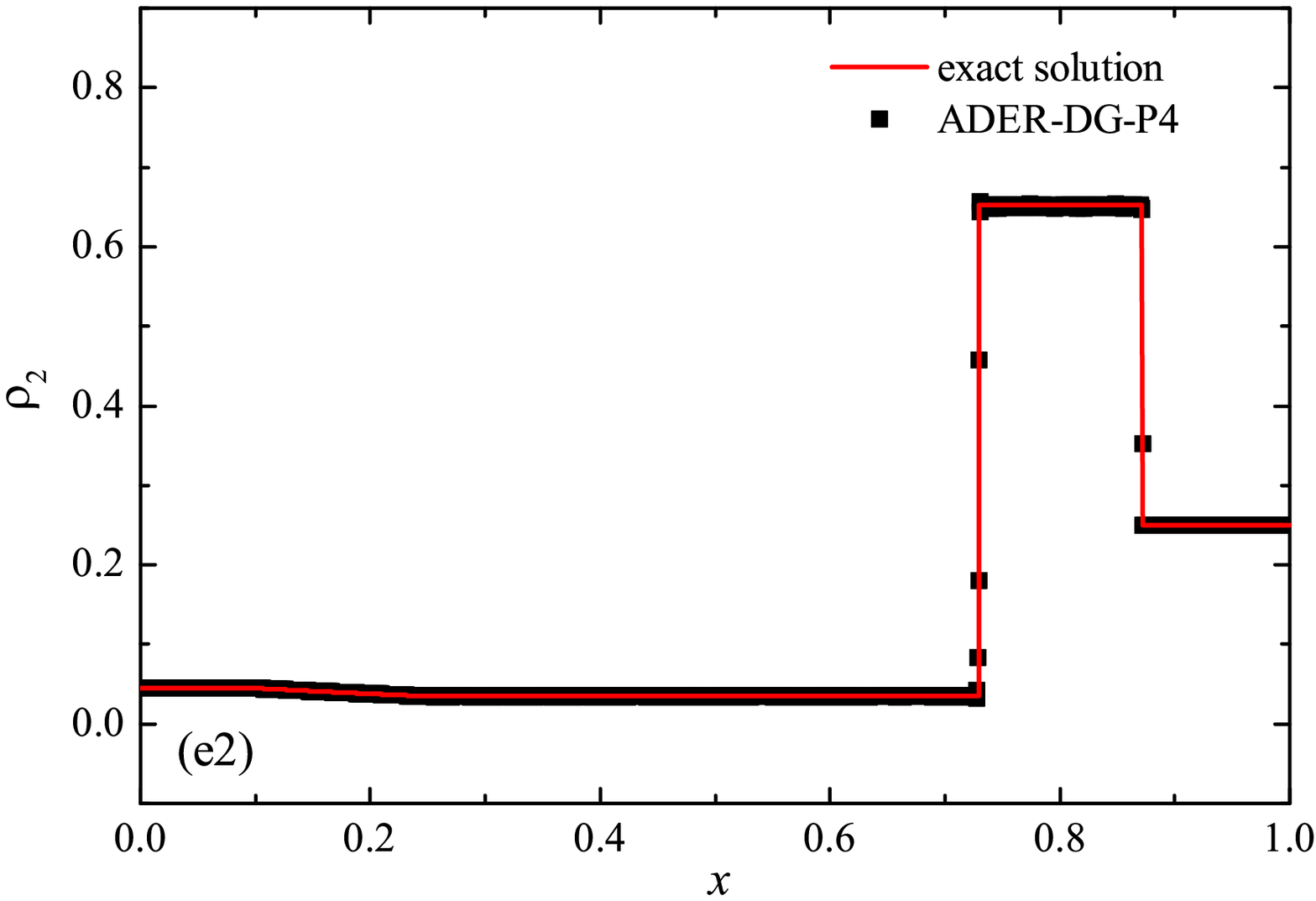}
\includegraphics[width=0.24\textwidth]{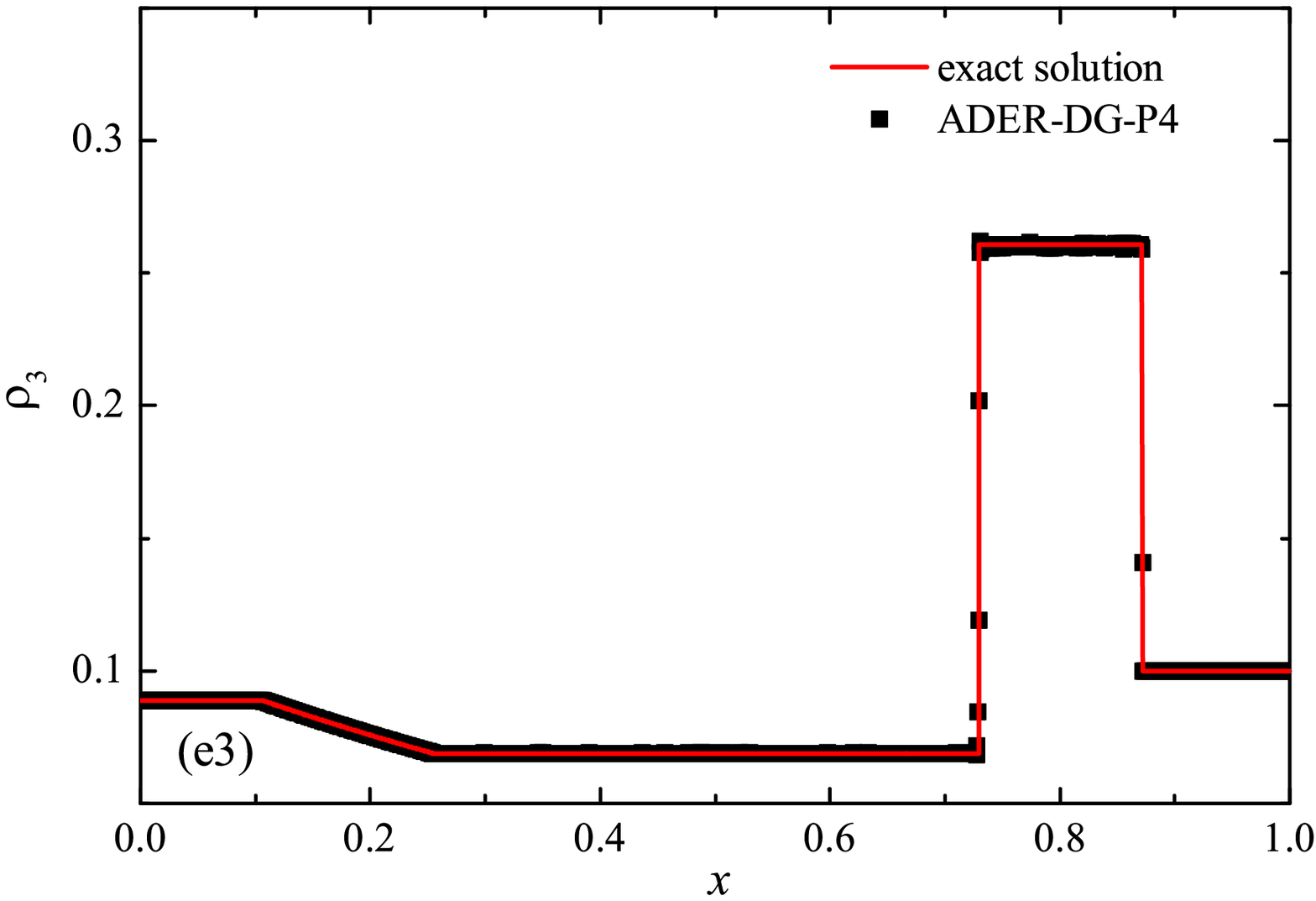}
\includegraphics[width=0.24\textwidth]{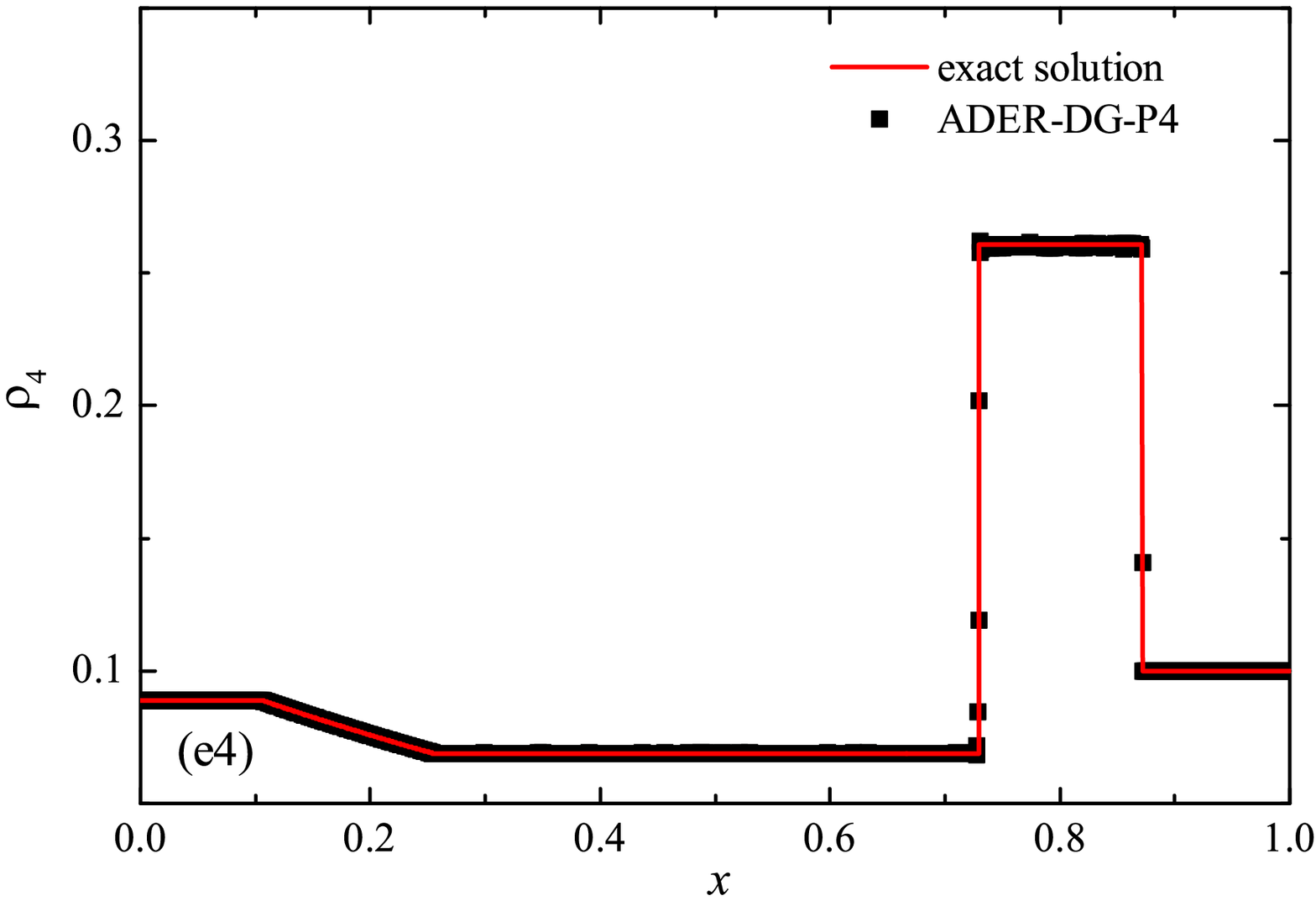}\\
\caption{%
	Numerical solution of the classical Lax problem for a multicomponent medium 
	(a detailed statement of the problem is presented in the text),
	using the computational scheme $\mathrm{ADER}$-$\mathrm{DG}$-$\mathbb{P}_4$ with a posteriori 
	limitation of the solution by a $\mathrm{ADER}$-$\mathrm{WENO}4$ finite volume limiter,
	on a coordinate mesh with $1800$ finite element cells.
	The graphs show the coordinate dependencies of pressure $p$ (a), density $\rho$ (b), flow velocity $u$ (c), 
	sound speed $c$ (d), and densities $\rho_{k} = \rho c_{k}$ (e1-e4) of individual components $k$ of the multicomponent medium,
	at the final time $t_{\rm final} = 0.15$. The black square symbols represent the numerical solution; 
	the red solid lines represents the exact analytical solution of the problem.
}
\label{fig:lax_test_fig}
\end{figure*}

\begin{figure*}
\centering
\includegraphics[width=0.24\textwidth]{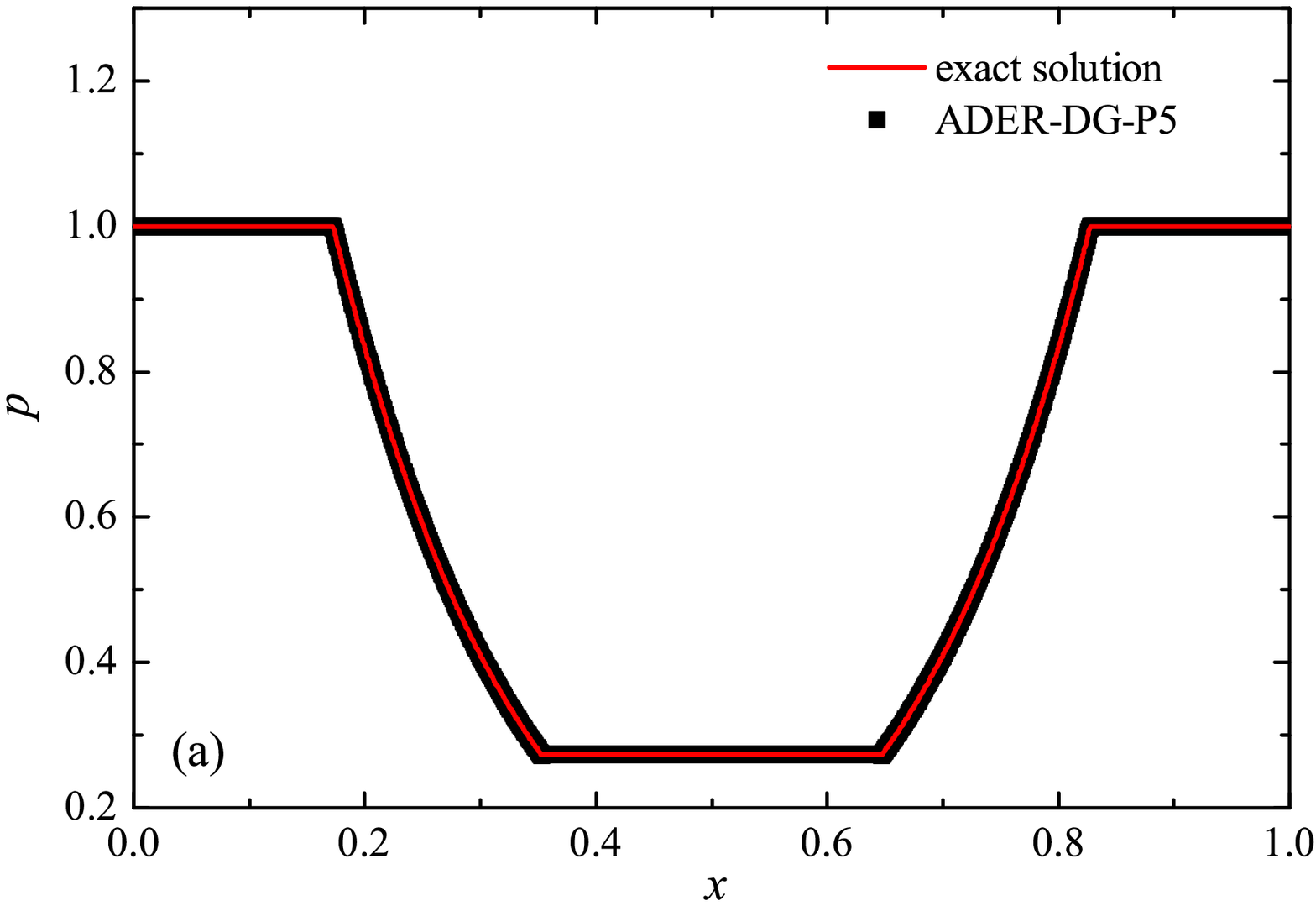}
\includegraphics[width=0.24\textwidth]{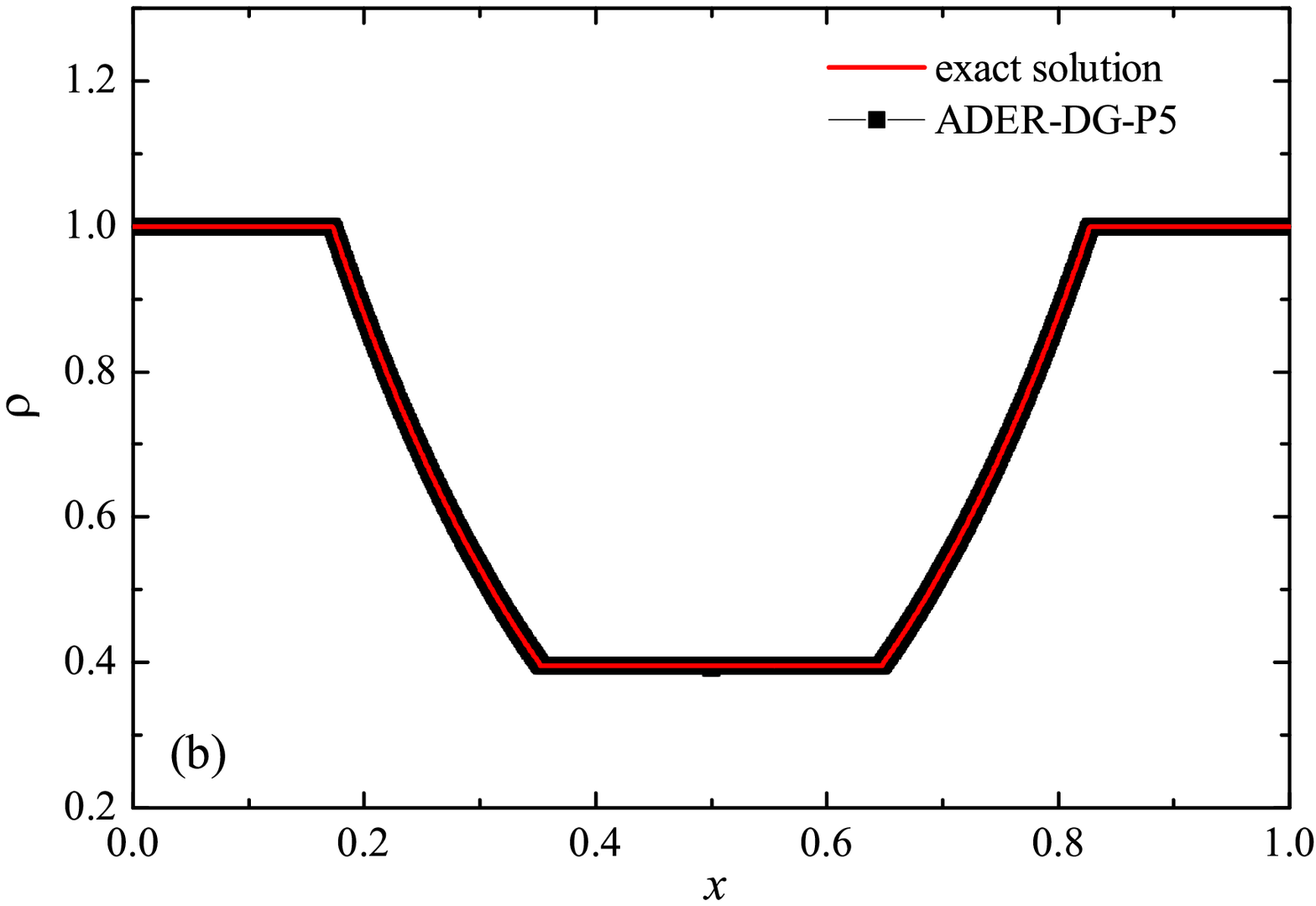}
\includegraphics[width=0.24\textwidth]{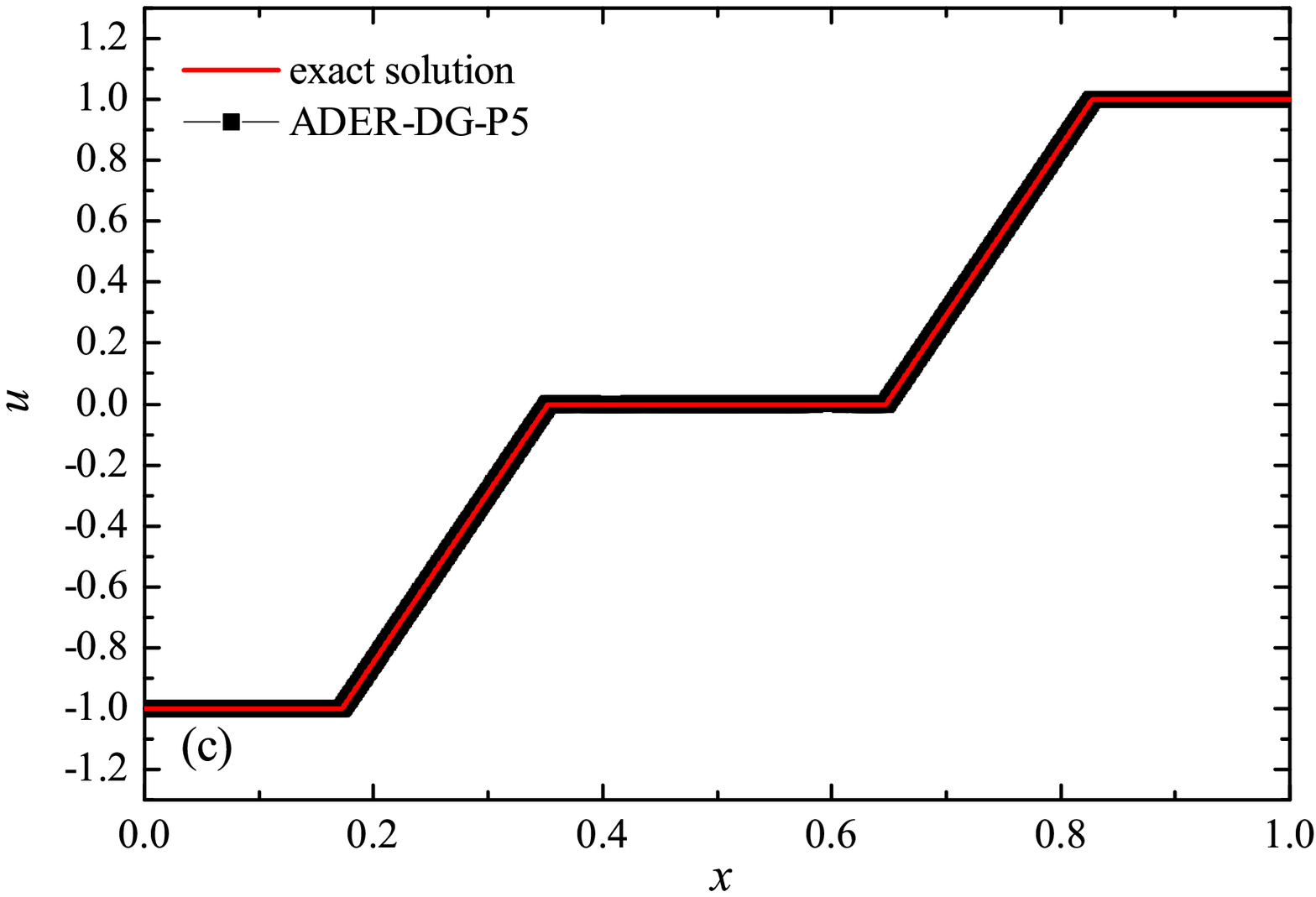}
\includegraphics[width=0.24\textwidth]{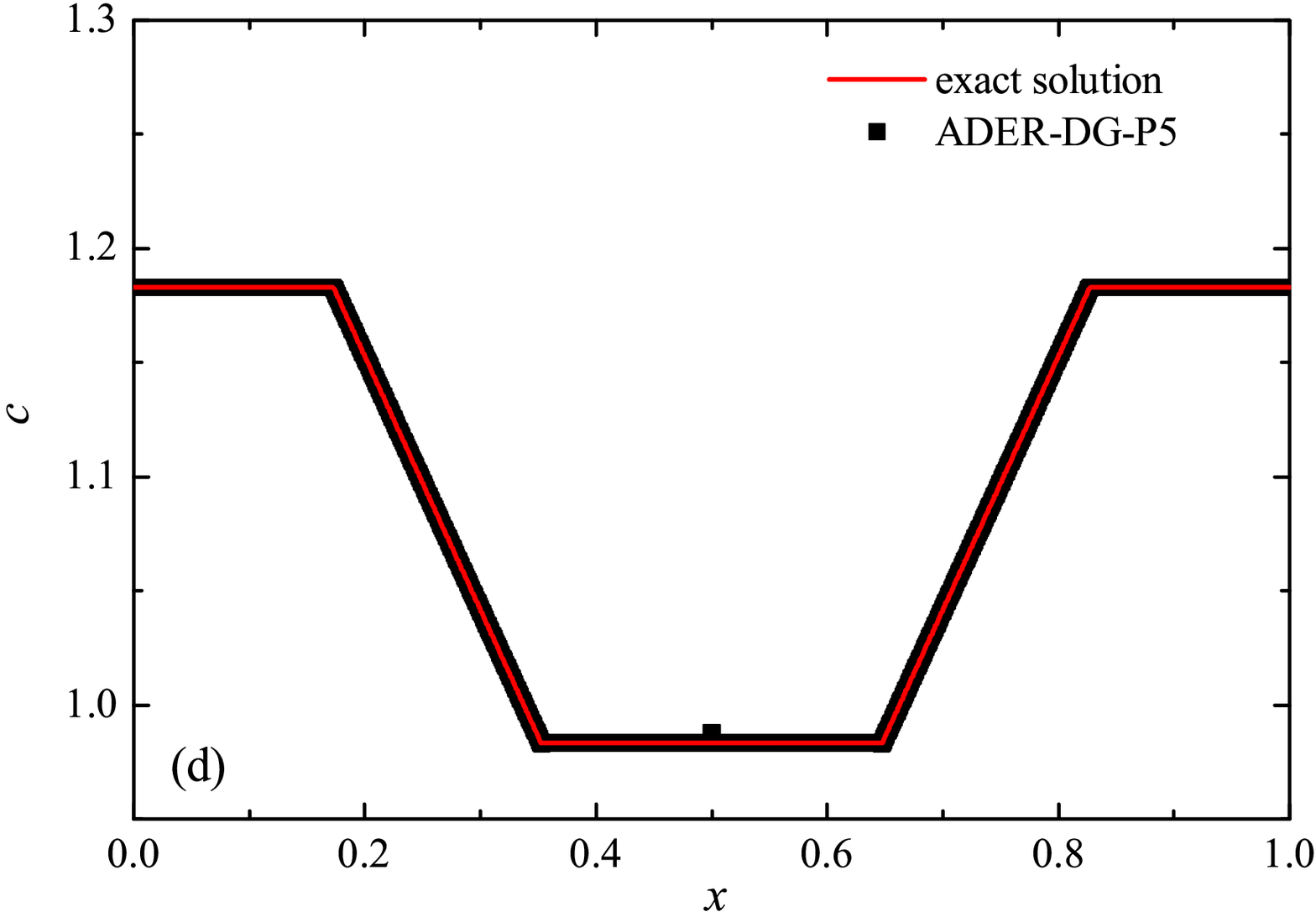}\\
\includegraphics[width=0.24\textwidth]{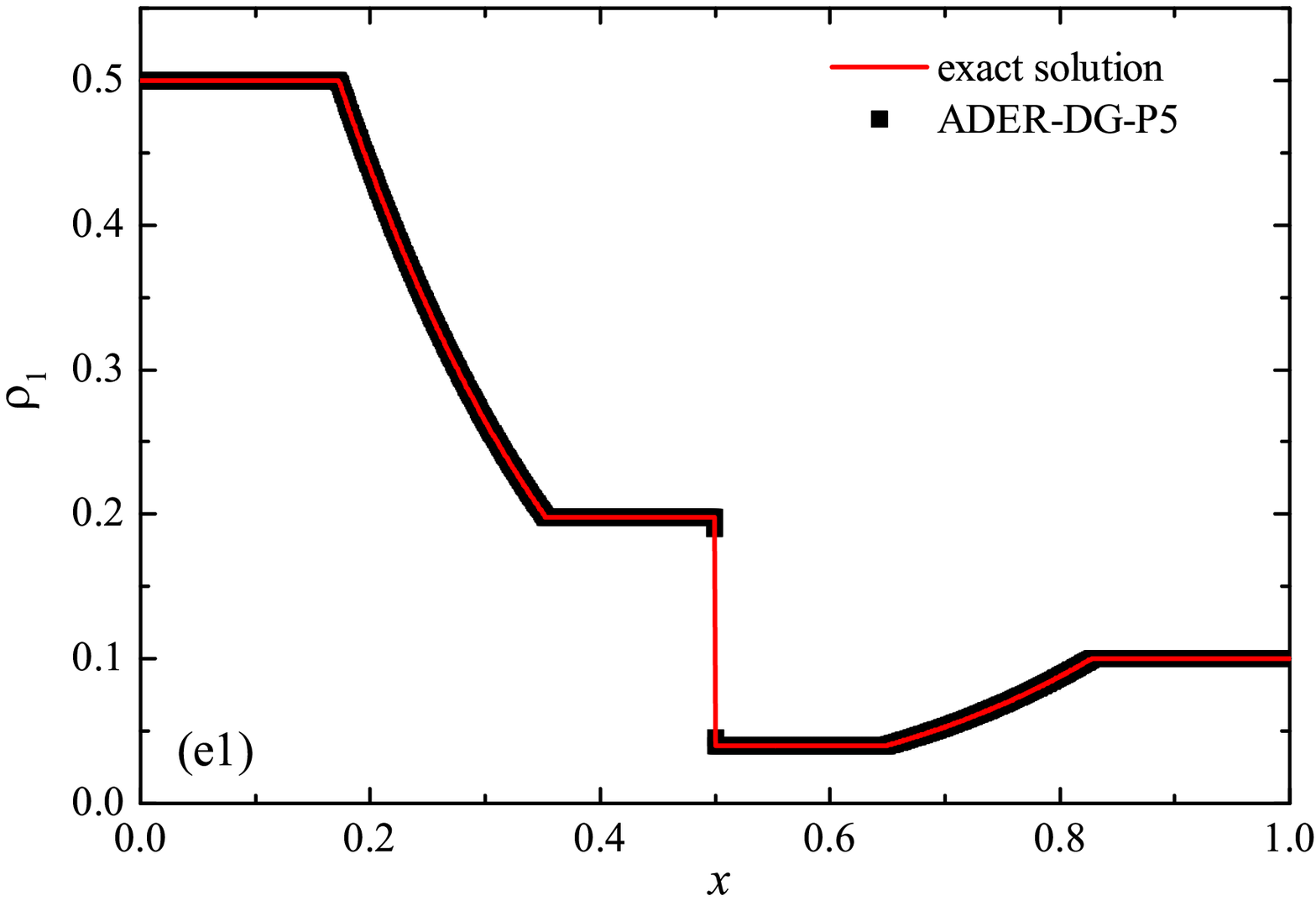}
\includegraphics[width=0.24\textwidth]{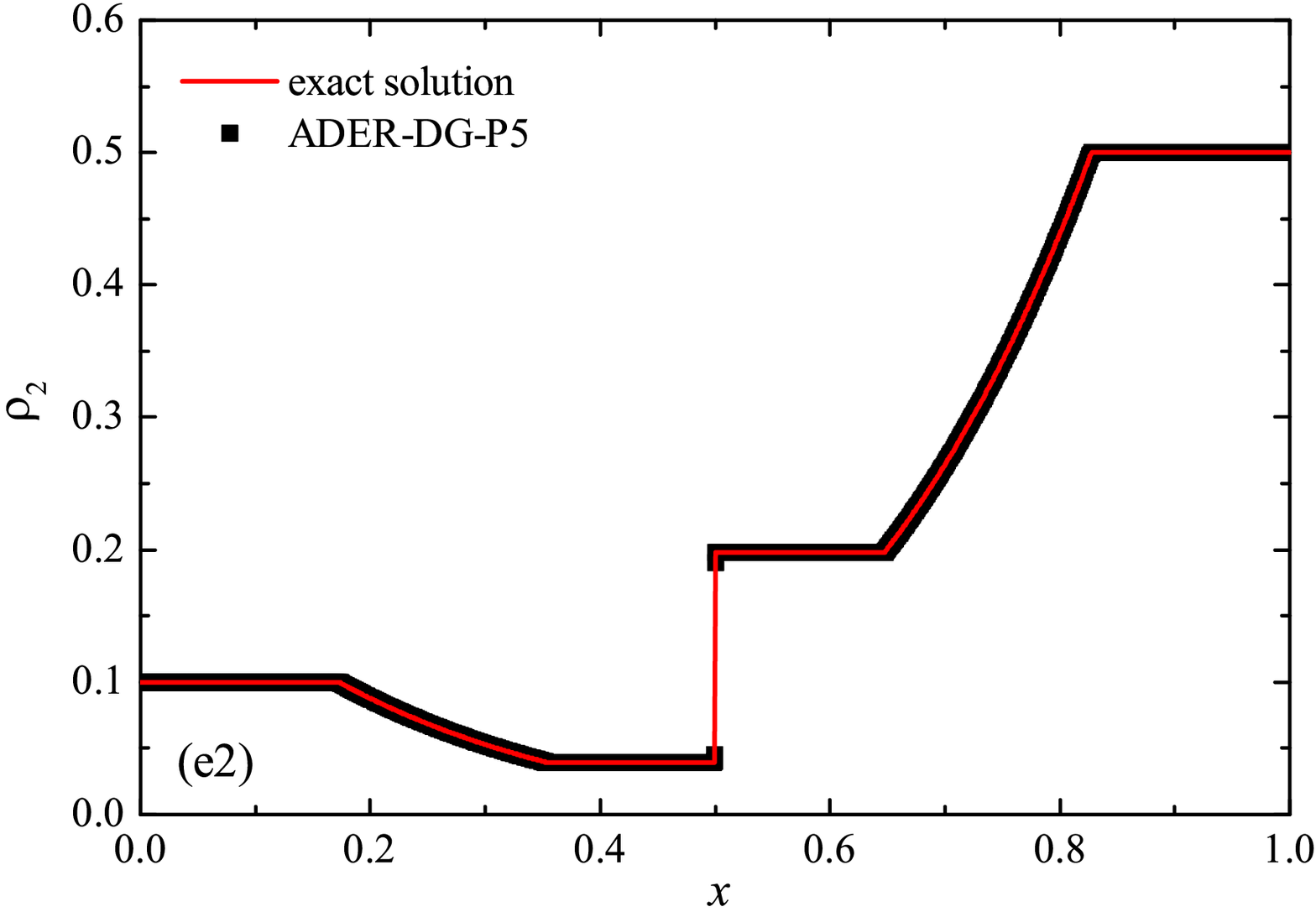}
\includegraphics[width=0.24\textwidth]{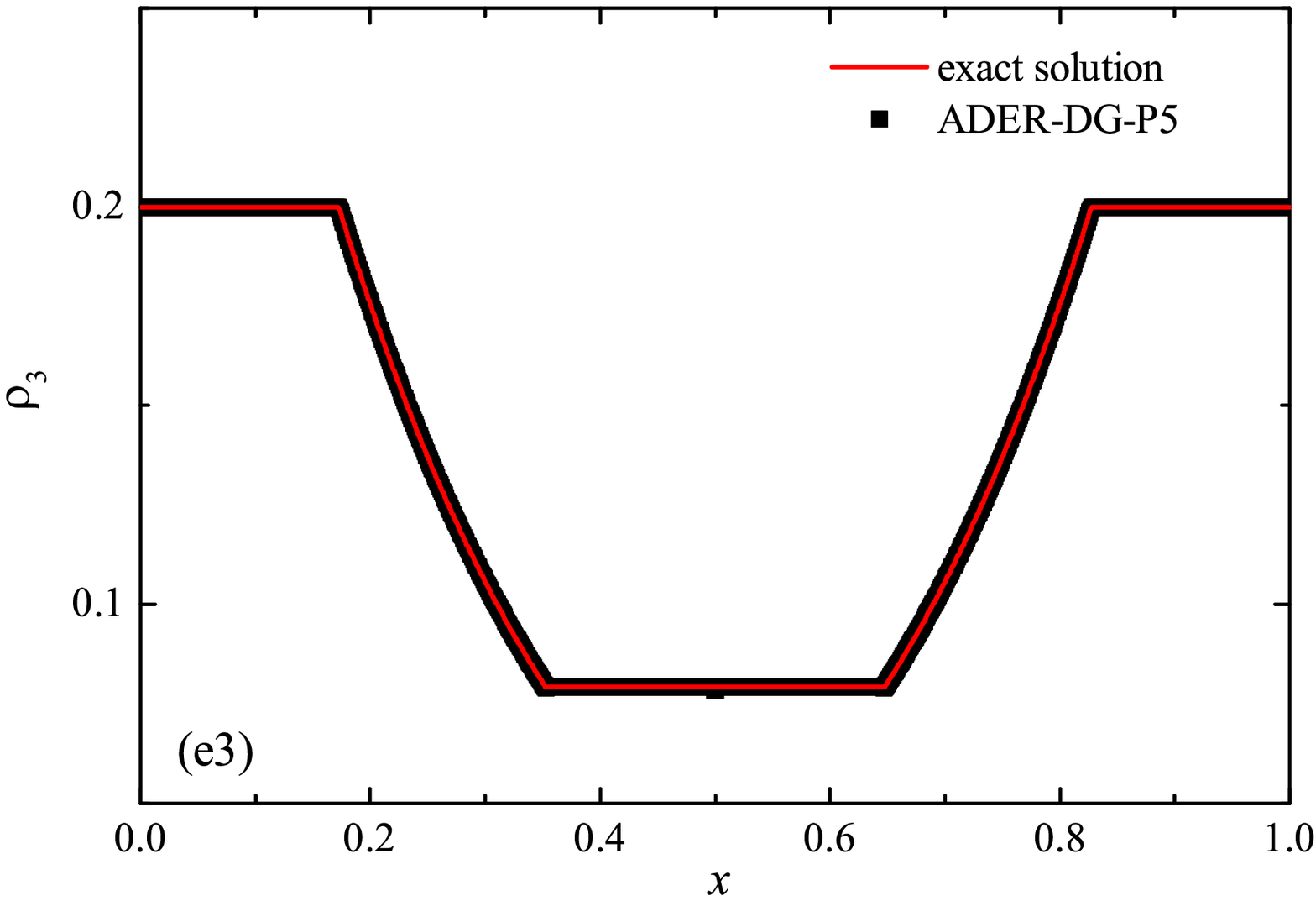}
\includegraphics[width=0.24\textwidth]{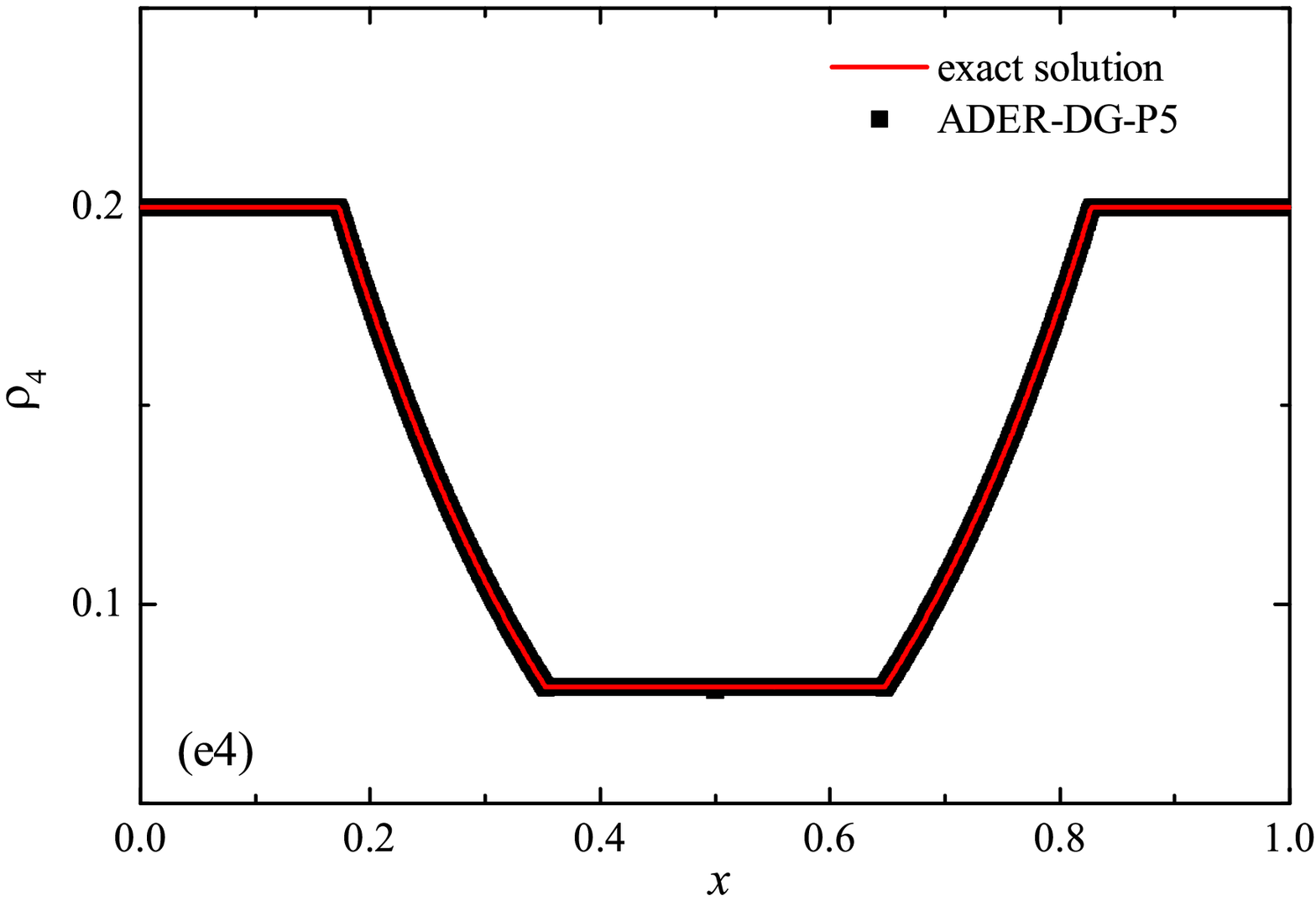}\\
\caption{%
	Numerical solution of the classical problem with two rarefaction waves for a multicomponent medium 
	(a detailed statement of the problem is presented in the text),
	using the computational scheme $\mathrm{ADER}$-$\mathrm{DG}$-$\mathbb{P}_5$ with a posteriori 
	limitation of the solution by a $\mathrm{ADER}$-$\mathrm{WENO}5$ finite volume limiter,
	on a coordinate mesh with $1800$ finite element cells.
	The graphs show the coordinate dependencies of pressure $p$ (a), density $\rho$ (b), flow velocity $u$ (c), 
	sound speed $c$ (d), and densities $\rho_{k} = \rho c_{k}$ (e1-e4) of individual components $k$ of the multicomponent medium,
	at the final time $t_{\rm final} = 0.15$. The black square symbols represent the numerical solution; 
	the red solid lines represents the exact analytical solution of the problem.
}
\label{fig:srw_test_fig}
\end{figure*}

\begin{figure*}
\centering
\includegraphics[width=0.24\textwidth]{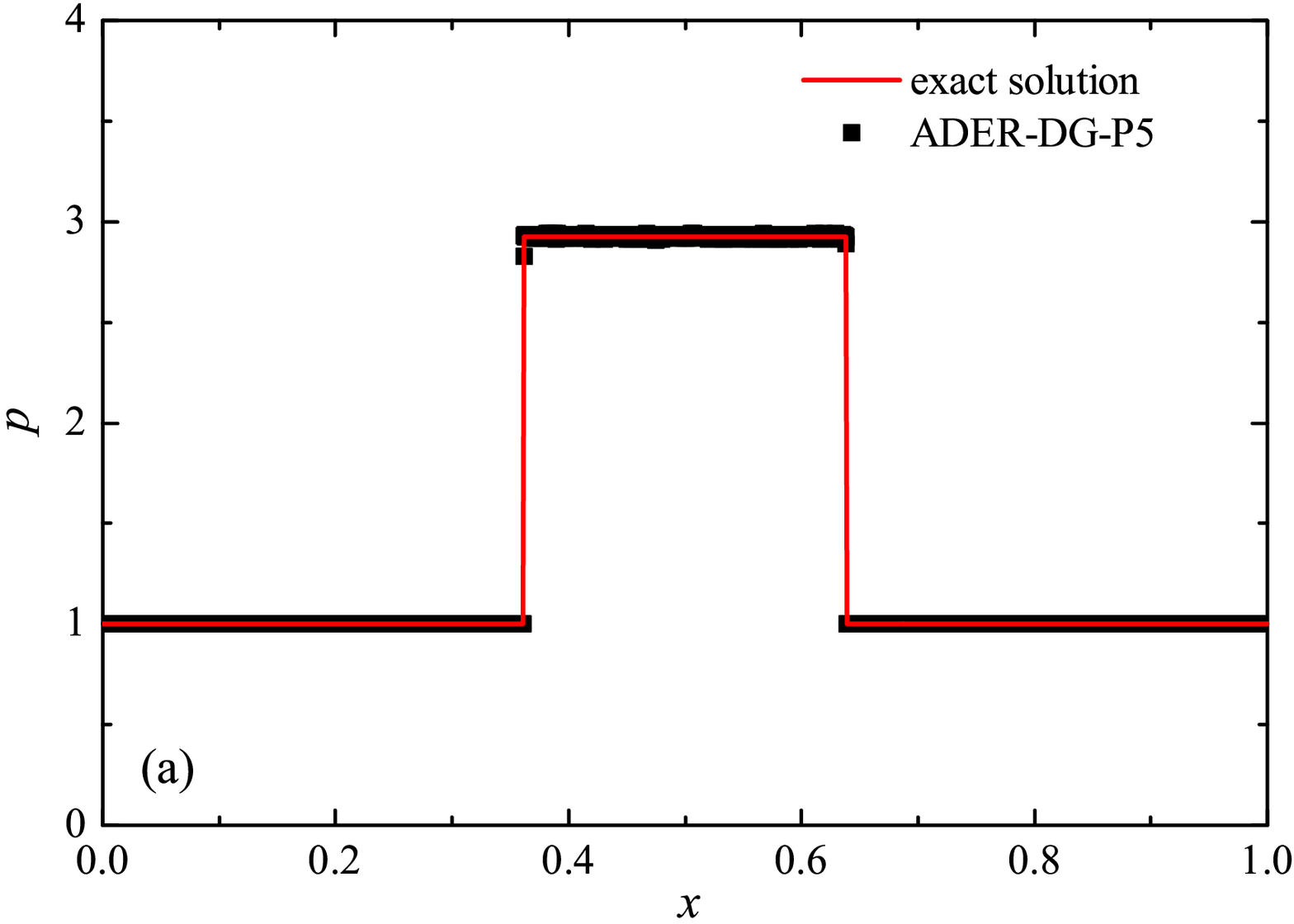}
\includegraphics[width=0.24\textwidth]{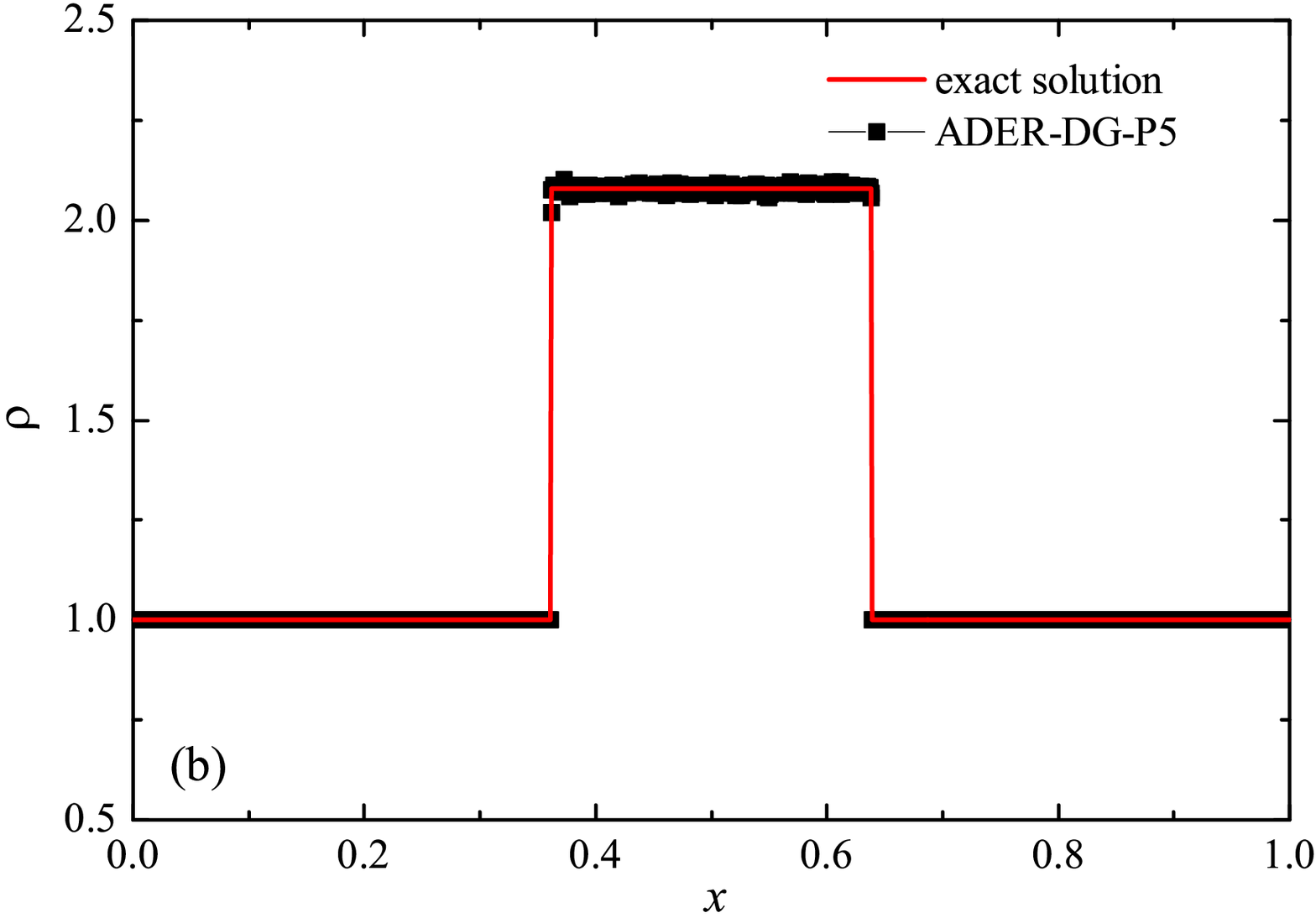}
\includegraphics[width=0.24\textwidth]{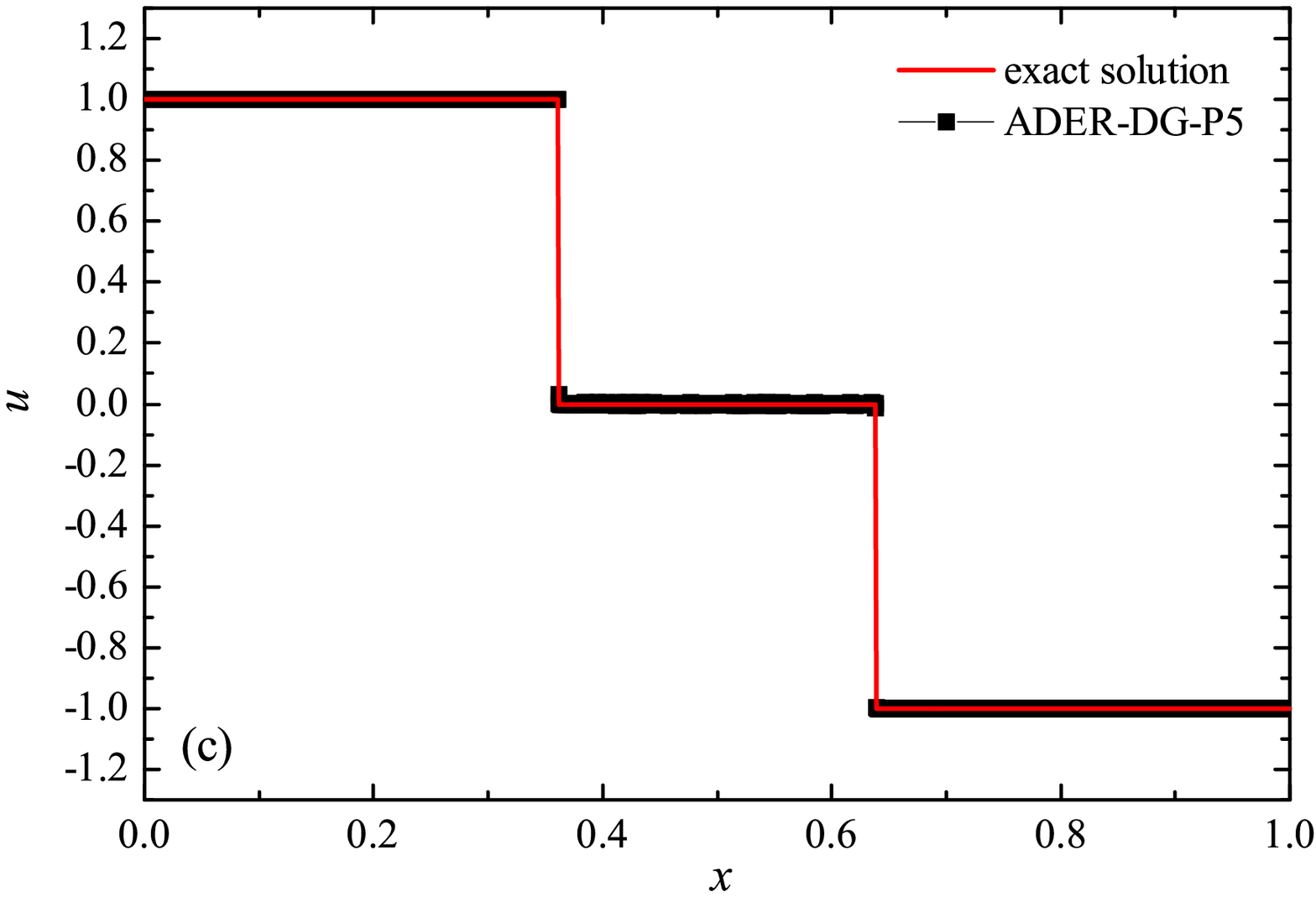}
\includegraphics[width=0.24\textwidth]{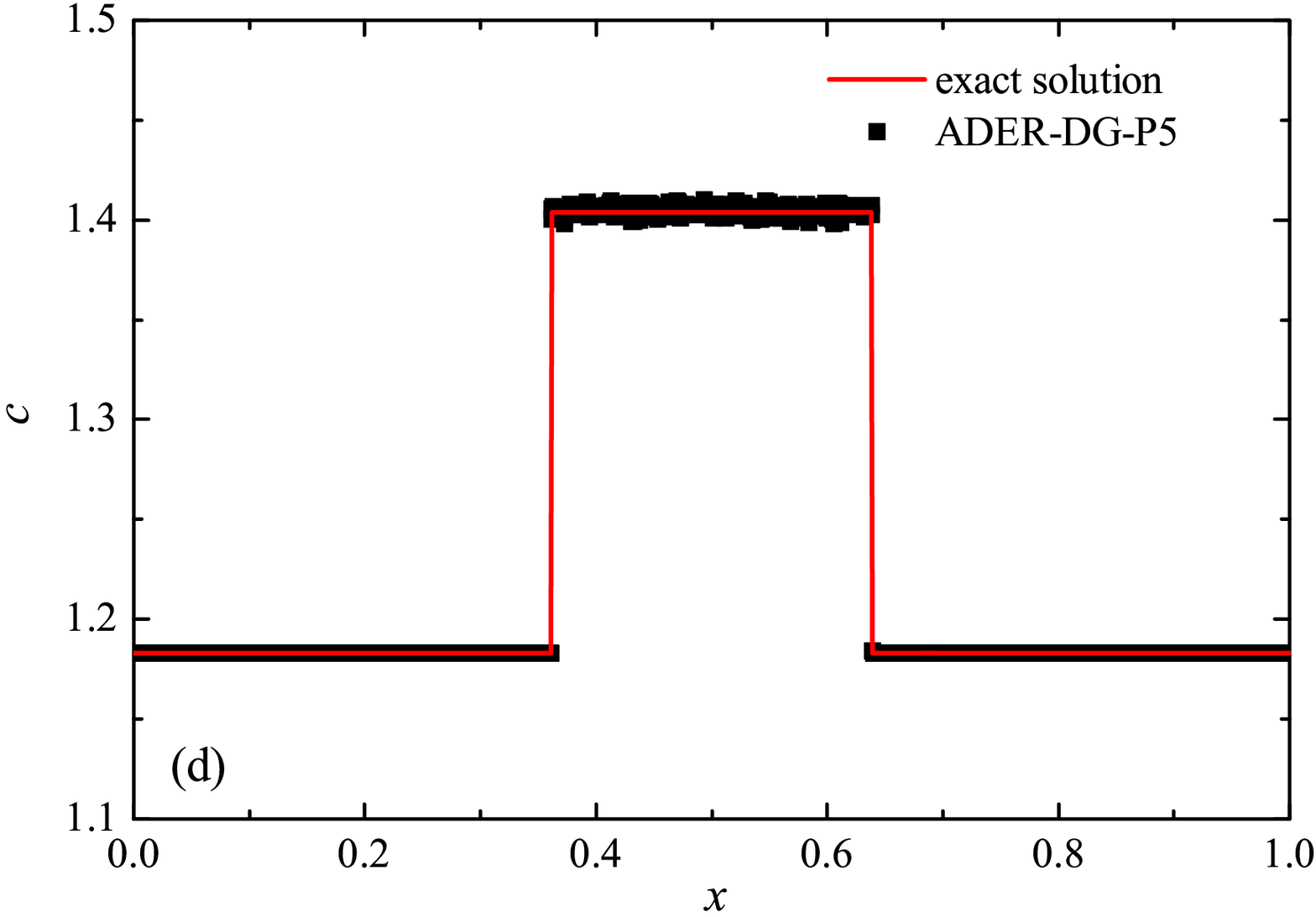}\\
\includegraphics[width=0.24\textwidth]{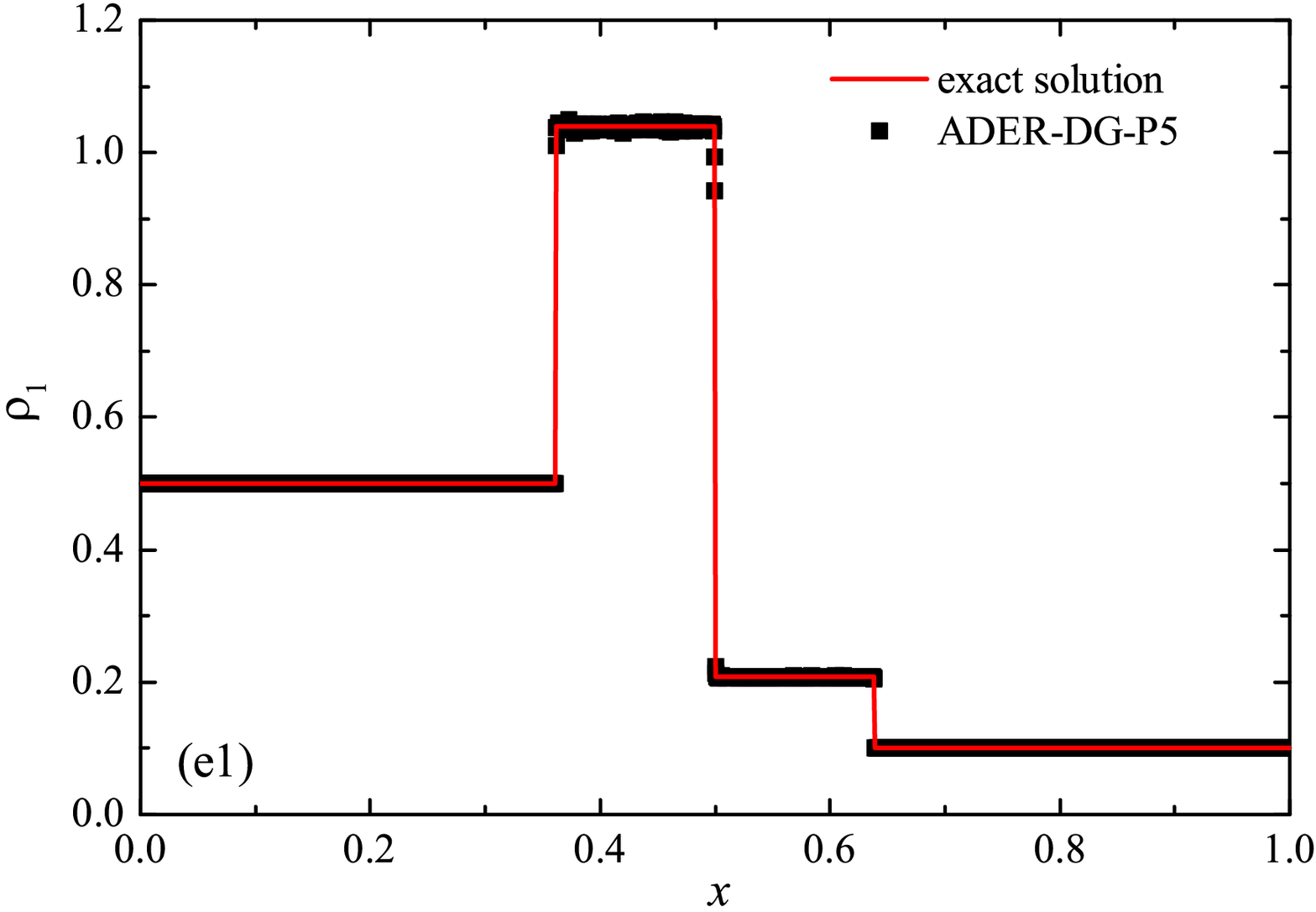}
\includegraphics[width=0.24\textwidth]{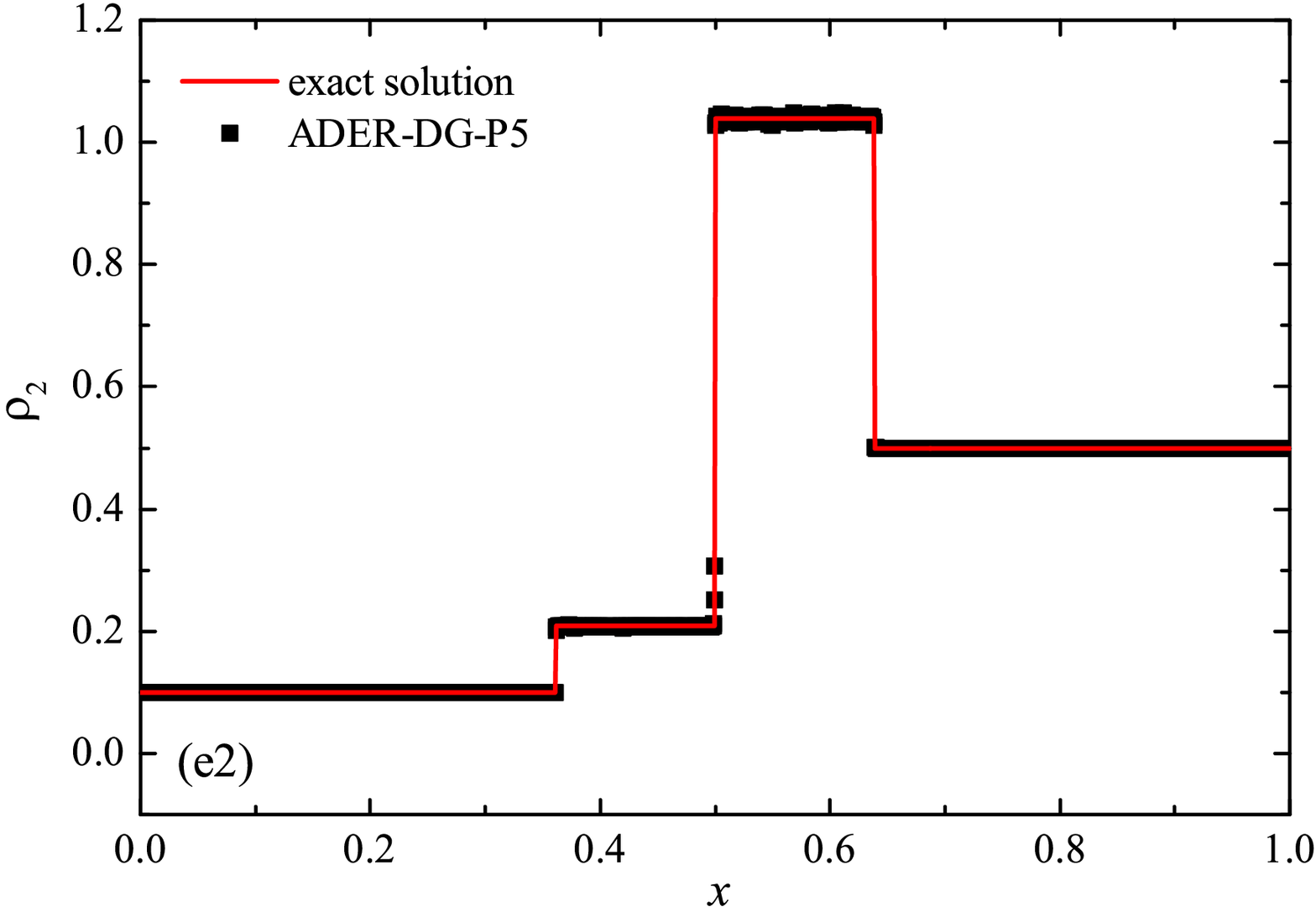}
\includegraphics[width=0.24\textwidth]{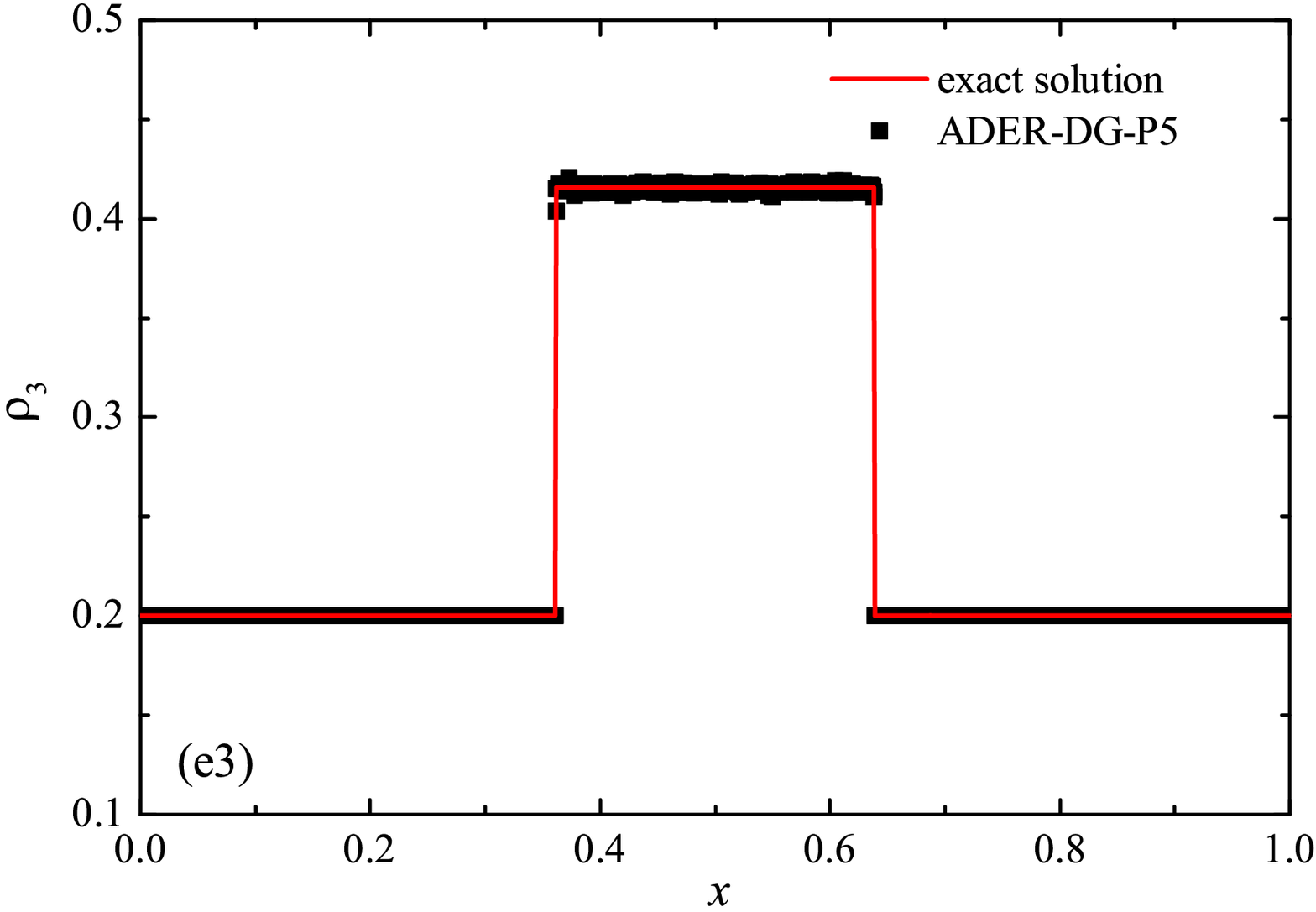}
\includegraphics[width=0.24\textwidth]{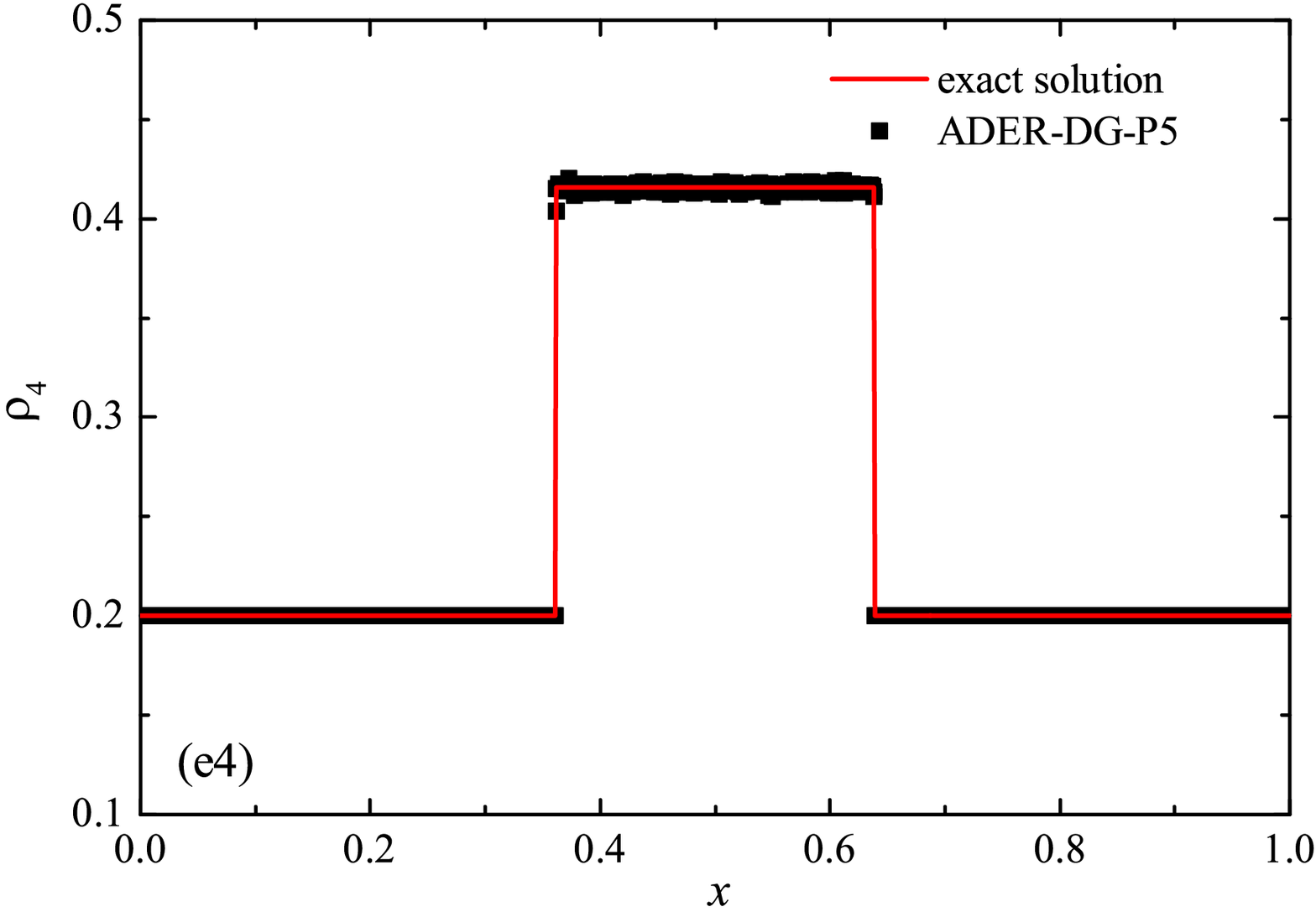}\\
\caption{%
	Numerical solution of the classical problem with two shock waves for a multicomponent medium 
	(a detailed statement of the problem is presented in the text),
	using the computational scheme $\mathrm{ADER}$-$\mathrm{DG}$-$\mathbb{P}_5$ with a posteriori 
	limitation of the solution by a $\mathrm{ADER}$-$\mathrm{WENO}5$ finite volume limiter,
	on a coordinate mesh with $1800$ finite element cells.
	The graphs show the coordinate dependencies of pressure $p$ (a), density $\rho$ (b), flow velocity $u$ (c), 
	sound speed $c$ (d), and densities $\rho_{k} = \rho c_{k}$ (e1-e4) of individual components $k$ of the multicomponent medium,
	at the final time $t_{\rm final} = 0.15$. The black square symbols represent the numerical solution; 
	the red solid lines represents the exact analytical solution of the problem.
}
\label{fig:csw_test_fig}
\end{figure*}

Verification and testing of the developed computer program in this work was carried out on four classical gas-dynamic tests~\cite{Toro_solvers_2009}: the classical Sod and Lax problems ($1$ and $2$ tests), the problem with two strong rarefaction waves ($3$ test), and the problem with two shock waves ($4$ test). The spatial domain of the flow was chosen as $x \in [0.0, 1.0]$. The initial discontinuity was located at the coordinate $x_{\rm c} = 0.5$. The final time of the simulation was chosen as $t_{\rm final} = 0.15$ for all four tests. Data for the parameter values $(\rho, u, p)$ of these four Riemann problem tests are presented in Table~\ref{tab:tests_table}: the parameter values $(\rho_{L}, u_{L}, p_{L})$ correspond to the state of the flow to the left of the discontinuity; the parameter values $(\rho_{R}, u_{R}, p_{R})$ correspond to the state of the flow to the right of the discontinuity. The case of a multicomponent medium consisting of four components was chosen. Data for the mass concentrations $\left(c_{1}, c_{2}, c_{3}, c_{4}\right)$ of each of the four components of the medium, that was used in all four test problems, are represented by the formula: 
\begin{eqnarray}\label{data_initial_mass_concentrations}
\left(c_{1}, c_{2}, c_{3}, c_{4}\right) \hspace{-0.5mm} = \hspace{-1mm}\left\{
\begin{array}{cl}
\left(0.5, 0.1, 0.2, 0.2\right),& \mathrm{if}\, x \in [0.0, 0.5],\\[2mm]
\left(0.1, 0.5, 0.2, 0.2\right),& \mathrm{if}\, x \in [0.5, 1.0],
\end{array}
\right.
\end{eqnarray}
The initial conditions for the first two components (with mass concentrations $c_{1}$ and $c_{2}$) were chosen in the form of a discontinuity in mass concentrations by a factor of $5$; for the other two components (with mass concentrations $c_{3}$ and $c_{4}$), continuous initial conditions were chosen --- this is due to the need for a detailed consideration of the correctness of the solution of the equations of convective transfer of the components of the medium, both under discontinuous initial conditions and under continuous ones (however, of course, hydrodynamic values $(\rho, u, p)$, in this case, undergo a discontinuity). The adiabatic exponents $\gamma_{k}$ of the components were chosen equal to each other and were equal to the adiabatic exponent $\gamma = 1.4$ of the medium. The Courant number for all selected tests was chosen $\mathtt{CFL}\_\mathtt{number} = 0.4$. The study of the numerical solution was carried out using a sequence of $9$ mesh sizes, with the number $N_{\rm cells}$ of finite element cells: $200$, $400$, $600$, $800$, $1000$, $1200$, $1400$, $1600$ and $1800$. Numerical schemes ADER-DG-$\mathbb{P}_{N}$ with the degrees $N = 1$, $2$, $3$, $4$ and $5$ (for the Lax problem, results are obtained for $N \leqslant 4$) of polynomials were investigated. On the boundaries of the spatial domain of the solution, the boundary conditions of the free boundary were set. The sub-cell limiter for the numerical scheme ADER-DG-$\mathbb{P}_{N}$ was the ADER-WENO ($\mathbb{P}_{N}$) finite-volume scheme. Numerical solutions of the tests problems for a multicomponent medium using the computational scheme $\mathrm{ADER}$-$\mathrm{DG}$-$\mathbb{P}_5$ with a posteriori limitation of the solution by a $\mathrm{ADER}$-$\mathrm{WENO}5$ finite volume limiter, on a coordinate mesh with $N_{\rm cells} = 1800$, are presented in Fig.~\ref{fig:sod_test_fig} for Sod problem, in Fig.~\ref{fig:lax_test_fig} for Lax problem, in Fig.~\ref{fig:srw_test_fig} for the problem with two strong rarefaction waves and in Fig.~\ref{fig:csw_test_fig} for the problem with two shock waves. 

\paragraph{Results}

The solution of the Sod problem, presented in Fig.~\ref{fig:sod_test_fig}, contains a shock wave propagating to the right, followed by a contact discontinuity, and a rarefaction wave propagating to the left. The numerical solution obtained by the numerical scheme correctly and accurately displays all the main components of the solution of the Sod problem. The shock wave front spreads into one or two finite element cells: in the case when the front is located inside the cell, then into one cell; in the case when the front goes to the interface of neighboring cells, then to two cells. The shock wave front in the numerical solution does not expand with time $t$. The position and velocity of the shock wave front are stably displayed in the numerical solution, in the coordinate dependencies of pressure $p$, density $\rho$, velocity $u$, speed of sound $c$ and the densities $\rho_{k}$ of the components of a multicomponent medium. The contact discontinuity in the numerical solution of the Sod problem also spreads out into one or two finite element cells and does not expand with time $t$, which is often a problem when using $\mathrm{TVD}$ numerical schemes~\cite{Kulikovskii}. The contact discontinuity is accurately displayed in the coordinate dependence of the density $\rho$ and speed of sound $c$ (as well as the densities $\rho_{k}$ of the components of a multicomponent medium), while it is not displayed in any way in the coordinate dependencies of pressure $p$ and velocity $u$. The rarefaction wave is reproduced quite accurately, and there are no significant oscillations at the leading front of the wave. It should be noted that with increase in the degree of polynomials $N$ used in the ADER-DG-$\mathbb{P}_N$ and $\mathrm{ADER}$-$\mathrm{WENO}$ ($\mathbb{P}_{N}$) numerical schemes, small oscillations of the numerical solution arise in the region of space between the front of the shock wave and the leading front of the rarefaction wave. The oscillations between the shock wave front and the contact discontinuity are larger in amplitude than the oscillations between the contact discontinuity and the leading front of the rarefaction wave. The amplitude of the oscillations does not change significantly as the polynomials $N$ used in the numerical schemes increase, starting from the degree of $3$. High-order ADER-DG-$\mathbb{P}_N$ numerical schemes were used in the work~\cite{ader_dg_ideal_flows}, but the ADER-WENO-FV schemes were limited to the $3$ order. The coordinate dependencies of the densities $\rho_{k}$ and mass concentrations $c_{k}$ of individual components of the medium in the numerical solution correctly and accurately reflect all the main components of the solution to the Sod problem. The quality of the obtained coordinate dependencies for the components of the medium corresponds to the quality of the results obtained for the main hydrodynamic quantities. 

The solution of the Lax problem, similarly to the solution of the Sod problem, contains a shock wave, a contact discontinuity, and a resolution wave. However, the solution to the Lax problem is more complex~\cite{Toro_solvers_2009}, due to the change in density difference and the non-zero flow velocity to the left of the discontinuity. This creates additional complexity for the numerical solution, so the solution of the problem usually reveals more peculiarities and artifacts of the numerical scheme. The solution of the Lax problem shown in Fig.~\ref{fig:lax_test_fig} demonstrates the behavior of the numerical solution, which is similar to the numerical solution of the Sod problem. However, the numerical solution of the Lax problem was obtained only for $N \leqslant 4$ degrees of polynomials. For higher degrees of polynomials $N$, a decrease in the Courant number $\mathtt{CFL}\_\mathtt{number} = 0.3$ and below was required. The numerical solution of the Lax problem demonstrates noticeable oscillations of the solution in the vicinity of the contact discontinuity, which also spreads into $4$-$5$ finite-element cells. The shock wave and rarefaction wave are accurately represented in the numerical solution, similar to the numerical solution of the Sod problem. In this case, the leading front of the rarefaction wave is also accurately reproduced, without significant oscillations of the numerical solution. The coordinate dependencies of the densities $\rho_{k}$ and mass concentrations $c_{k}$ of the components of the medium in terms of the quality of the numerical solution are, in general, similar to the numerical solution for the main hydrodynamic quantities. 

The numerical solution of the problem with two rarefaction waves, presented in Fig.~\ref{fig:srw_test_fig}, demonstrates the sufficient capabilities of the numerical scheme to obtain a solution in areas with strong rarefaction. The solution of the problem contains symmetric rarefaction waves, which are correctly and accurately resolved in the numerical solution. In the vicinity of rarefaction waves fronts closest to the center, small oscillations of the numerical solution, typical for this problem, are observed. The coordinate dependencies of the density $\rho$ and speed of sound $c$ demonstrate a small ``carbuncle'' in the center $x_{\rm c} = 0.5$ of the computational domain. The coordinate dependencies of the densities $\rho_{k}$ and mass concentrations $c_{k}$ of the medium components quite accurately repeat the peculiarities of the numerical solution of the problem for the main hydrodynamic quantities. The initial discontinuity in the concentrations of components $1$ and $2$ spreads out into $2$-$3$ finite element cells. 

The solution to the problem with two counterflows, presented in Fig.~\ref{fig:csw_test_fig}, contains two symmetrical divergent shock waves in the solution. The numerical solution of this test problem, in contrast to the three previous test problems, is characterized by the strongest oscillations of the numerical solution, in the region behind the shock waves. The intensity of non-physical oscillations increases with the degree $N$ of polynomials used in the numerical scheme, starting from the degree $N = 3$. This is connected, first of all, with the oscillations generated by the WENO scheme, with the order of the polynomials in the reconstruction $N \geqslant 3$. The coordinate dependencies of the densities $\rho_{k}$ and mass concentrations $c_{k}$ of the medium components quite accurately repeat the peculiarities of the numerical solution of the problem for the main hydrodynamic quantities.

As a result of the analysis of the obtained numerical solution of these four test problems, it can be concluded that the use of the space-time adaptive ADER-DG finite element method 
with LST-DG predictor and a posteriori sub-cell WENO finite-volume limiting for simulation of non-stationary compressible multicomponent reactive flows does not lead to the appearance of additional non-physical peculiarities and numerical artifacts in the spatial and temporal dependencies of the components of the medium, except for those that are already characteristic of the original numerical method~\cite{ader_dg_ideal_flows, ader_dg_dev_1, ader_dg_dev_2, ader_weno_lstdg, ader_dg_diss_flows, ader_dg_ale, ader_dg_grmhd, ader_dg_gr_prd, ader_dg_PNPM, PNPM_DG, ader_dg_eff_impl, fron_phys, exahype, ader_dg_hpc_impl_1, ader_dg_hpc_impl_2, ader_dg_hpc_impl_3, ader_dg_hpc_impl_4}.

\subsection{ZND-detonation waves}
\label{sec:3}

\paragraph{Adaptive change in the time step}
\label{subsec:3:1}

Obtaining a correct numerical solution using the computational scheme used in this work described above in the case of strongly stiffness of source terms associated with reaction kinetics and energy yield is generally problematic --- large and sensitive changes in the reaction rate constants and concentration factors in the reaction rates significantly slow down the convergence of the iterative process of obtaining the discrete space-time solution by the LST-DG predictor and quite often leads to \texttt{nan}, due to a significant change in the values of the source terms even with small variations in the medium parameters --- composition $c_{k}$, density $\rho$, temperature $T$ (and therefore pressure $p$).

An approach similar to the adaptive change in the time step presented in the work~\cite{chem_kin_hrs_weno} was used (a description of the use of such techniques is also presented in the review~\cite{chem_kin_hrs_rev}). The approach of adaptive change in the time step was applied in the work~\cite{chem_kin_hrs_weno} to the method of the Strang splitting by physical processes, when the full time step was split into two steps --- the hydrodynamic step, in which the hydrodynamic characteristics were calculated using a high-precision finite-volume WENO scheme, and the reaction kinetic step, within which the equations of chemical kinetics were solved by the classical third-order total variation diminishing Runge-Kutta method. It is clear that within a single step, this approach of adaptive change in the time step can be implemented, as in the work~\cite{chem_kin_hrs_weno}. However, in this work, the splitting method was not used --- in the framework of a full one-step scheme, the entire initial system of equations was solved. Therefore, using the adaptive time step approach requires some modifications to it. Stiffness of the terms of the problem is associated with a significant difference in the time (and spatial) scales of hydrodynamic processes and reaction kinetics, so the scaling factor should be focused on the relative time scales of the processes.

Relative characteristic times $\tau^{k}_{R}$ of the source terms $\mathbf{S}$, normalized to conserved variables $\mathbf{U}$, were chosen as the ``rates'' of source terms:
\begin{equation}
\left(\tau^{k}_{R}\right)^{-1} = \frac{[\mathbf{S}]_{k}}{[\mathbf{U}]_{k}},
\end{equation}
where $[\mathbf{U}]_{k}$ and $[\mathbf{S}]_{k}$ is the $k$-th component of the vectors of conserved variables $\mathbf{U}$ and source terms $\mathbf{S}$; there are no source terms for mass and momentum in the system of equations (\ref{system_of_equations}), so the time scales are determined only by the relative rate of energy yield $S_{e}$ of the reactions and the relative rates of the reactions $\mathbf{S}_{r}$ in the medium. From the obtained values of the relative characteristic times $\tau^{k}_{R}$, the minimum value $\tau^{min}_{R}$ was chosen on the full spatial grid, corresponding to the locally fastest process. The calculation of the minimum value $\tau^{min}_{R}$ of the relative characteristic time was carried out for all sub-cells of each cell of the spatial grid, and not by analyzing the extrema of the DG-representations of the solution, as in the cases of verifying the DMP (within the framework of the NAC) and calculating the time steps $\tau^{n}_{DG}$ for the ADER-DG scheme. It should be noted that the relative time scales were chosen from the ratio of the local increment of values to their current state, and not to the rate of hydrodynamic processes (which are determined by flux terms $\mathbf{F}$), so no indirect slowdown of processes is expected. The adaptive factor $\alpha$ can be obtained by the following formula (modified expression from work~\cite{chem_kin_hrs_weno}):
\begin{equation}\label{adaptive_factor}
\alpha\left(\frac{\tau^{n}_{DG}}{\tau^{min}_{R}}\right) = \left(N_{R} - 1\right)\left[1 - \exp\left(-A \frac{\tau^{n}_{DG}}{\tau^{min}_{R}}\right)\right] + 1,
\end{equation}
where the parameter $N_{R}$ defines the limit value for reducing the time step; the parameter $A > 0$ is chosen for reasons of stability --- the higher the parameter, the faster the step decreases with an increase in the rate of source processes, such as reactions and their energy yield (typical value $A = 10^{1} - 10^{3}$ in~\cite{chem_kin_hrs_weno}). The expression (\ref{adaptive_factor}) for the factor allows you to accurately track the area of change in the character of processes in the system: in the case of sufficiently slow processes described by the source terms, the condition $\tau^{n}_{DG} \ll \tau^{min}_{R}$ is satisfied, which corresponds to the limiting case $\alpha(0) = 1$; in the case of fast processes described by the source terms, when the problem becomes stiff, the condition $\tau^{n}_{DG} \gg \tau^{min}_{R}$ is satisfied, which corresponds to the limiting case $\alpha(\infty) = N_{R}$. The dimensionless parameter $A$ is used to set the width of the transition region of the adaptive factor change. The resulting value of the adapted time step $\tau^{n}_{res}$ was calculated using the value of the adaptive factor $\alpha$:
\begin{equation}
\tau^{n}_{res} = \frac{\tau^{n}_{DG}}{\alpha\left(\frac{\tau^{n}_{DG}}{\tau^{min}_{R}}\right)}.
\end{equation}
The resulting value of the adapted time step $\tau^{n}_{res}$ was used both for the ADER-DR scheme and for the ADER-WENO-FV scheme. To obtain the results of calculations of detonation waves in two-component medium with model kinetics presented below, the values of the parameters $A = 1.0$ and $N_{R} = 100$ were used.

It is necessary to add information that, in the case of using the strategy of adaptive mesh refinement~\cite{ader_dg_ideal_flows, ader_dg_dev_1, ader_dg_dev_2, ader_weno_lstdg, ader_dg_diss_flows, ader_dg_ale, ader_dg_grmhd, ader_dg_gr_prd, ader_dg_PNPM, PNPM_DG, ader_dg_eff_impl, fron_phys, exahype, ader_dg_hpc_impl_1, ader_dg_hpc_impl_2, ader_dg_hpc_impl_3, ader_dg_hpc_impl_4}, the use of adaptive change in the time step can be realized within the framework of the strategy of local time stepping; at the same time, information about the strongly stiffness of source terms of local flow regions can be used within the framework of an indicator function that determines the performance of procedures for refining/coarse grid cells.

\paragraph{Formulation of the problem and the results obtained}
\label{subsec:3:2}

\begin{figure*}
\centering
\includegraphics[width=0.32\textwidth]{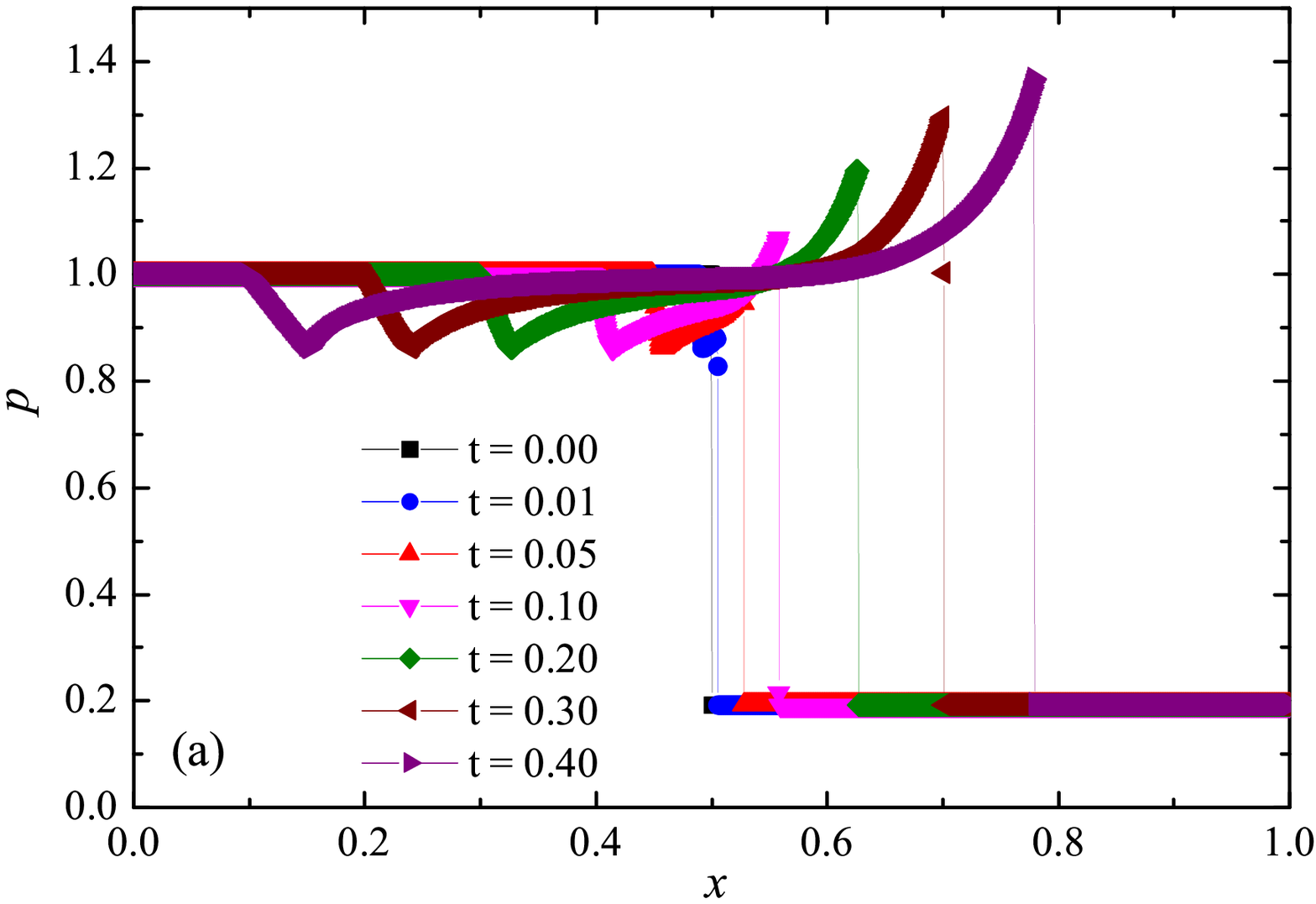}
\includegraphics[width=0.32\textwidth]{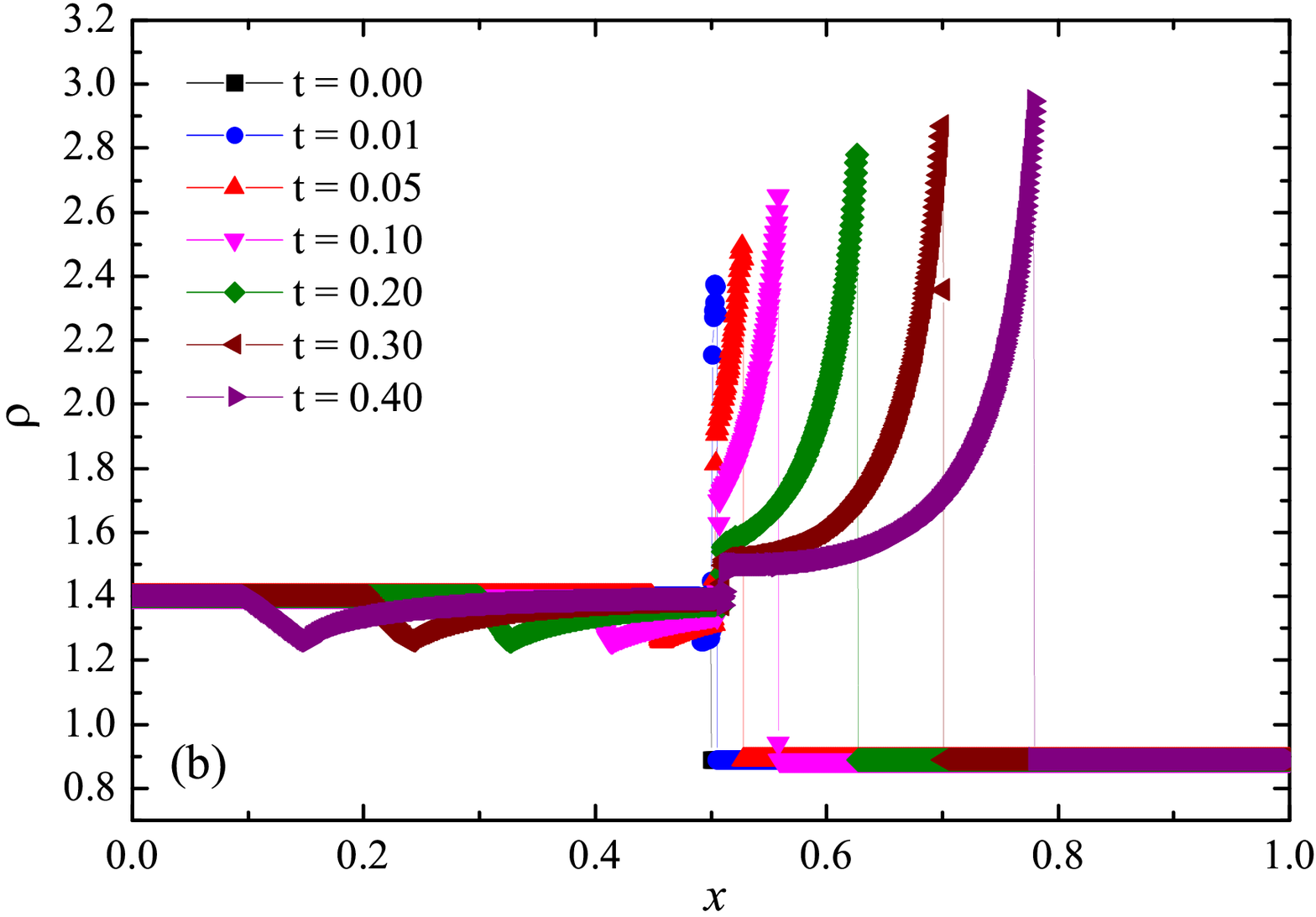}
\includegraphics[width=0.32\textwidth]{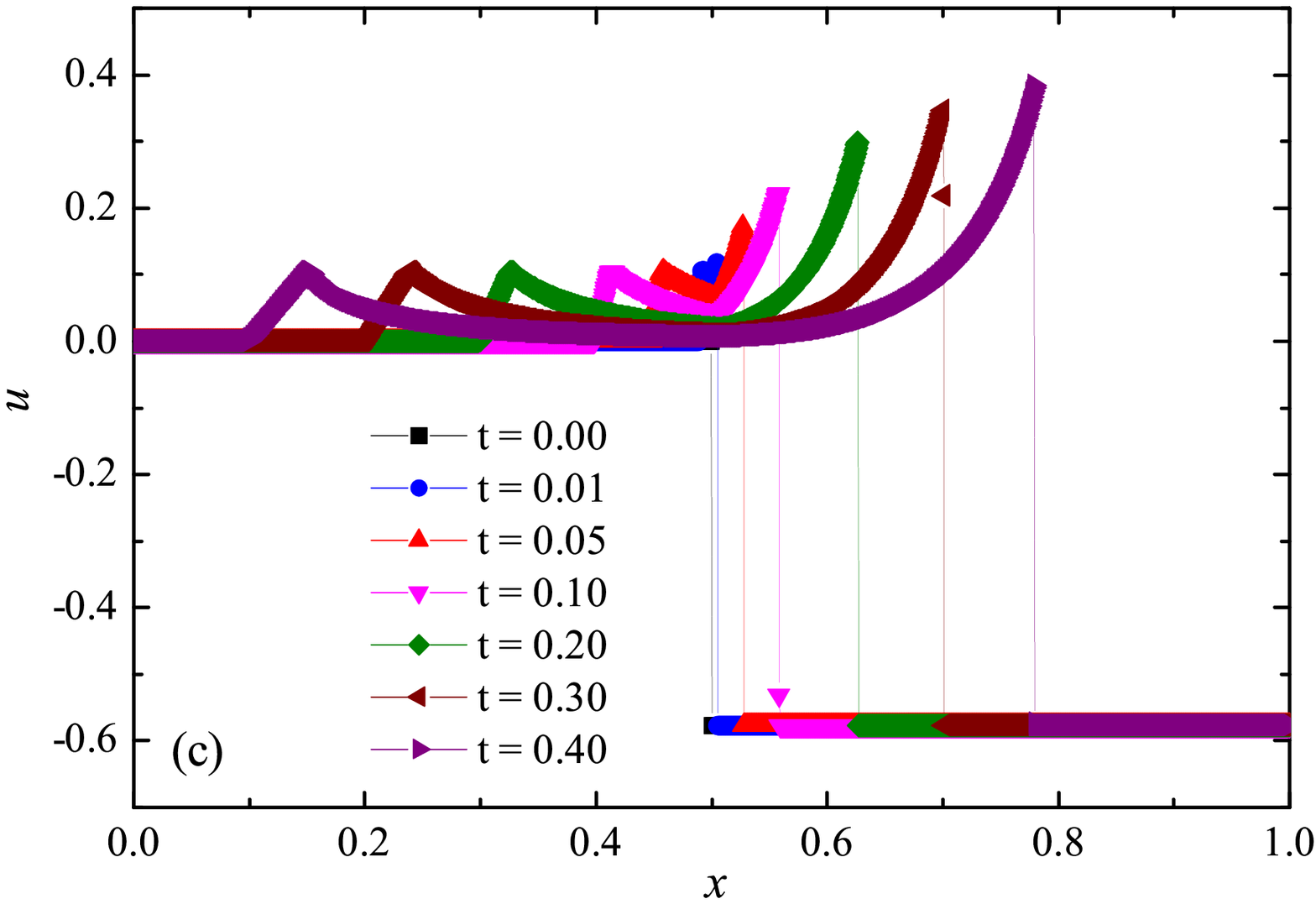}\\
\includegraphics[width=0.32\textwidth]{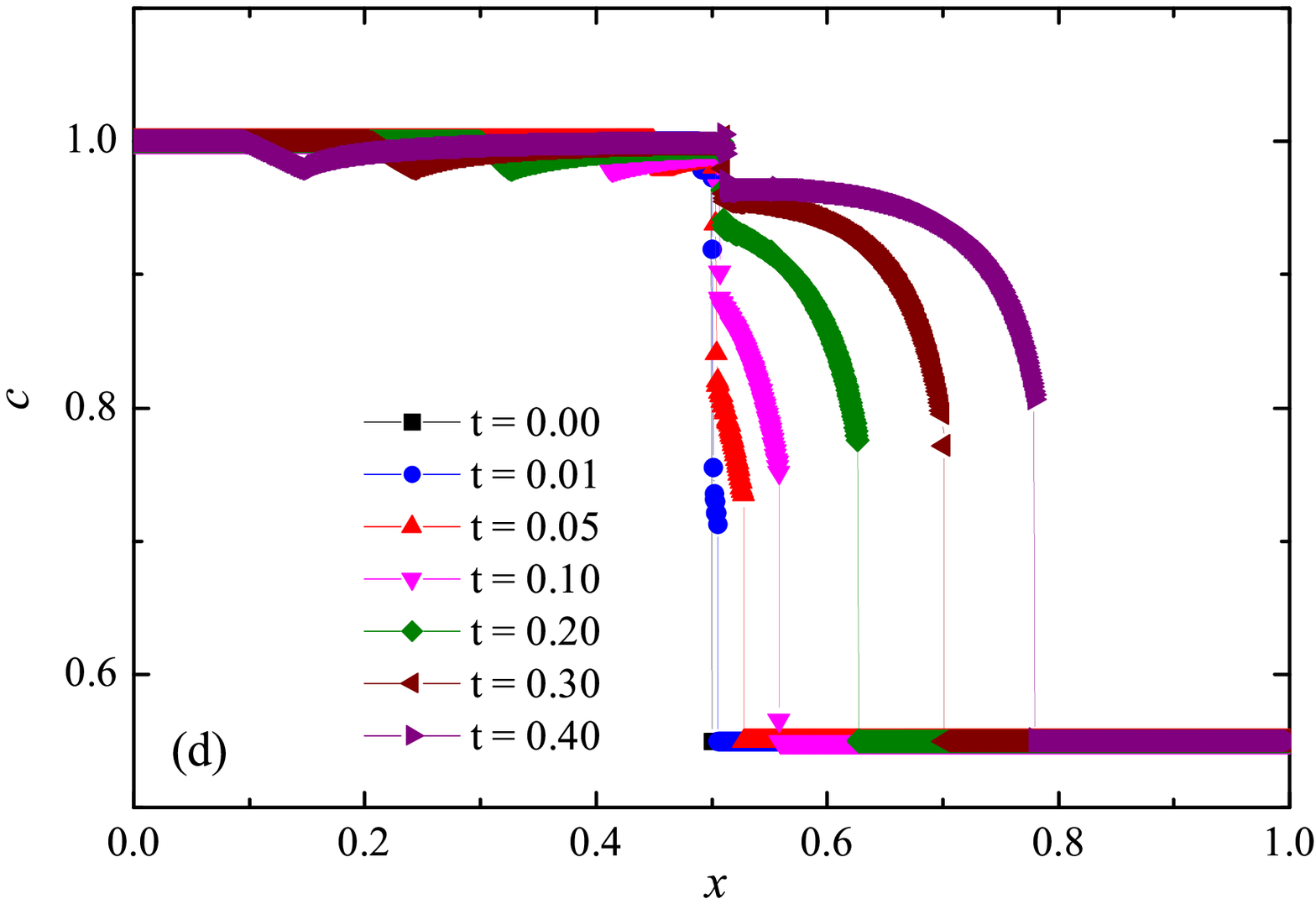}
\includegraphics[width=0.32\textwidth]{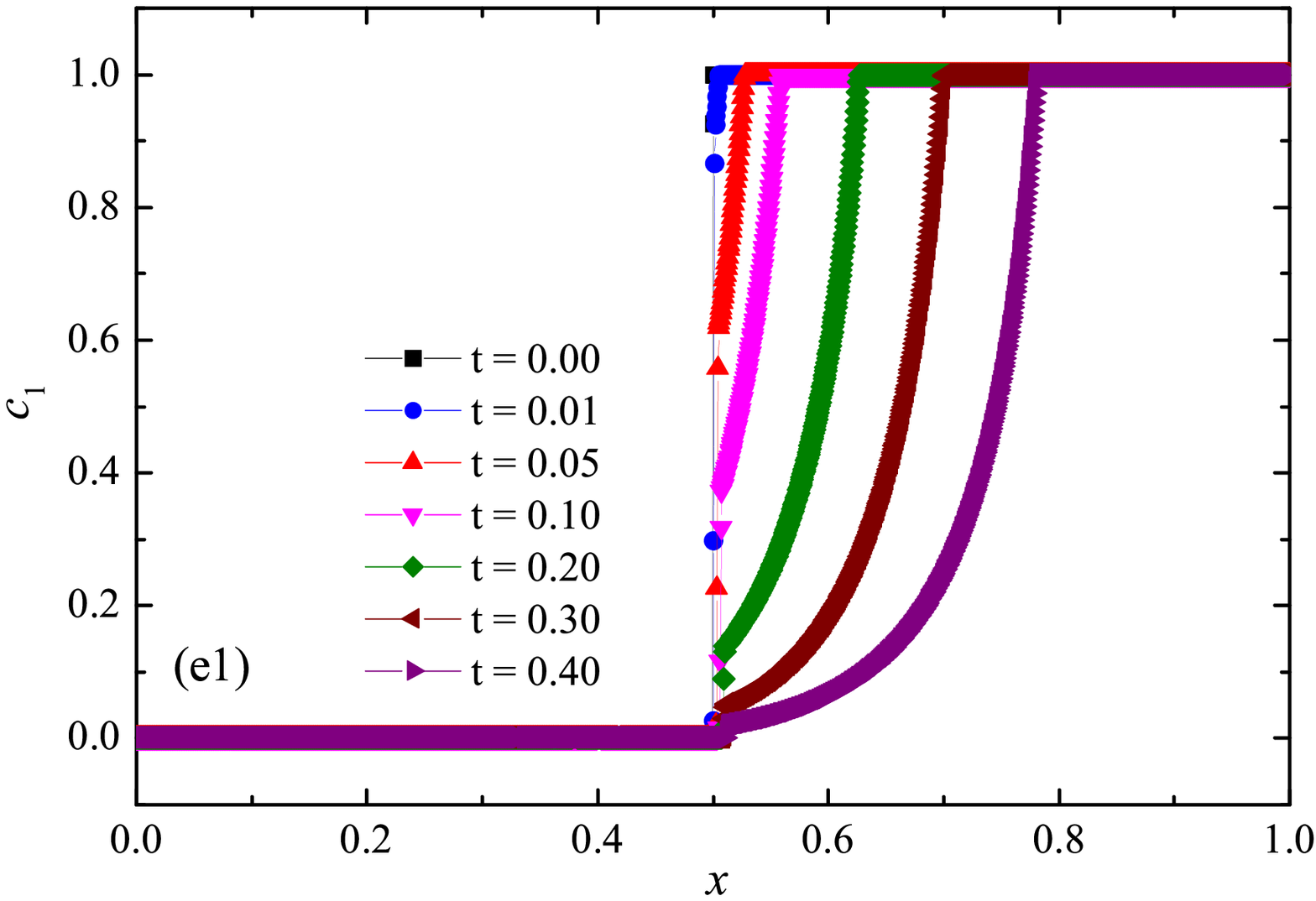}
\includegraphics[width=0.32\textwidth]{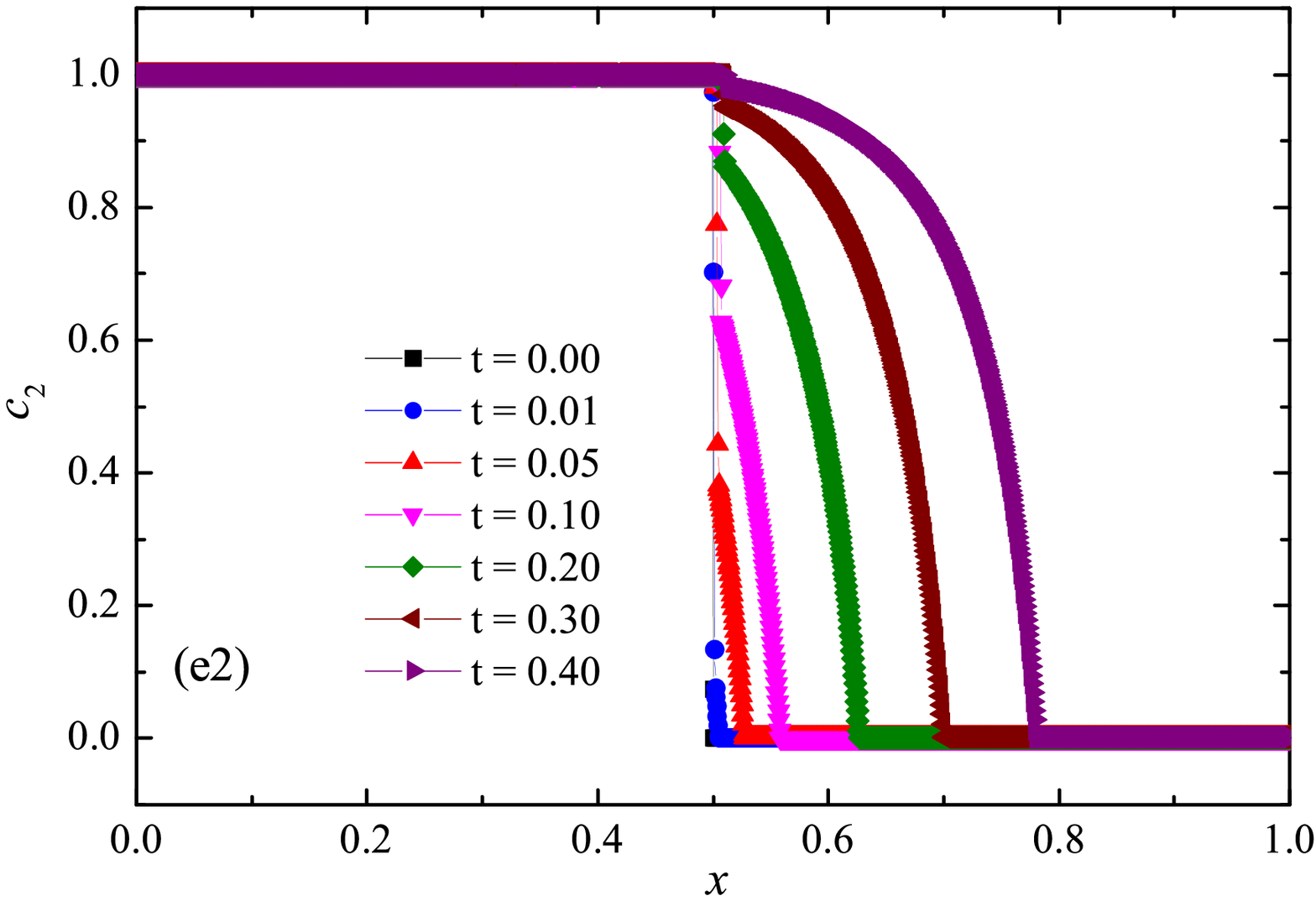}
\caption{%
	Reference solution of the problem of the formation of a detonation wave in a two-component medium with a ``slow'' reaction:
	the solution is obtained using the numerical scheme $\mathrm{ADER}$-$\mathrm{DG}$-$\mathbb{P}_1$ with $3200$ cells.
	The graphs show the coordinate dependencies of pressure $p$ (a), density $\rho$ (b), flow velocity $u$ (c), 
	sound speed $c$ (d) and mass concentrations $c_{k}$ (e1-e2) 
	of individual components $k$ of the two-component medium,
	at the times $t = 0$ (initial conditions), $0.01,\ 0.05,\ 0.10,\ 0.20,\ 0.30$ and $0.4$ ($t_{\rm final}$).
}
\label{fig:figs_s_dwt_f_ref_solution}
\end{figure*} 

\begin{figure*}
\centering
\includegraphics[width=0.32\textwidth]{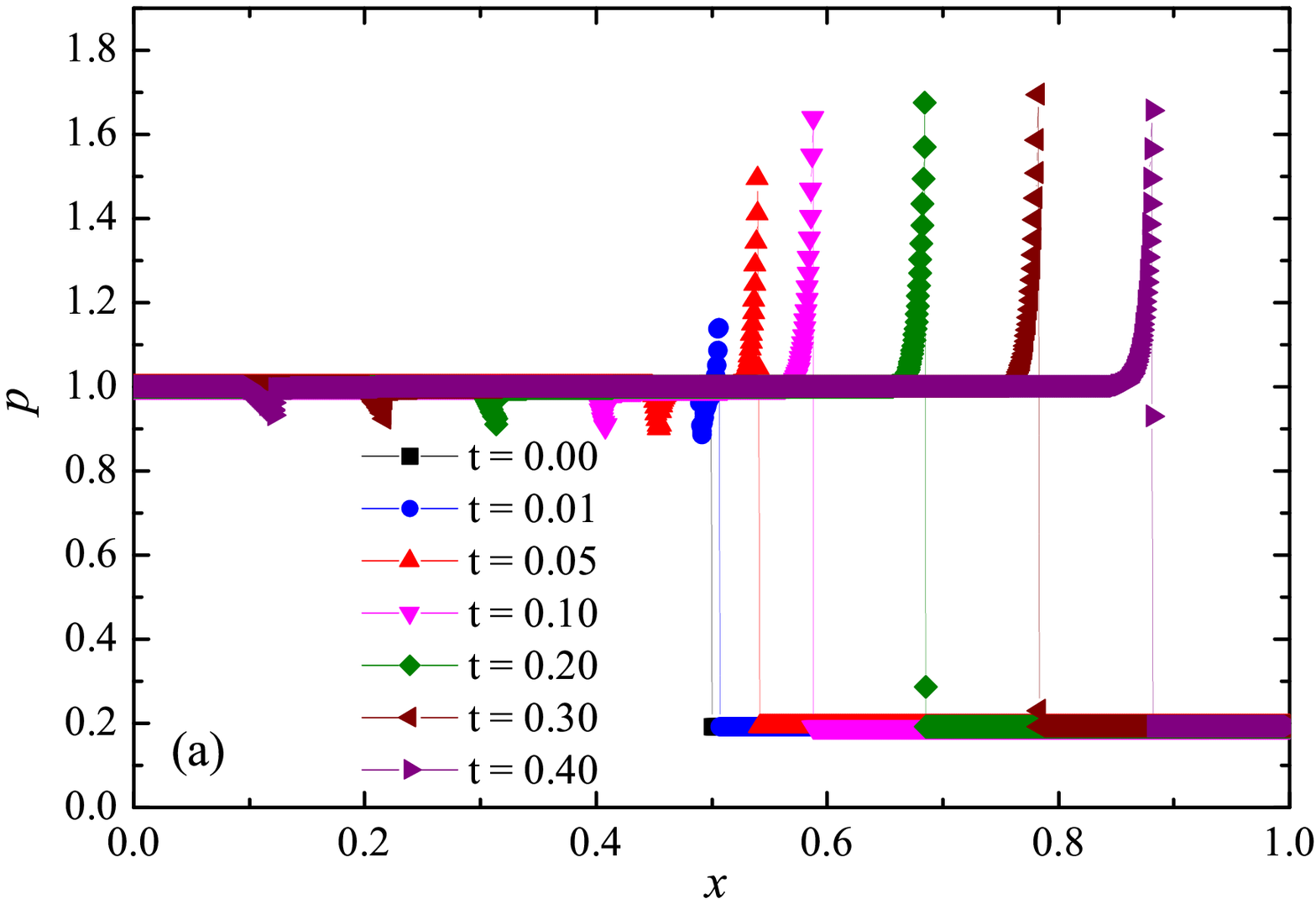}
\includegraphics[width=0.32\textwidth]{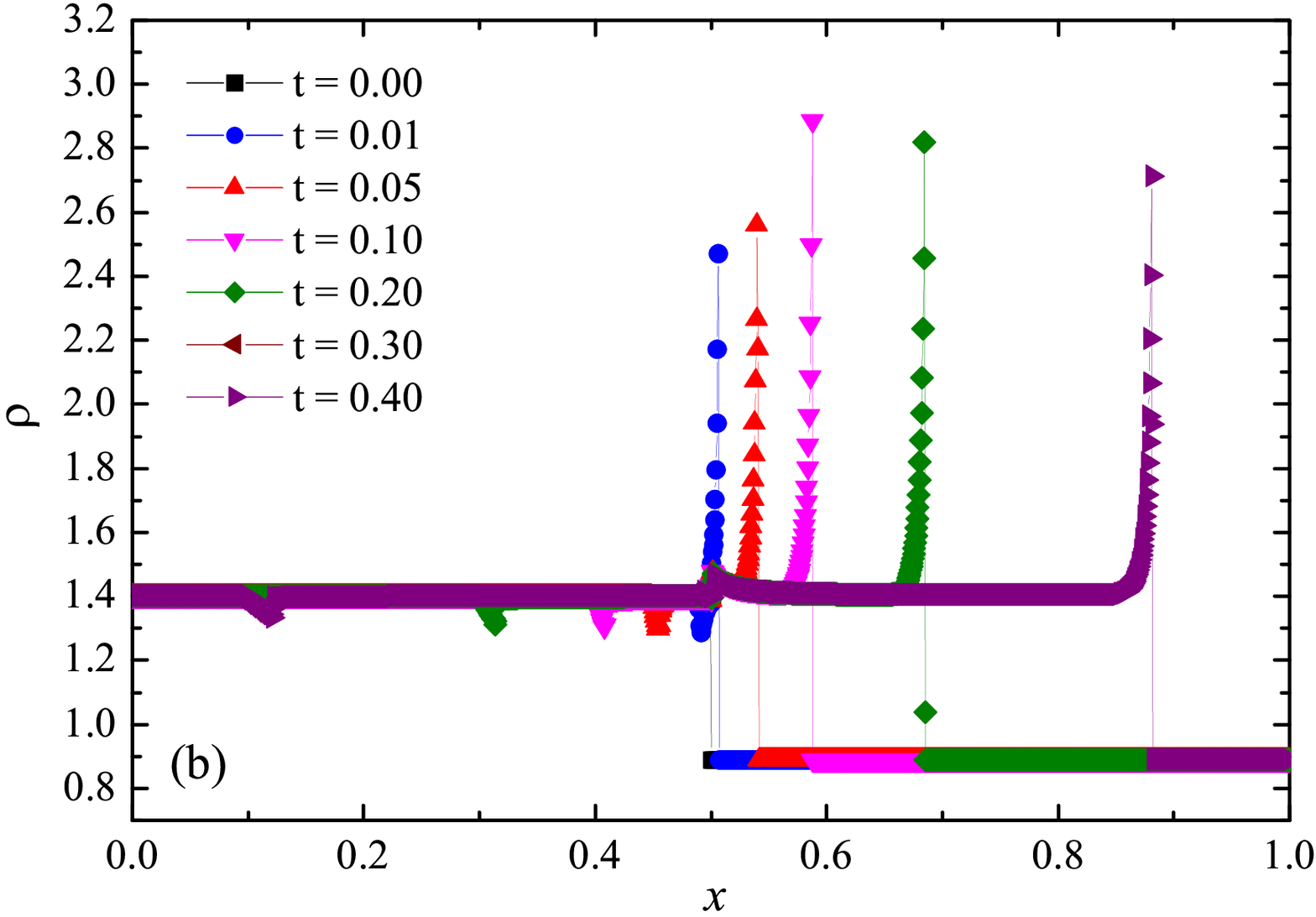}
\includegraphics[width=0.32\textwidth]{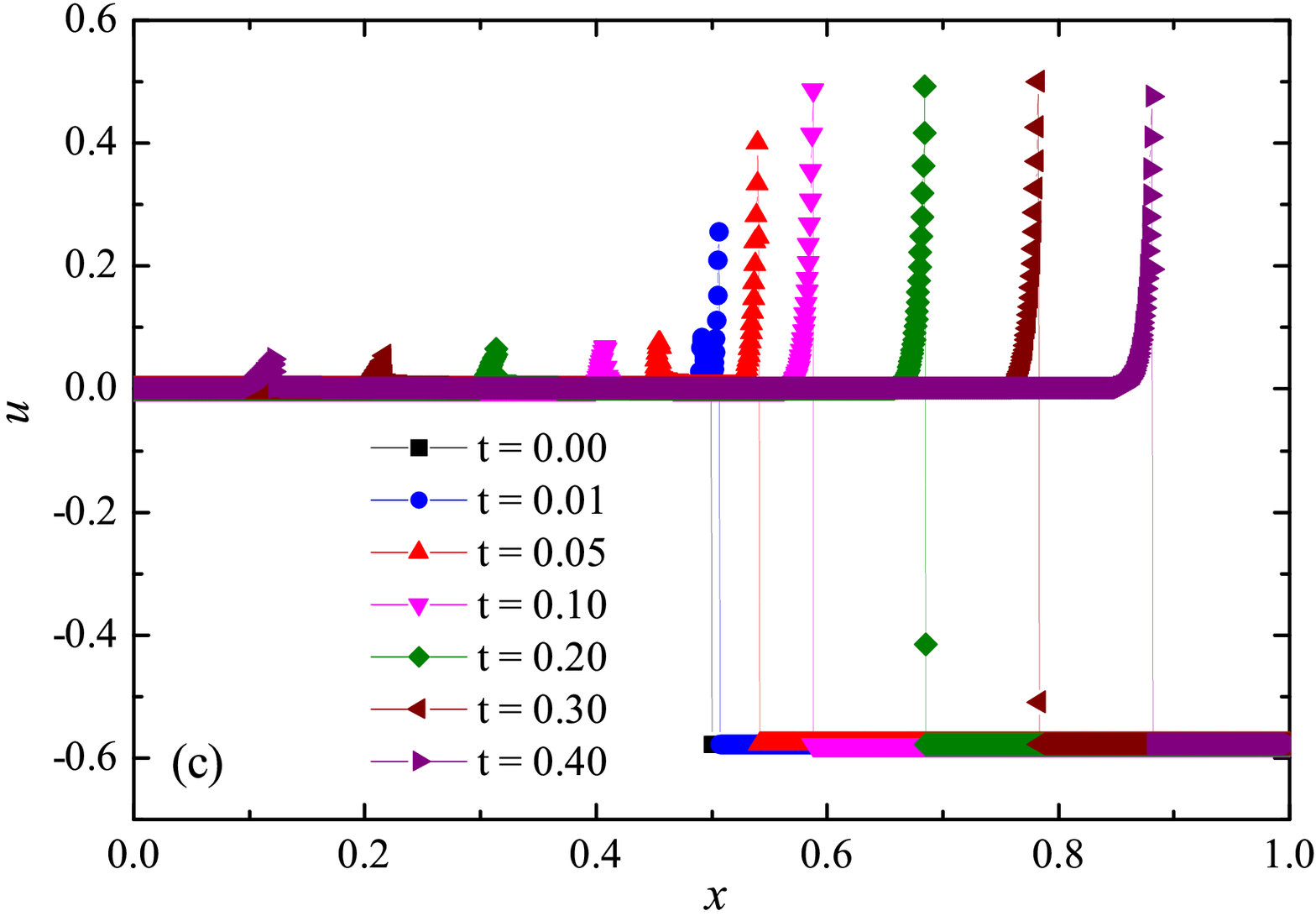}\\
\includegraphics[width=0.32\textwidth]{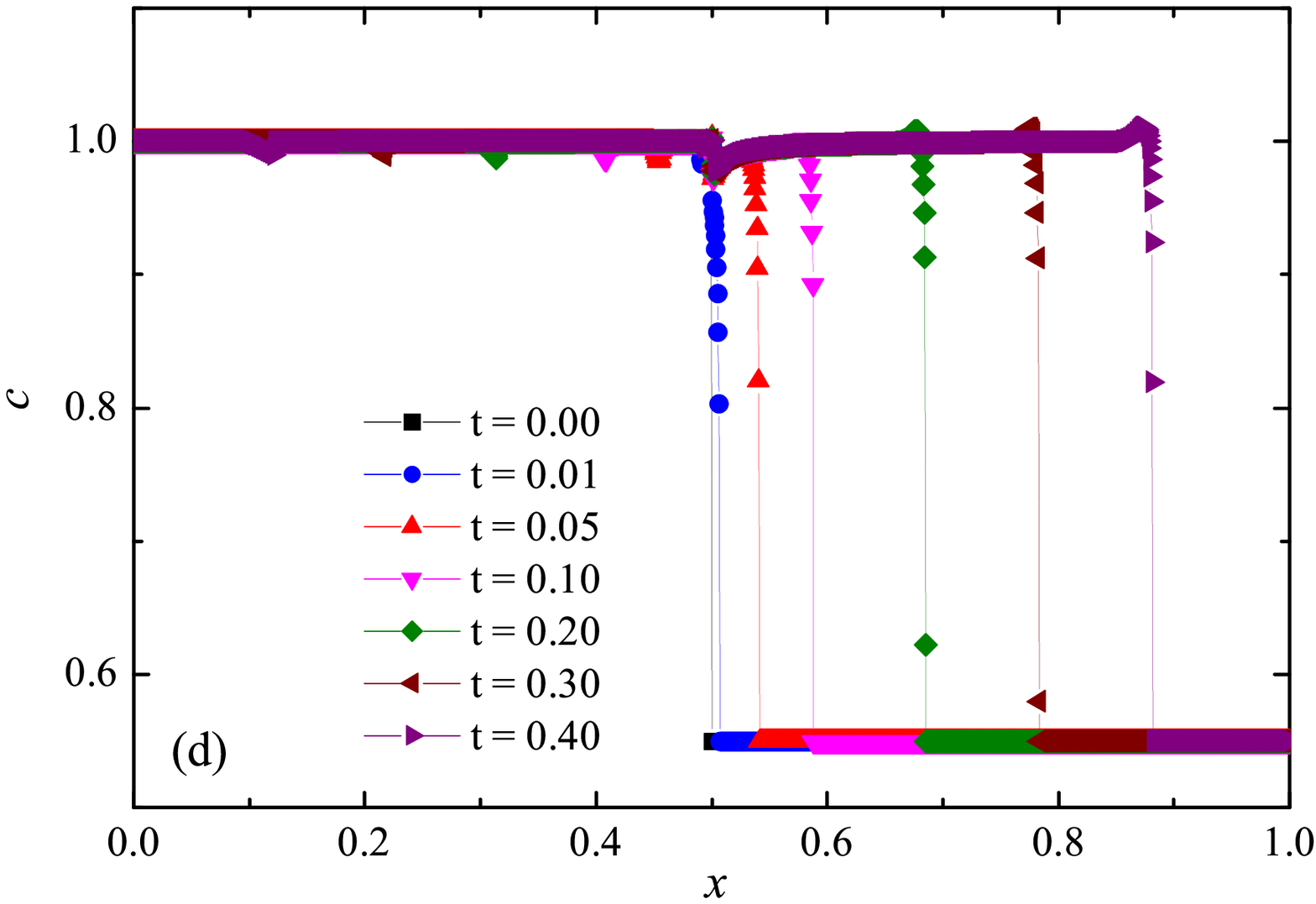}
\includegraphics[width=0.32\textwidth]{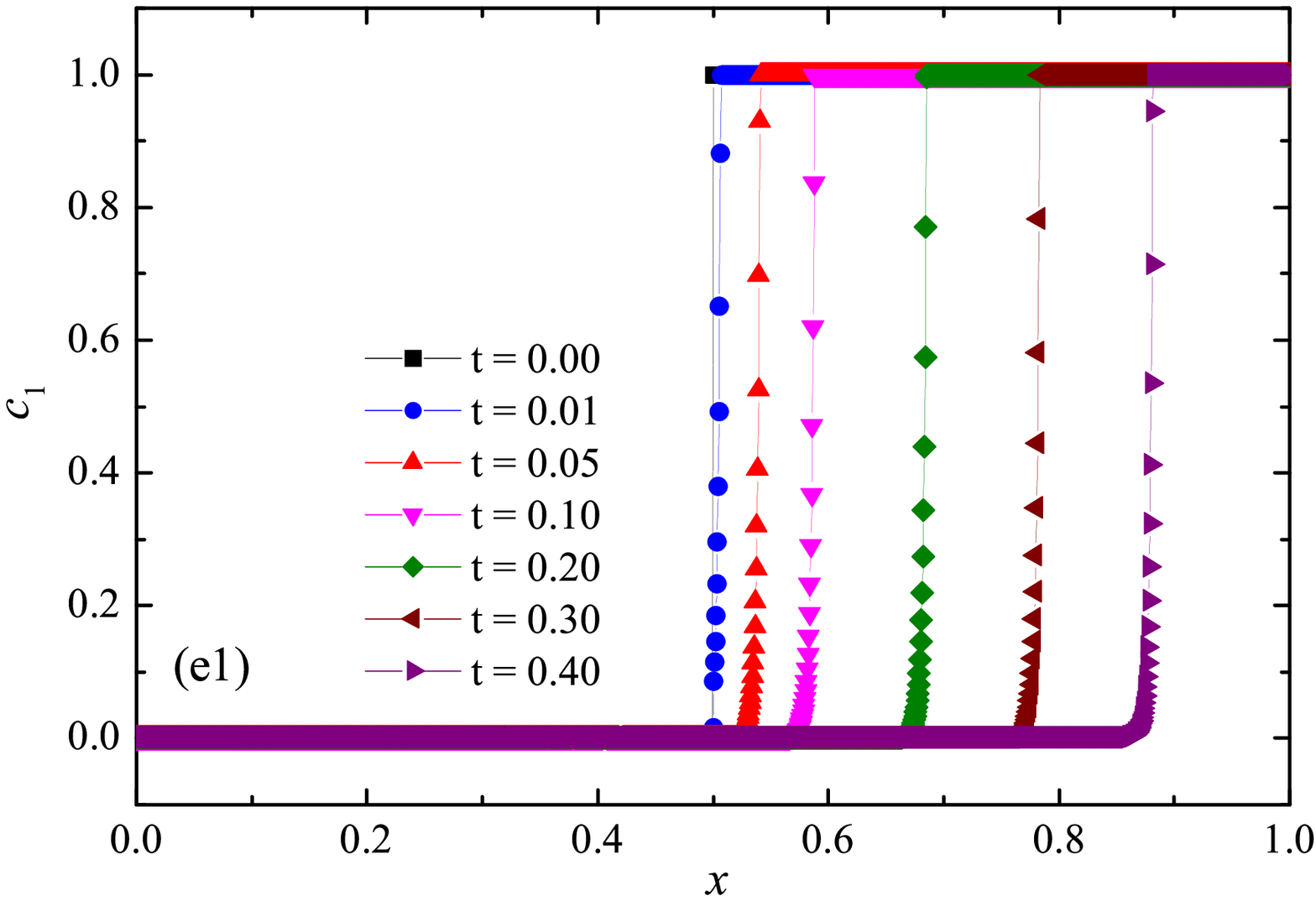}
\includegraphics[width=0.32\textwidth]{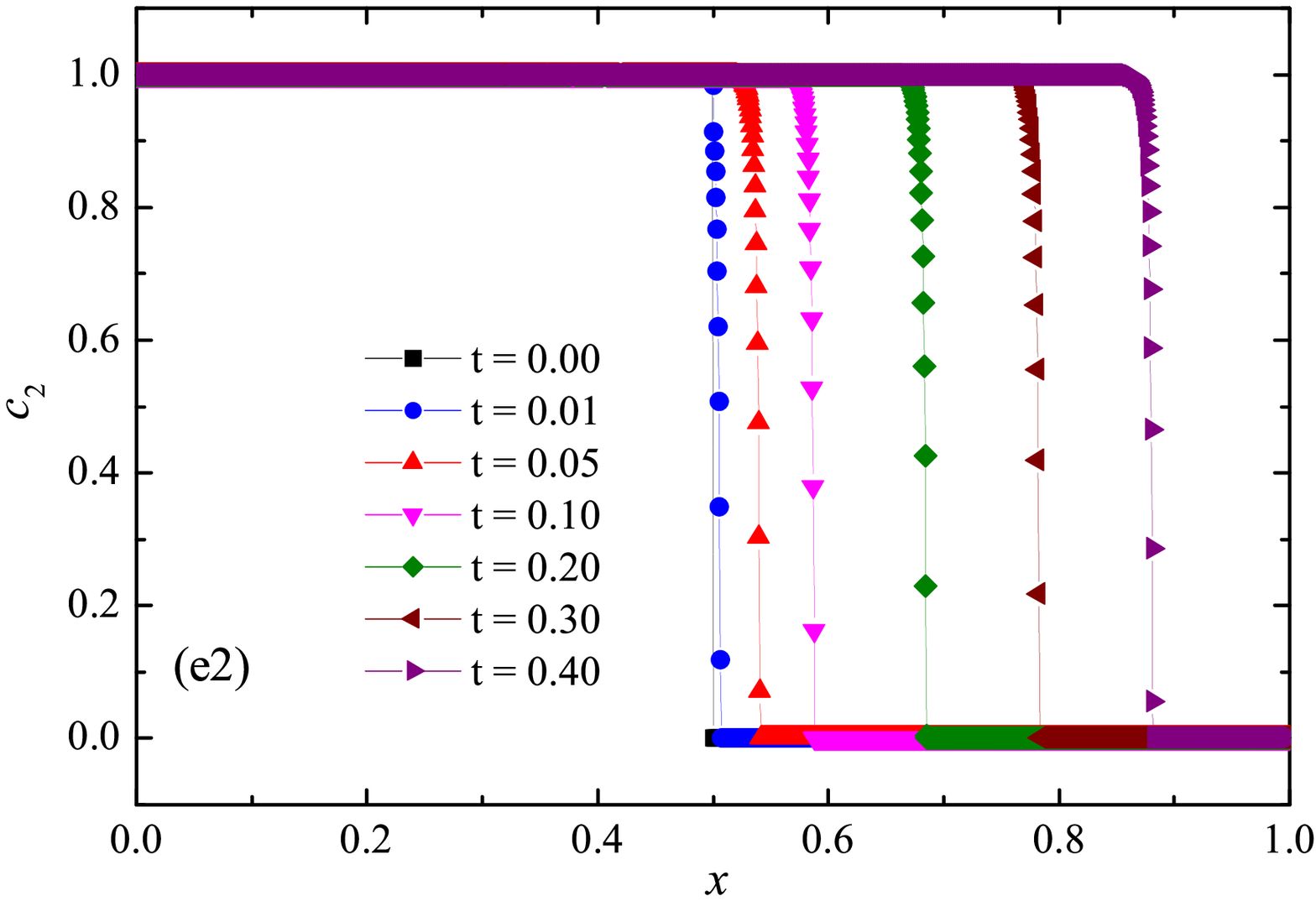}
\caption{%
	Reference solution of the problem of the formation of a detonation wave in a two-component medium with a ``fast'' reaction:
	the solution is obtained using the numerical scheme $\mathrm{ADER}$-$\mathrm{DG}$-$\mathbb{P}_1$ with $3200$ cells.
	The graphs show the coordinate dependencies of pressure $p$ (a), density $\rho$ (b), flow velocity $u$ (c), 
	sound speed $c$ (d) and mass concentrations $c_{k}$ (e1-e2) 
	of individual components $k$ of the two-component medium,
	at the times $t = 0$ (initial conditions), $0.01,\ 0.05,\ 0.10,\ 0.20,\ 0.30$ and $0.4$ ($t_{\rm final}$).
}
\label{fig:figs_s_dwt_s_ref_solution}
\end{figure*} 

\begin{figure*}
\centering
\includegraphics[width=0.24\textwidth]{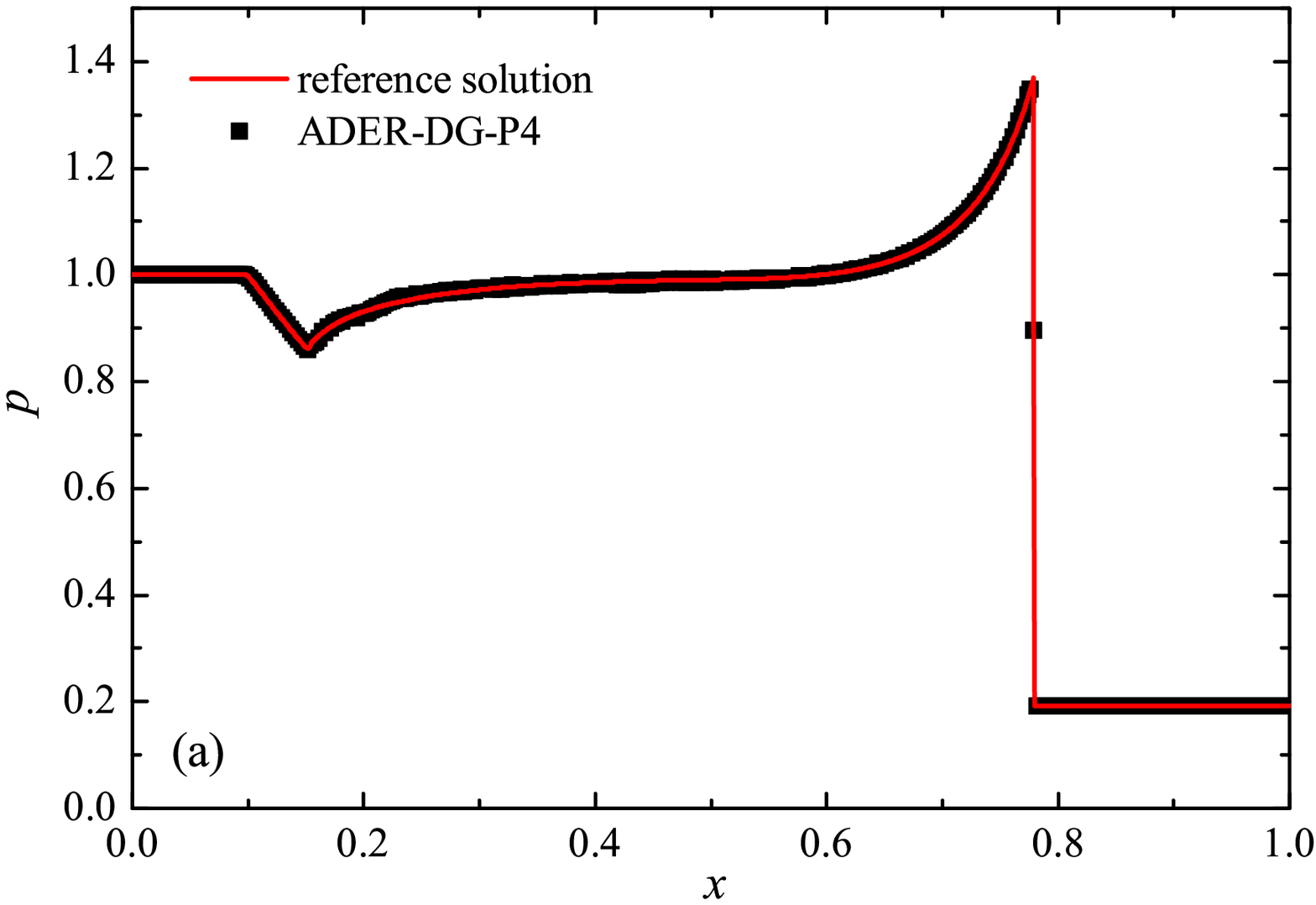}
\includegraphics[width=0.24\textwidth]{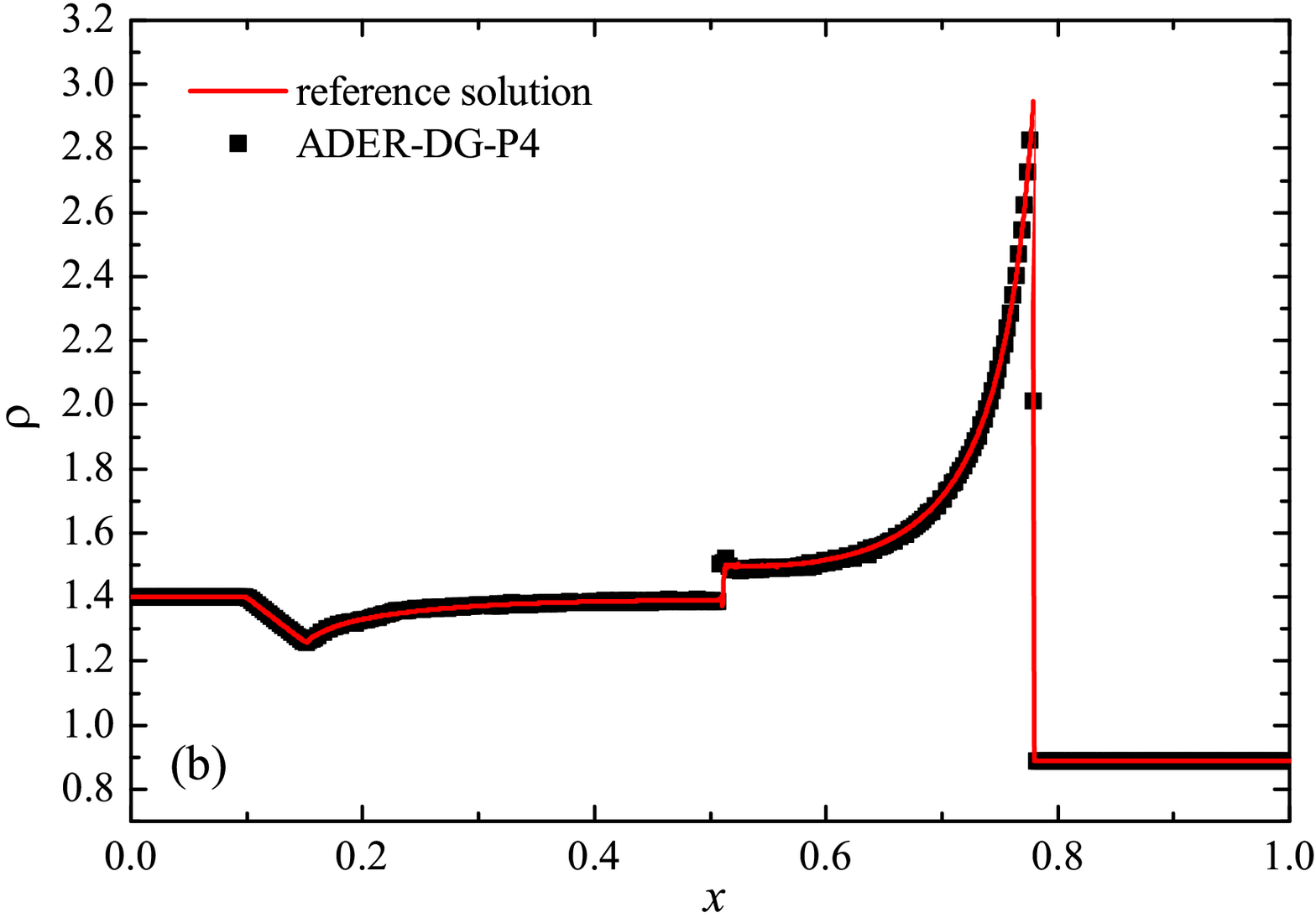}
\includegraphics[width=0.24\textwidth]{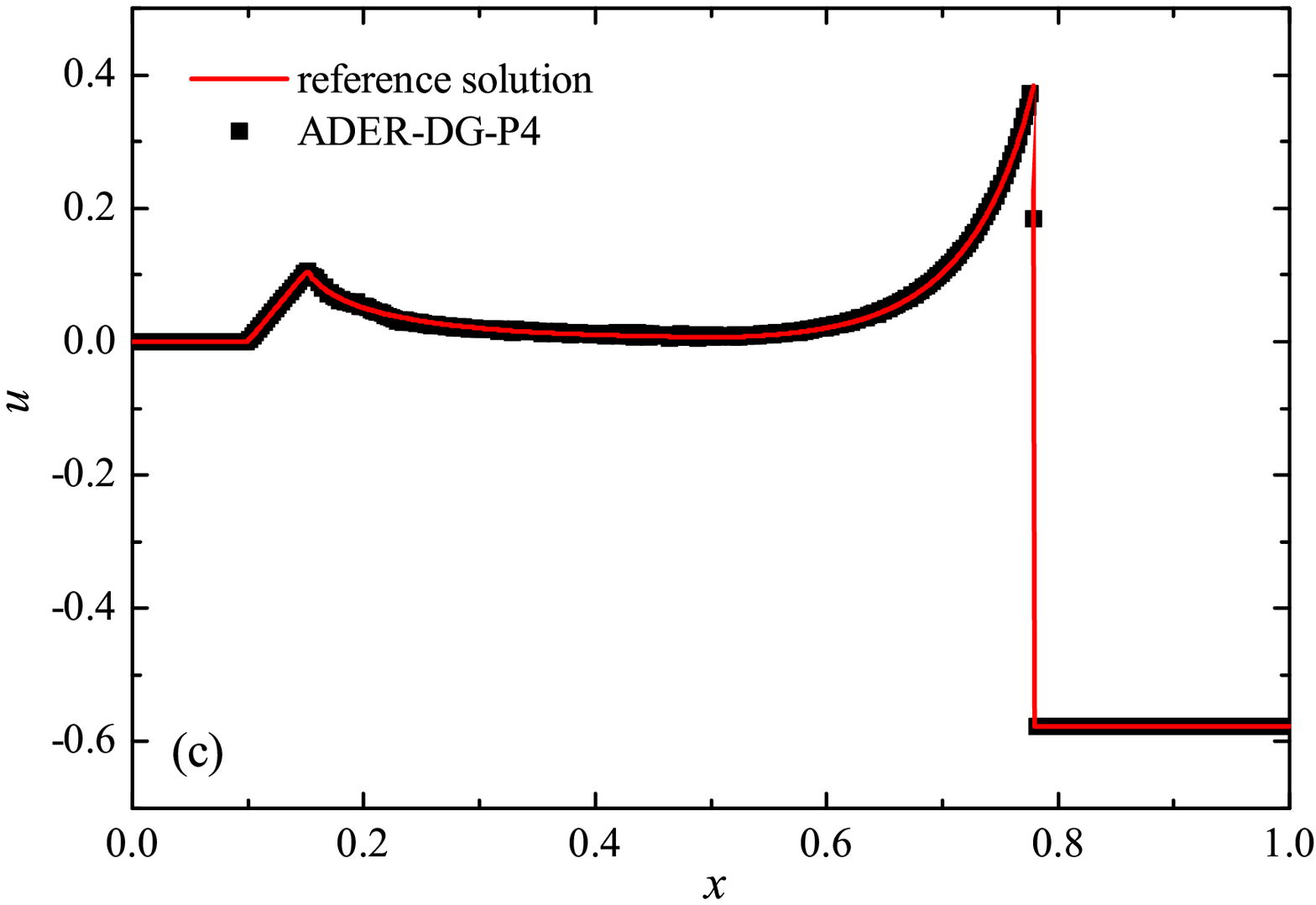}
\includegraphics[width=0.24\textwidth]{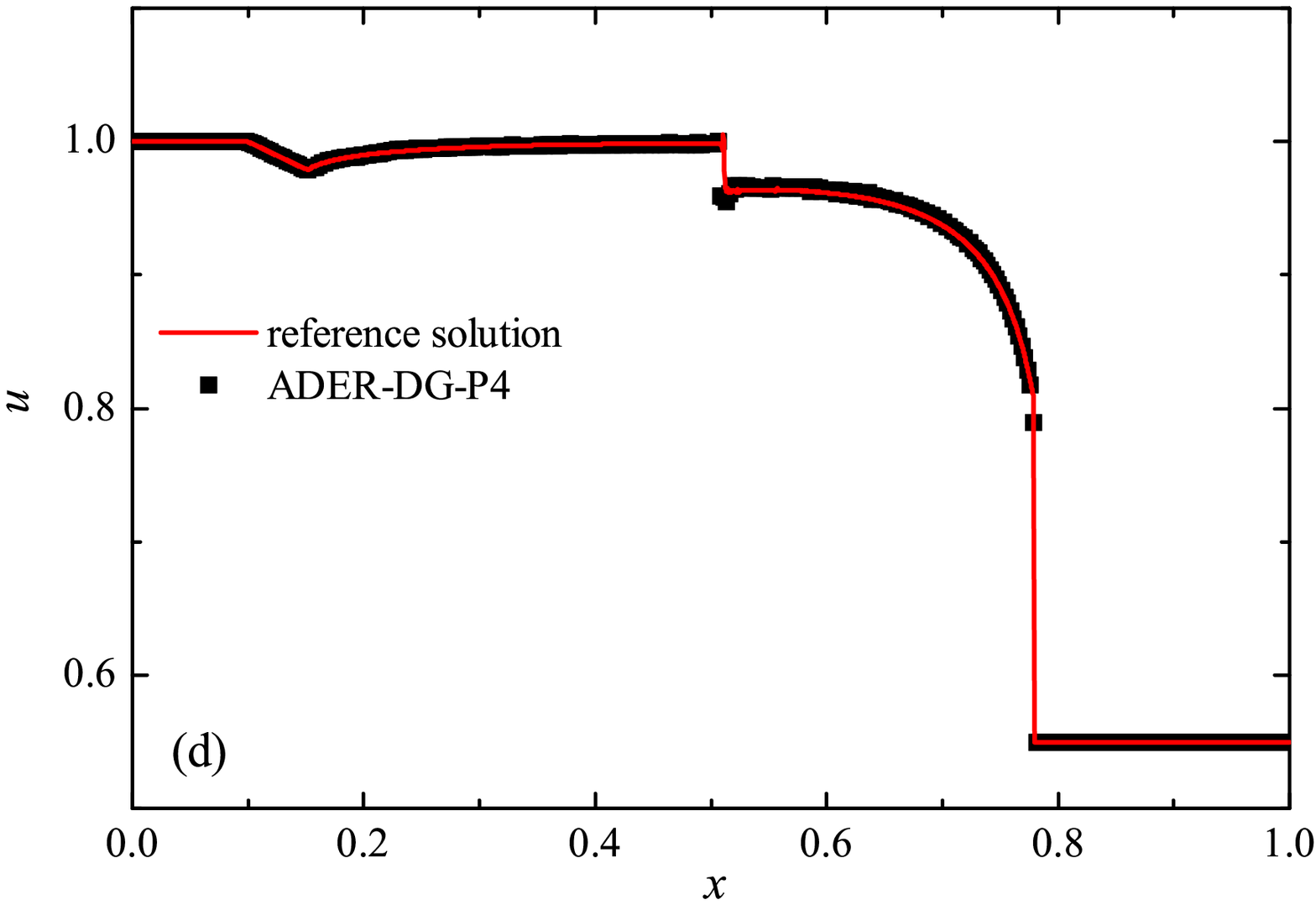}\\
\includegraphics[width=0.24\textwidth]{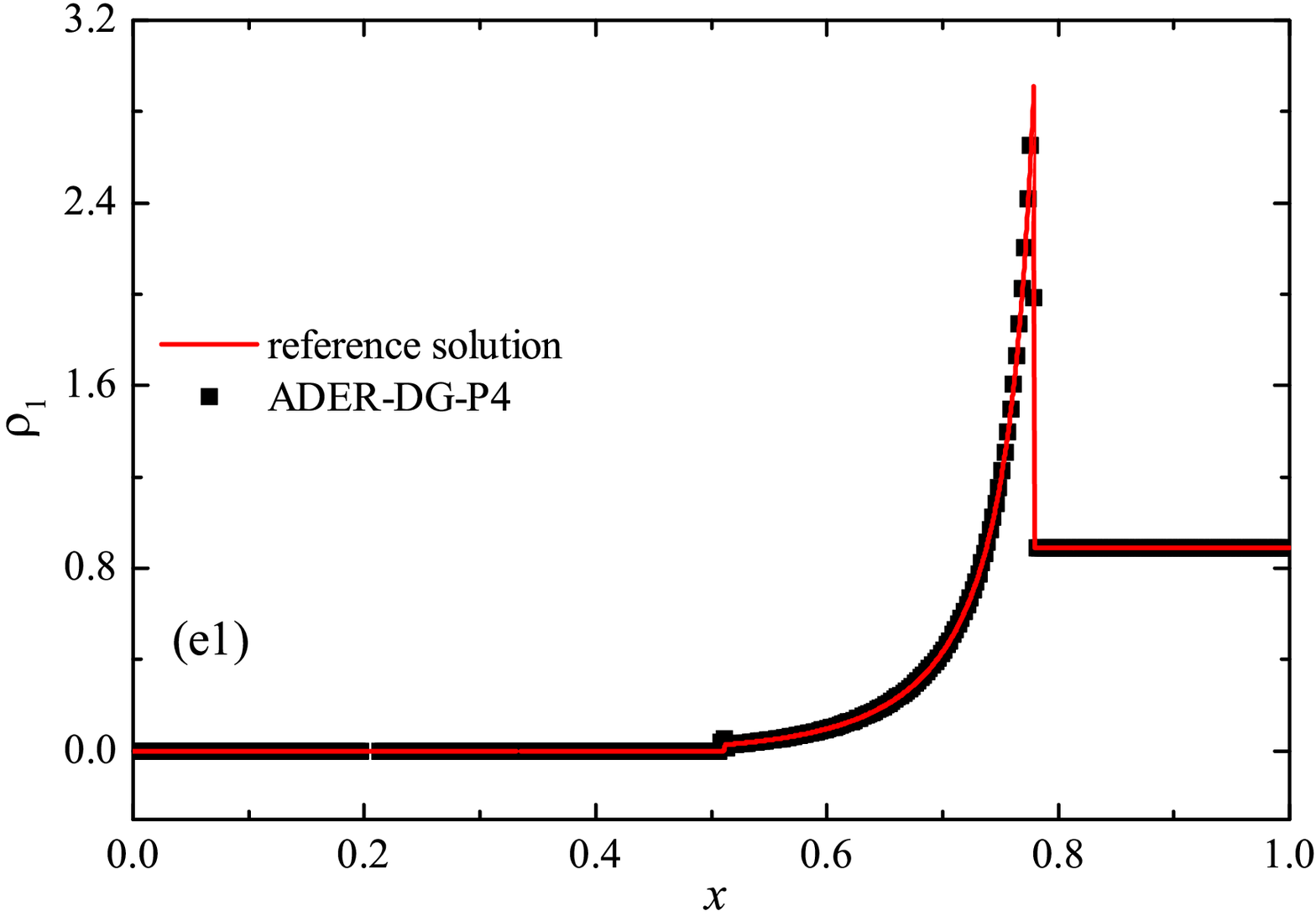}
\includegraphics[width=0.24\textwidth]{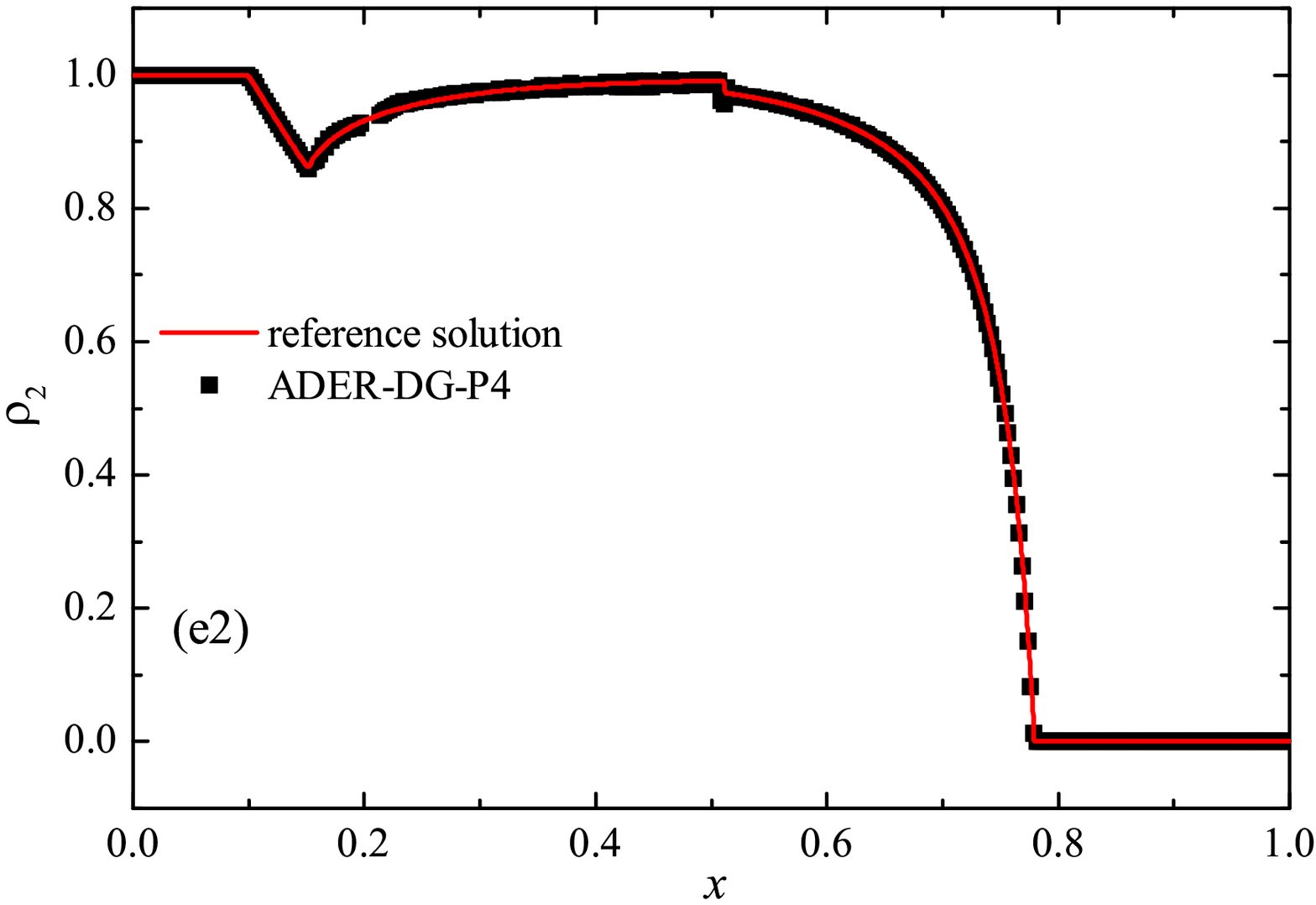}
\includegraphics[width=0.24\textwidth]{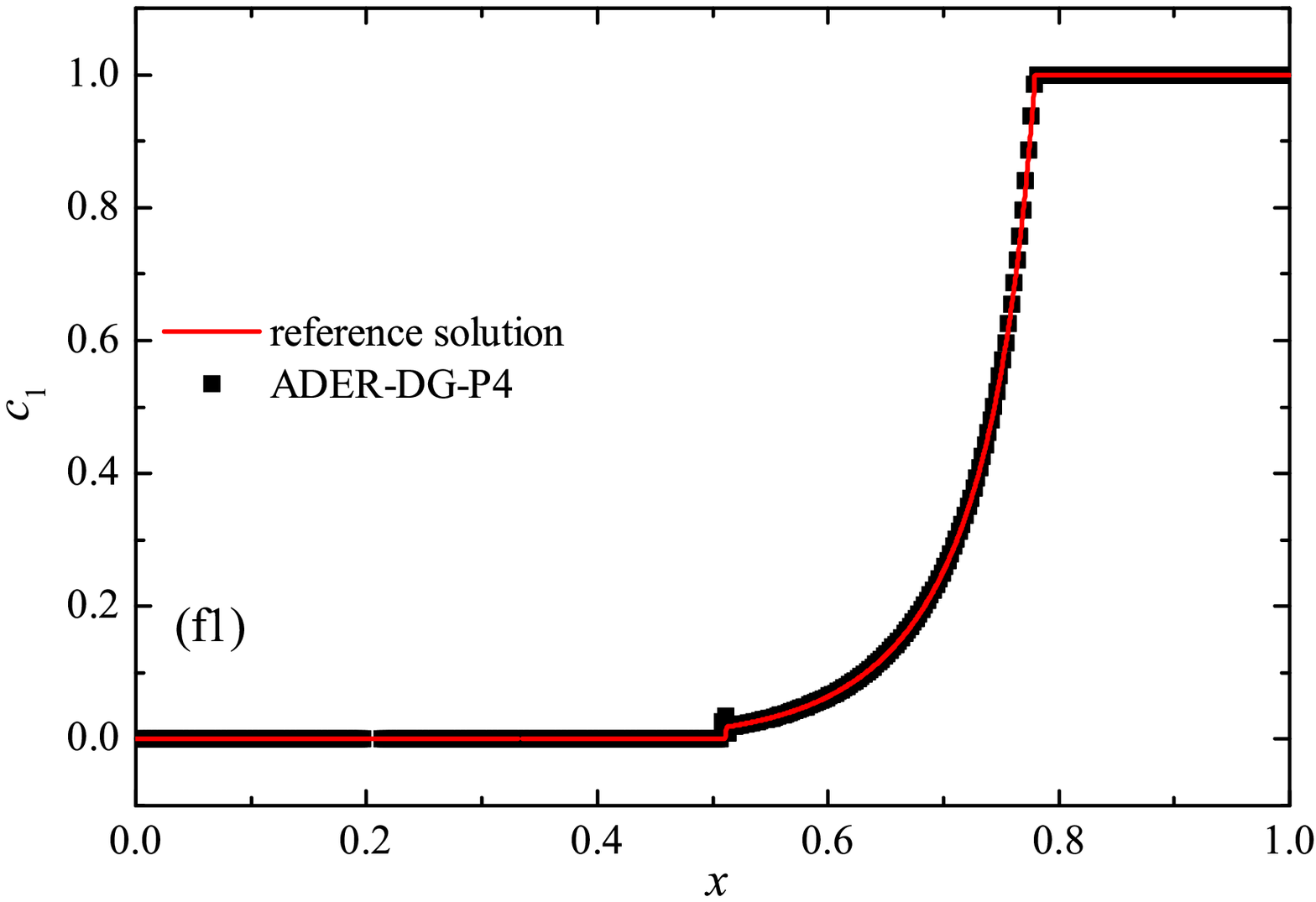}
\includegraphics[width=0.24\textwidth]{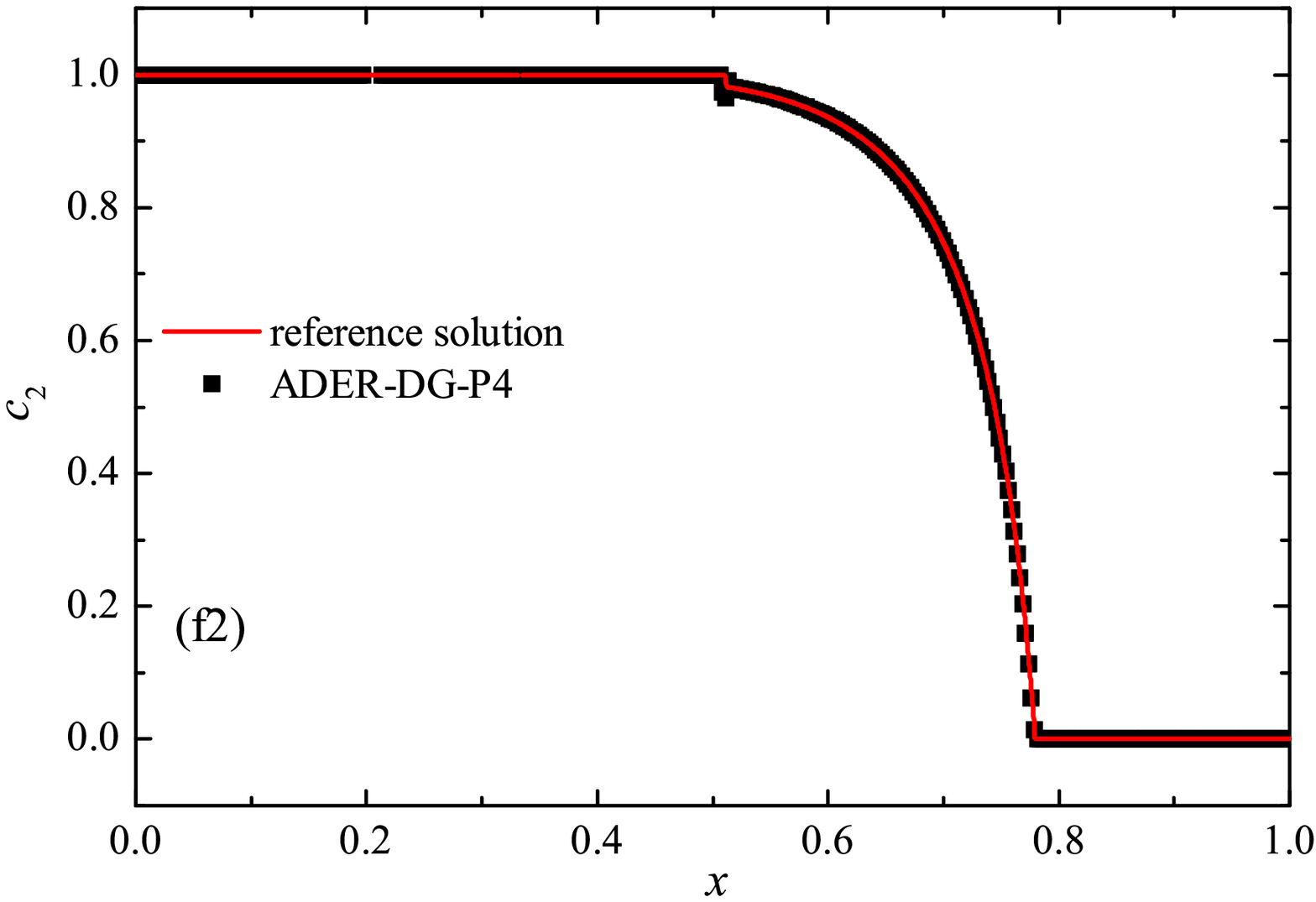}\\
\caption{%
	Numerical solution of the problem of the formation of a detonation wave in a two-component medium with a ``slow'' reaction
	(weak stiff case, a detailed statement of the problem is presented in the text),
	using the computational scheme $\mathrm{ADER}$-$\mathrm{DG}$-$\mathbb{P}_5$ with a posteriori 
	limitation of the solution by a $\mathrm{ADER}$-$\mathrm{WENO}4$ finite volume limiter,
	on a coordinate mesh with $400$ finite element cells.
	The graphs show the coordinate dependencies of pressure $p$ (a), density $\rho$ (b), flow velocity $u$ (c), 
	sound speed $c$ (d), densities $\rho_{k} = \rho c_{k}$ (e1-e2) and mass concentrations $\rho_{k} = \rho c_{k}$ (f1-f2) 
	of individual components $k$ of the two-component medium,
	at the final time $t_{\rm final} = 0.4$. The black square symbols represent the numerical solution; 
	the red solid lines represents the exact reference solution of the problem.
}
\label{fig:s_dwt_s_numerical}
\end{figure*} 

\begin{figure*}
\centering
\includegraphics[width=0.24\textwidth]{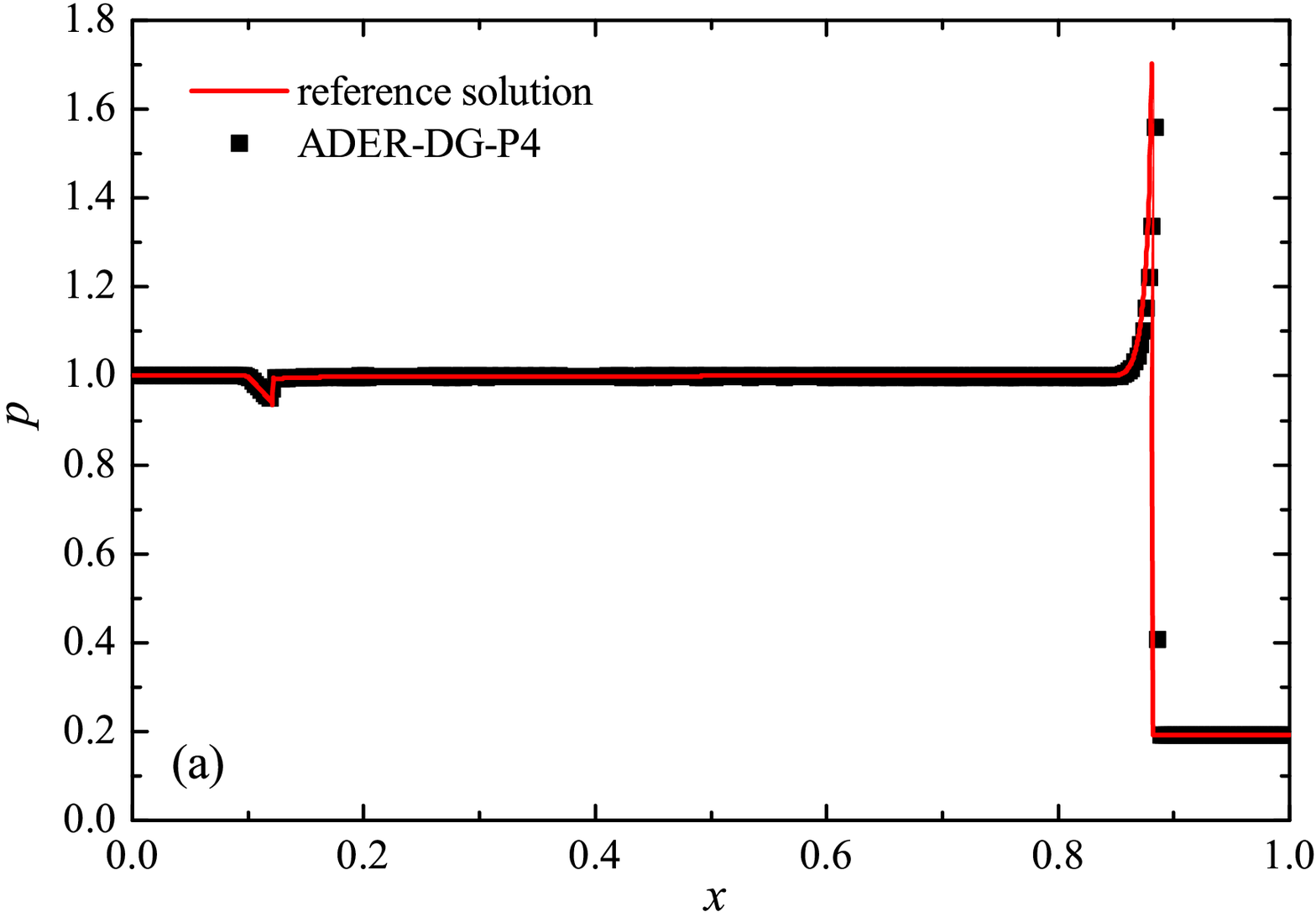}
\includegraphics[width=0.24\textwidth]{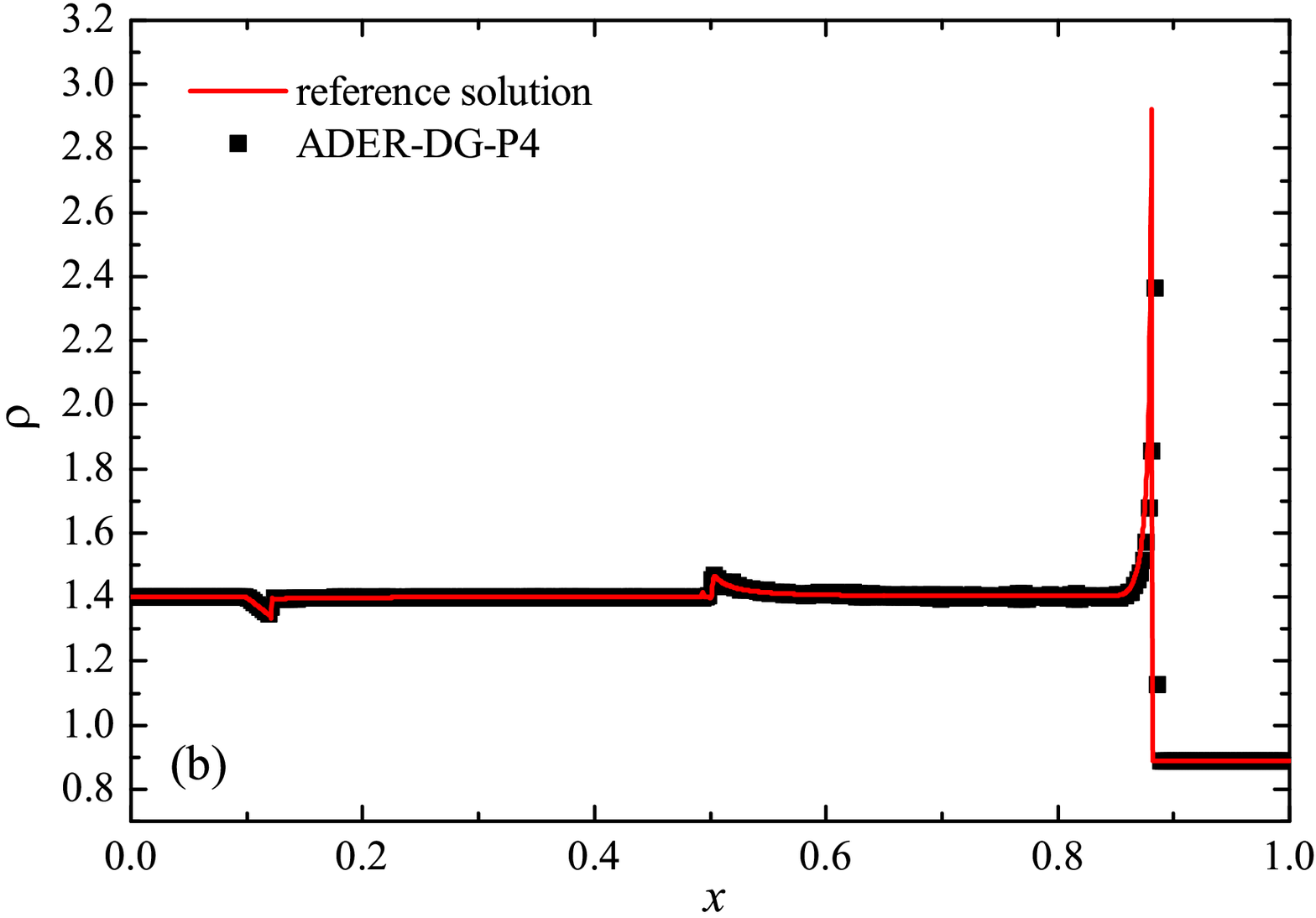}
\includegraphics[width=0.24\textwidth]{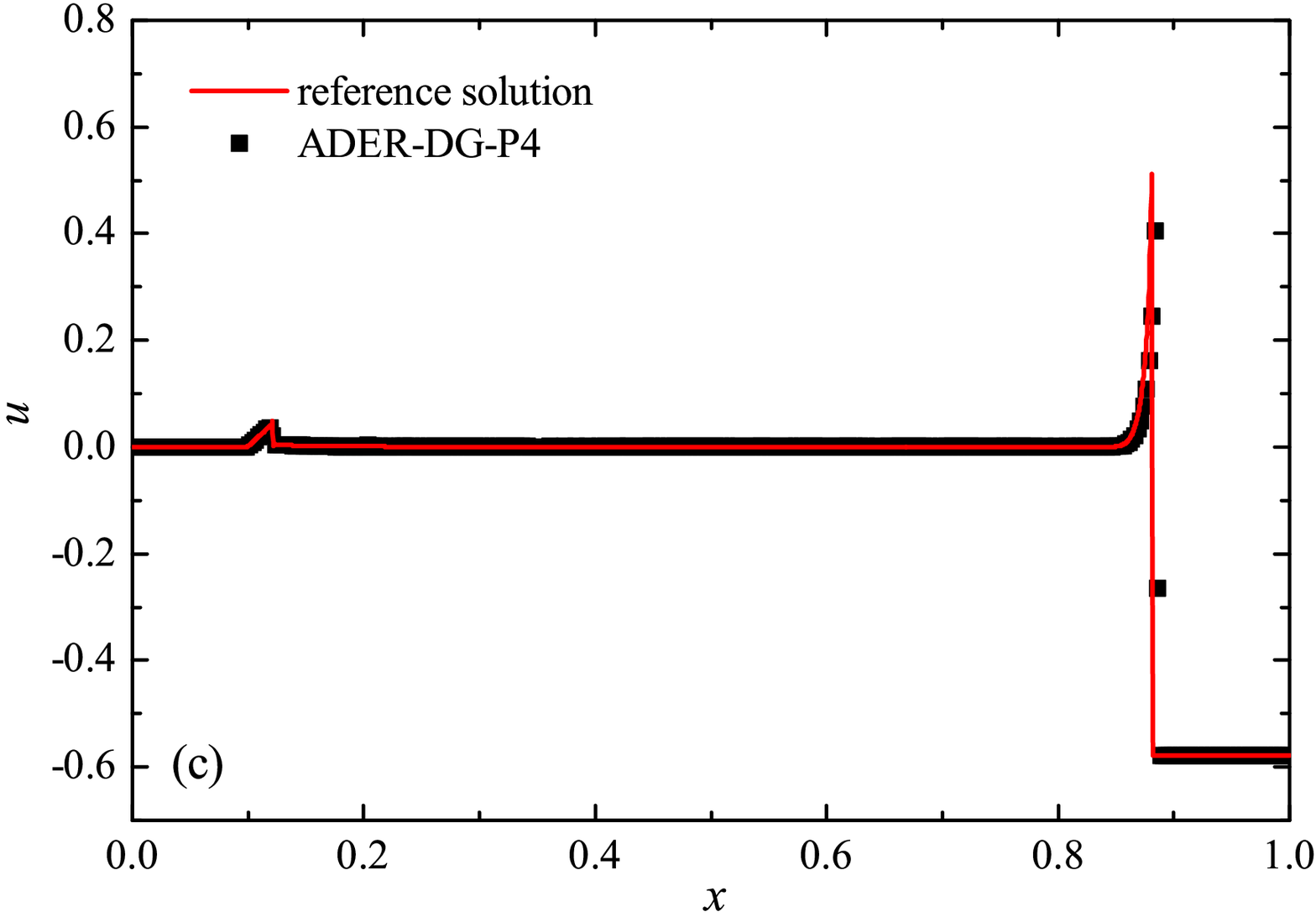}
\includegraphics[width=0.24\textwidth]{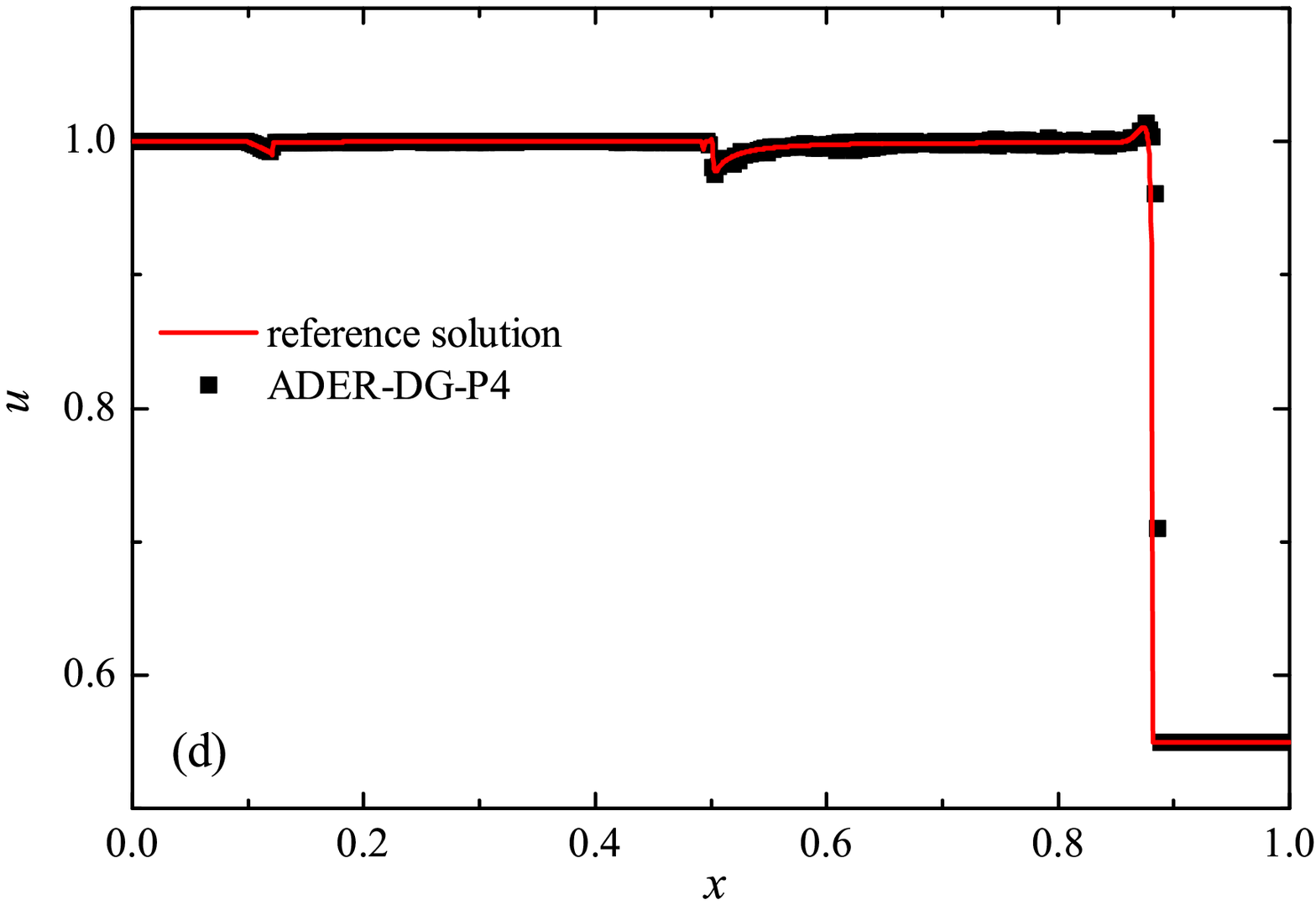}\\
\includegraphics[width=0.24\textwidth]{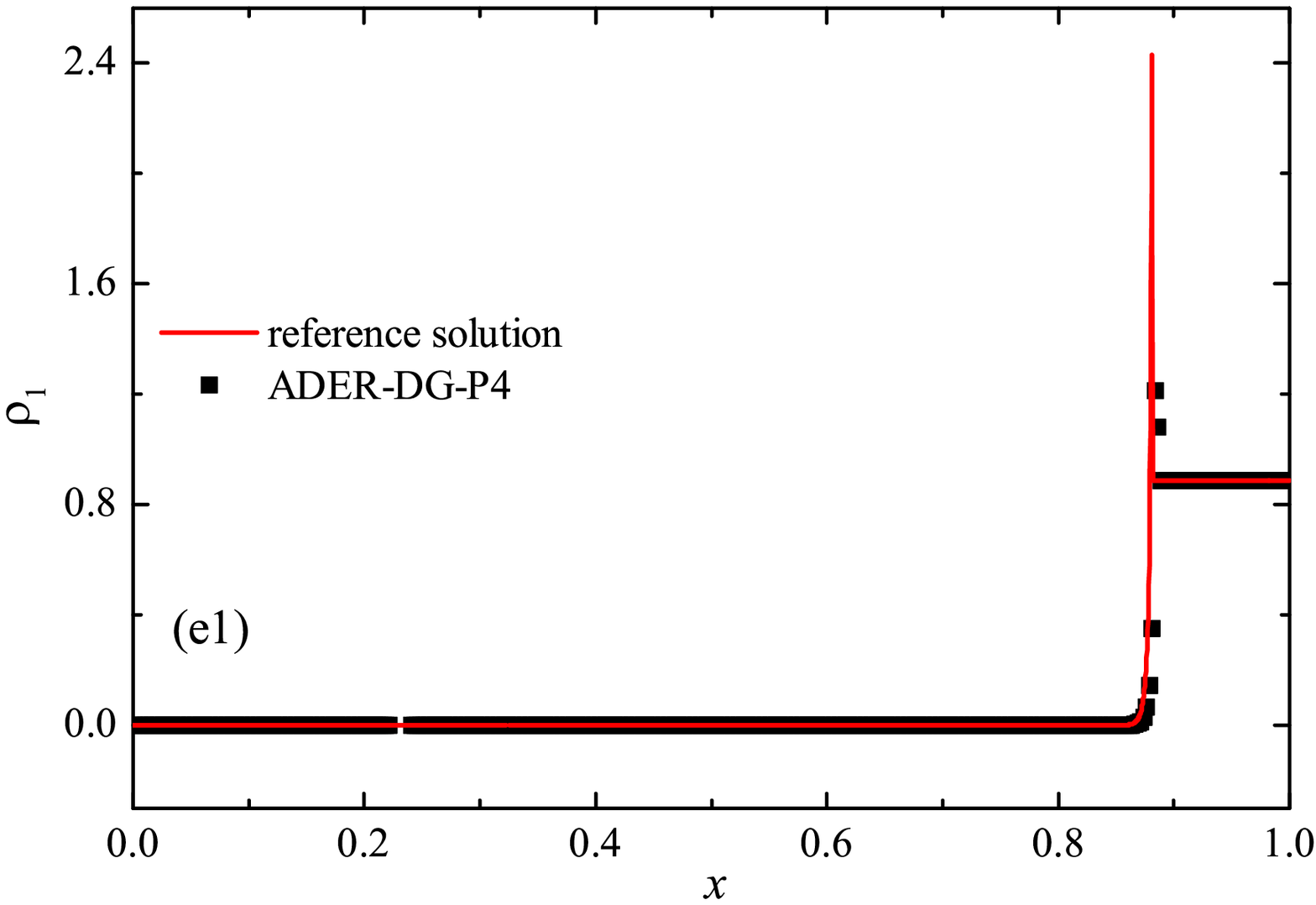}
\includegraphics[width=0.24\textwidth]{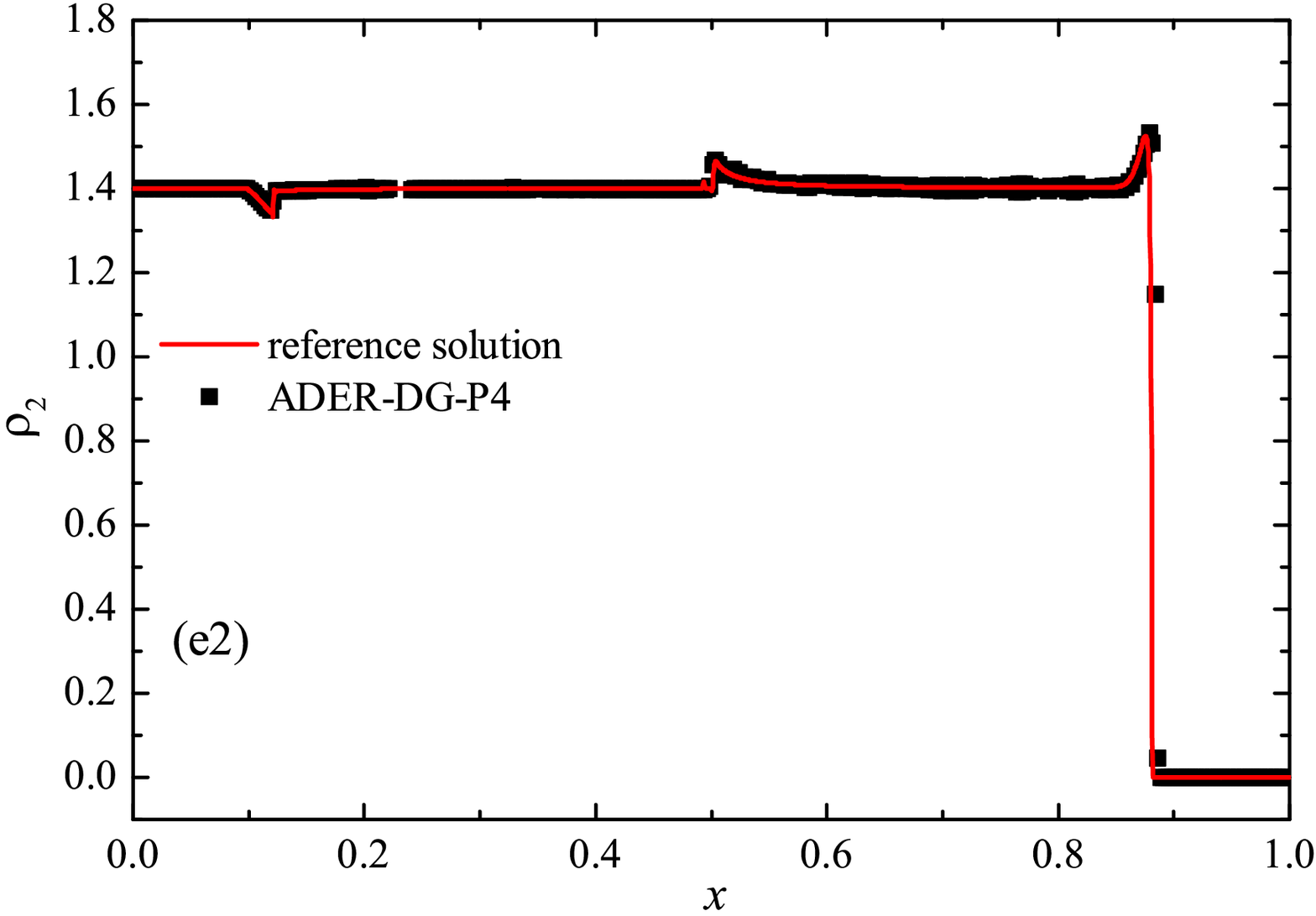}
\includegraphics[width=0.24\textwidth]{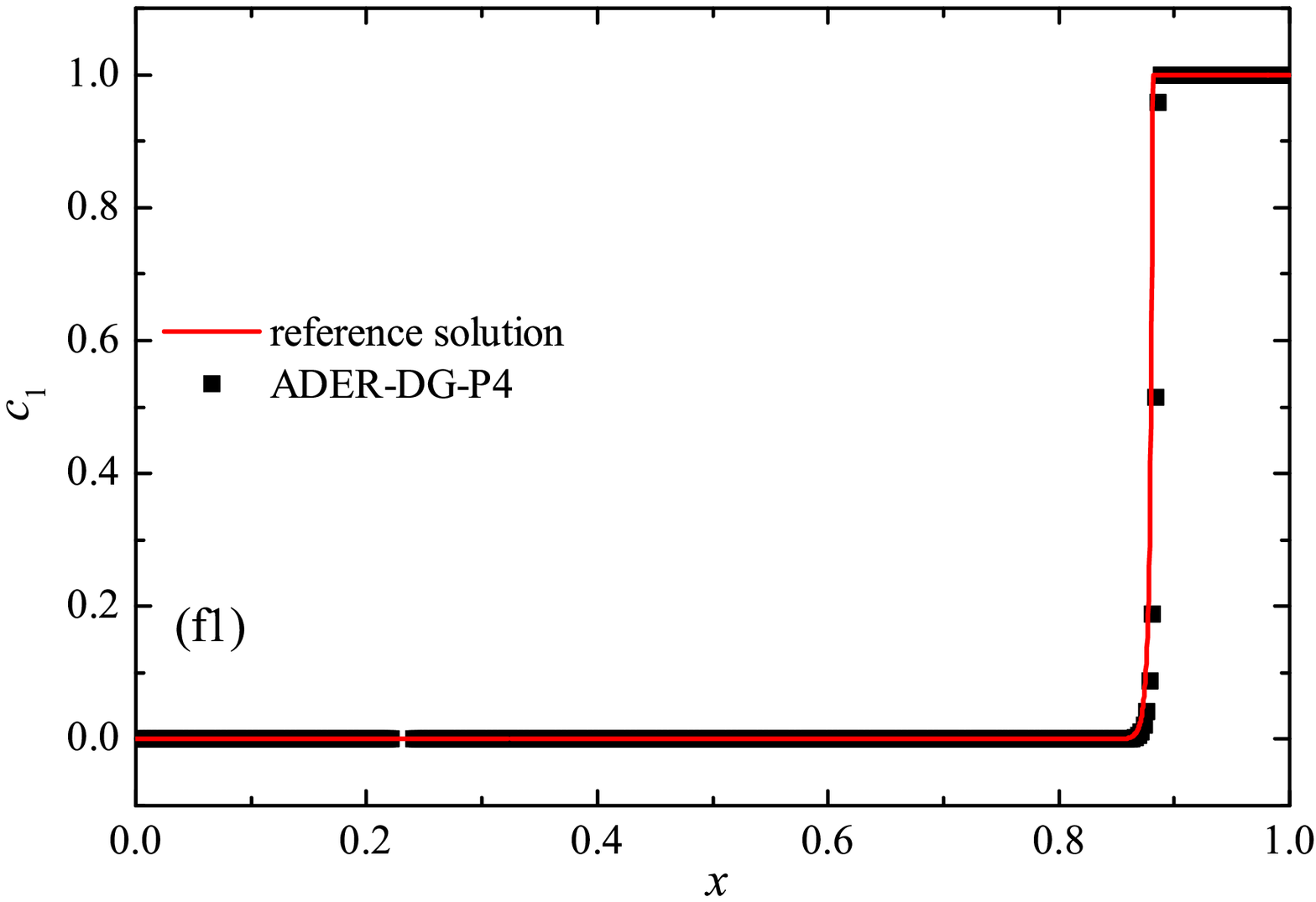}
\includegraphics[width=0.24\textwidth]{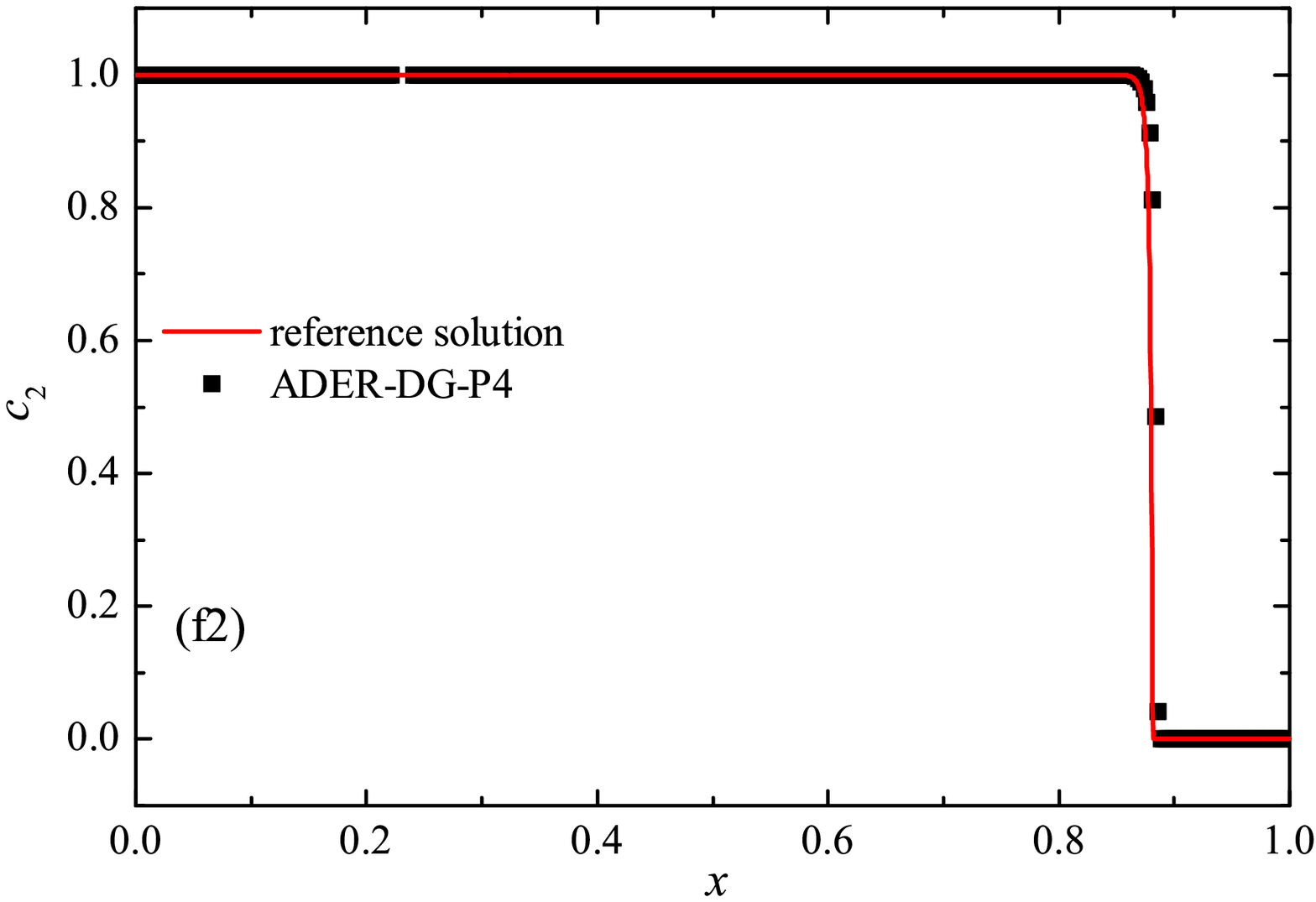}\\
\caption{%
	Numerical solution of the problem of the formation of a detonation wave in a two-component medium with a ``fast'' reaction
	(strong stiff case, a detailed statement of the problem is presented in the text),
	using the computational scheme $\mathrm{ADER}$-$\mathrm{DG}$-$\mathbb{P}_5$ with a posteriori 
	limitation of the solution by a $\mathrm{ADER}$-$\mathrm{WENO}4$ finite volume limiter,
	on a coordinate mesh with $400$ finite element cells.
	The graphs show the coordinate dependencies of pressure $p$ (a), density $\rho$ (b), flow velocity $u$ (c), 
	sound speed $c$ (d), densities $\rho_{k} = \rho c_{k}$ (e1-e2) and mass concentrations $\rho_{k} = \rho c_{k}$ (f1-f2) 
	of individual components $k$ of the two-component medium,
	at the final time $t_{\rm final} = 0.4$. The black square symbols represent the numerical solution; 
	the red solid lines represents the exact reference solution of the problem.
}
\label{fig:s_dwt_f_numerical}
\end{figure*}

\begin{figure*}
\centering
\includegraphics[width=0.24\textwidth]{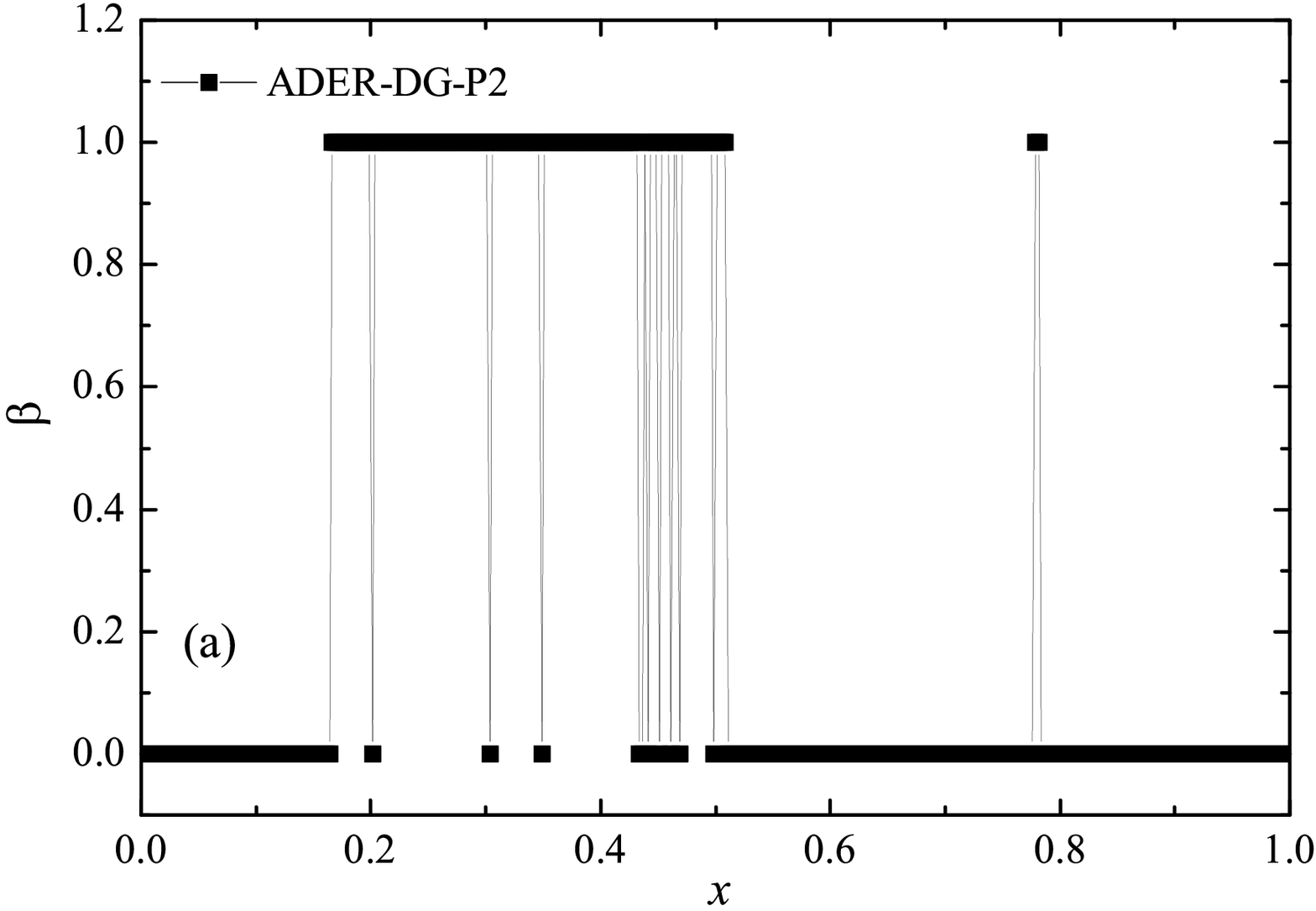}
\includegraphics[width=0.24\textwidth]{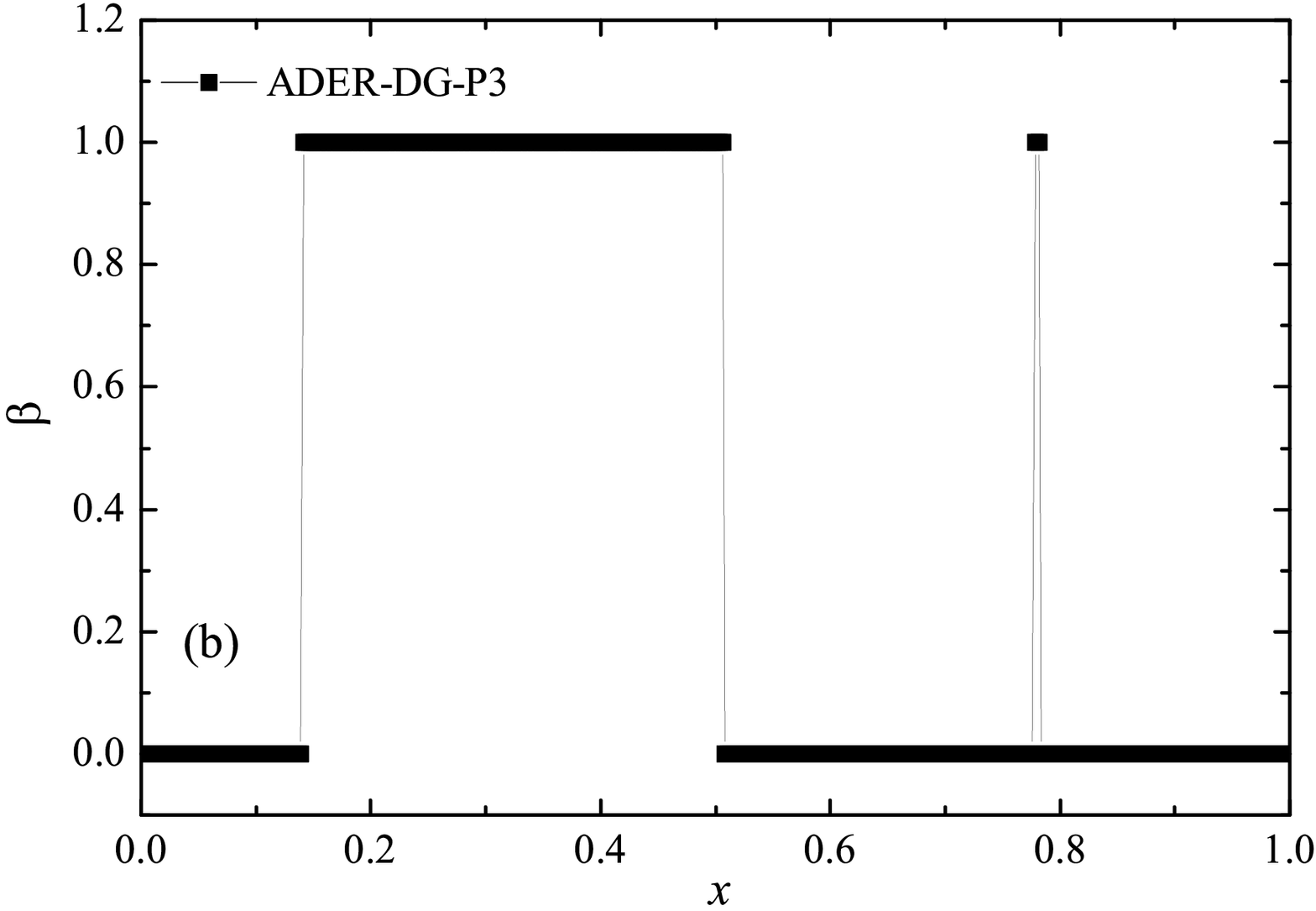}
\includegraphics[width=0.24\textwidth]{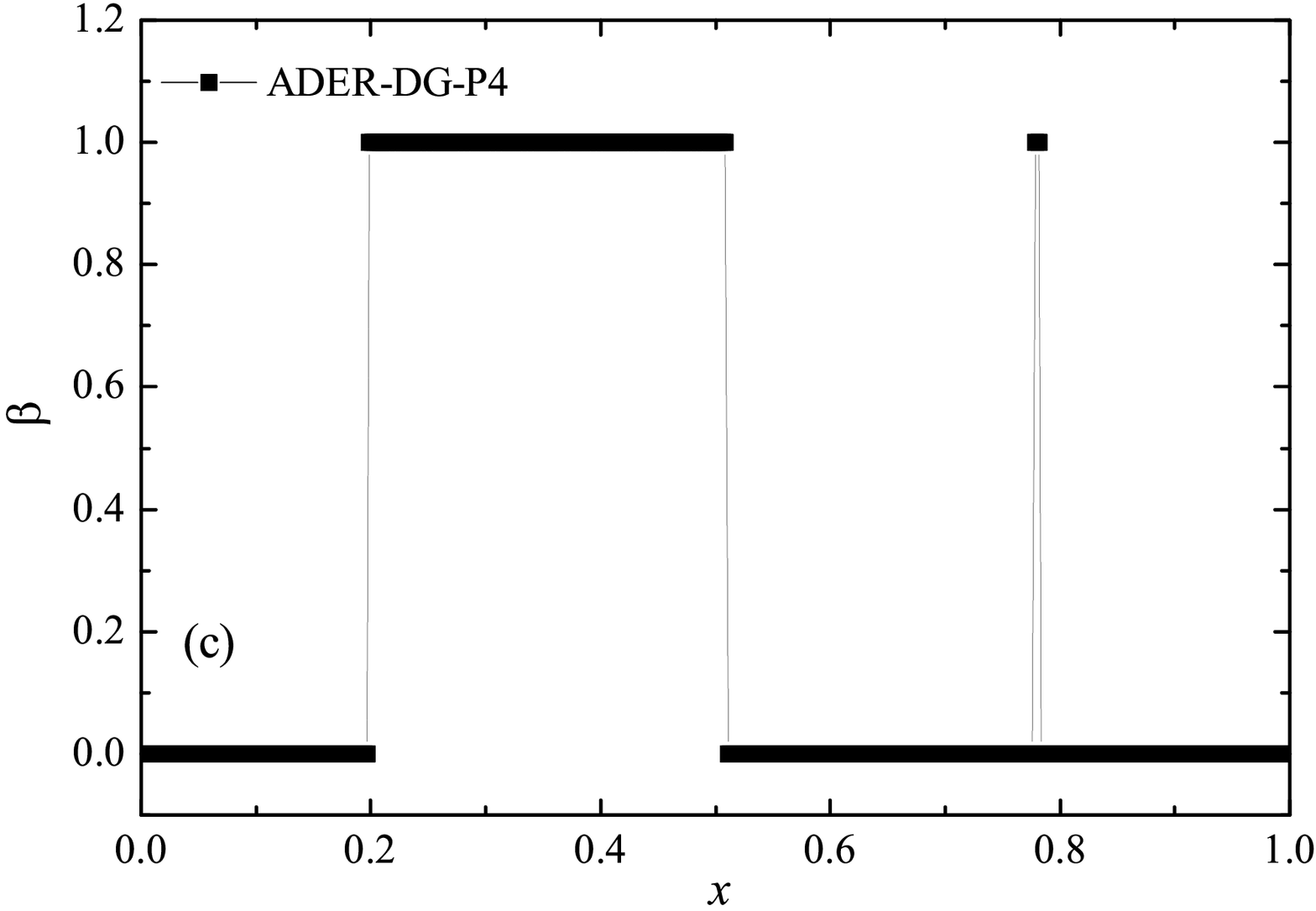}
\includegraphics[width=0.24\textwidth]{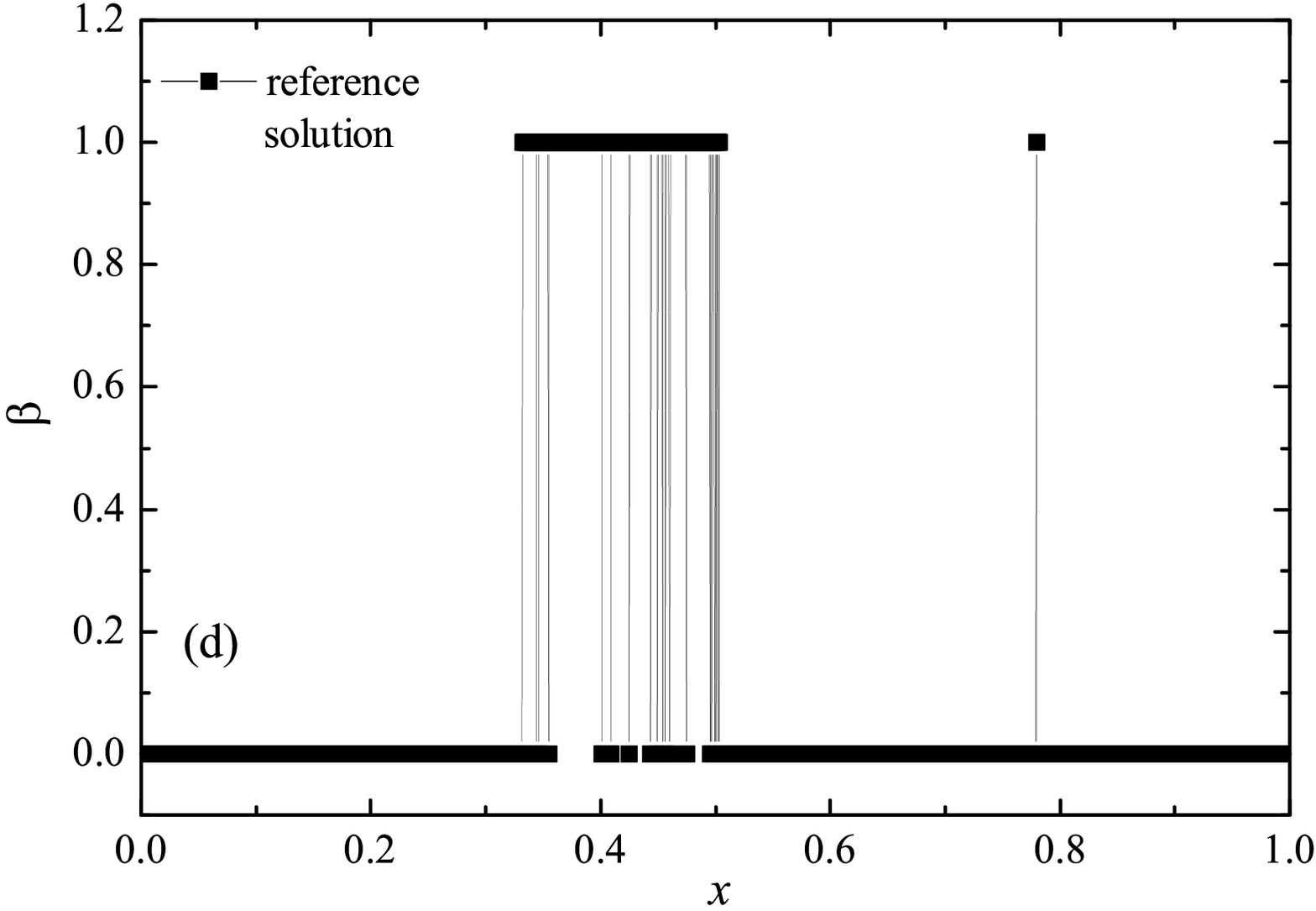}
\caption{%
	Troubled cells indicator $\beta$: $0$ --- normal cell, $1$ --- troubled cell,
	for numerical solution of the problem of the formation of a detonation wave in a 
	two-component medium with a ``slow'' reaction (weak stiff case).
	The graphs show the coordinate dependencies of troubled cells indicator $\beta$
	at the final time $t_{\rm final} = 0.4$ for computational scheme $\mathrm{ADER}$-$\mathrm{DG}$-$\mathbb{P}_2$ (a),
	$\mathrm{ADER}$-$\mathrm{DG}$-$\mathbb{P}_3$ (b), $\mathrm{ADER}$-$\mathrm{DG}$-$\mathbb{P}_4$ (c)
	and for reference solution (c): $\mathrm{ADER}$-$\mathrm{DG}$-$\mathbb{P}_1$ with $3200$ cells.
}
\label{fig:s_dwt_s_limiter}
\end{figure*} 

\begin{figure*}
\centering
\includegraphics[width=0.24\textwidth]{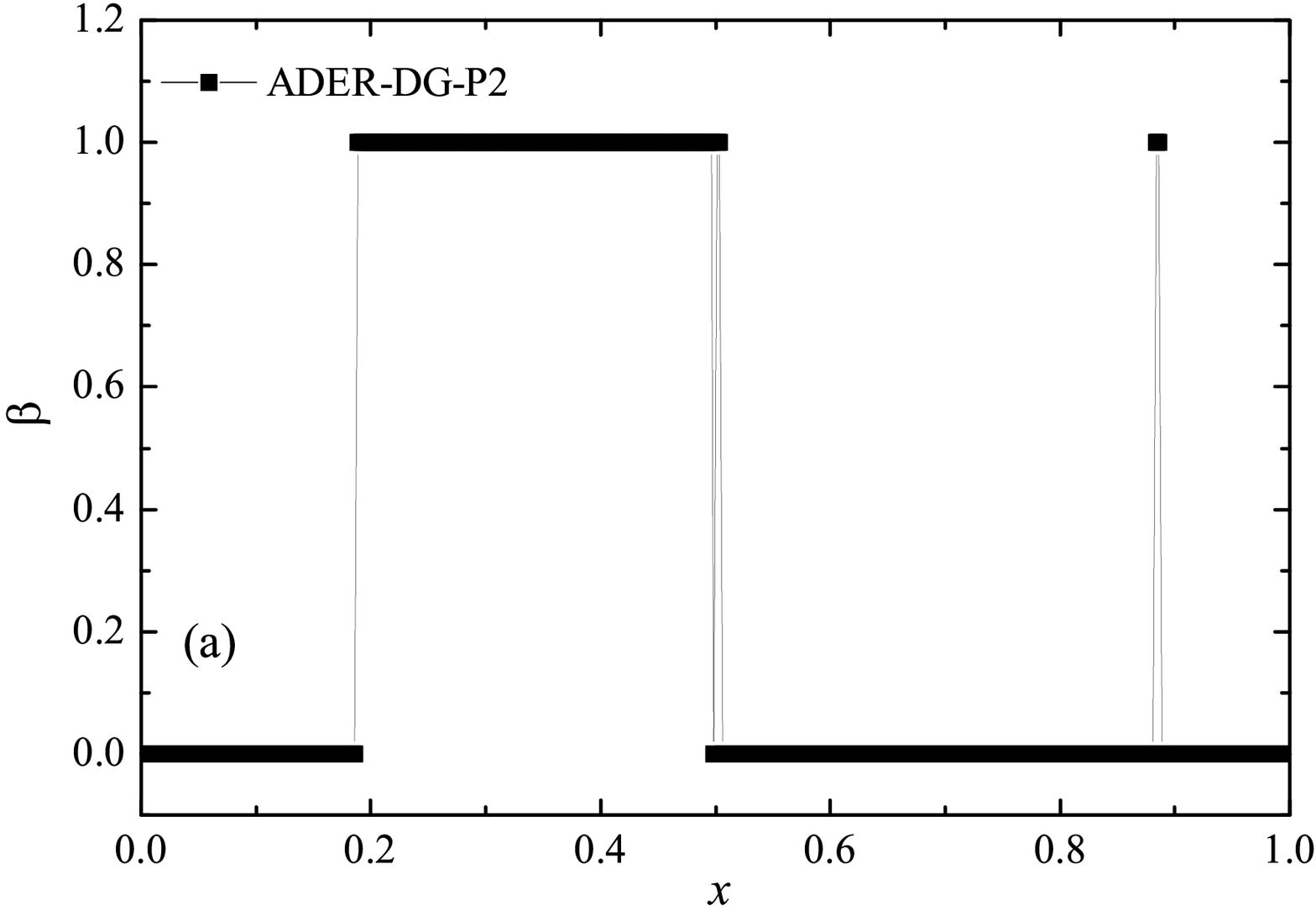}
\includegraphics[width=0.24\textwidth]{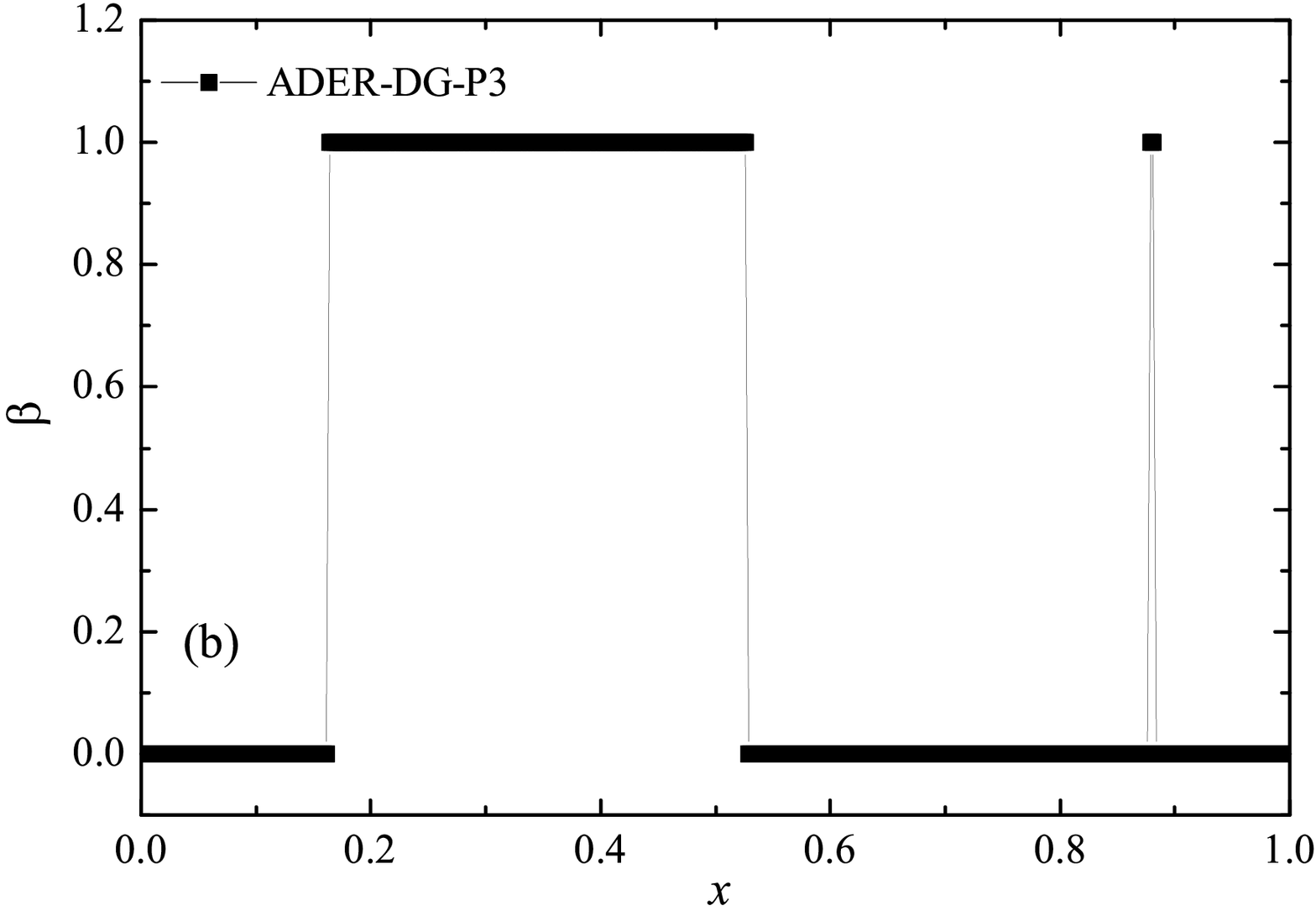}
\includegraphics[width=0.24\textwidth]{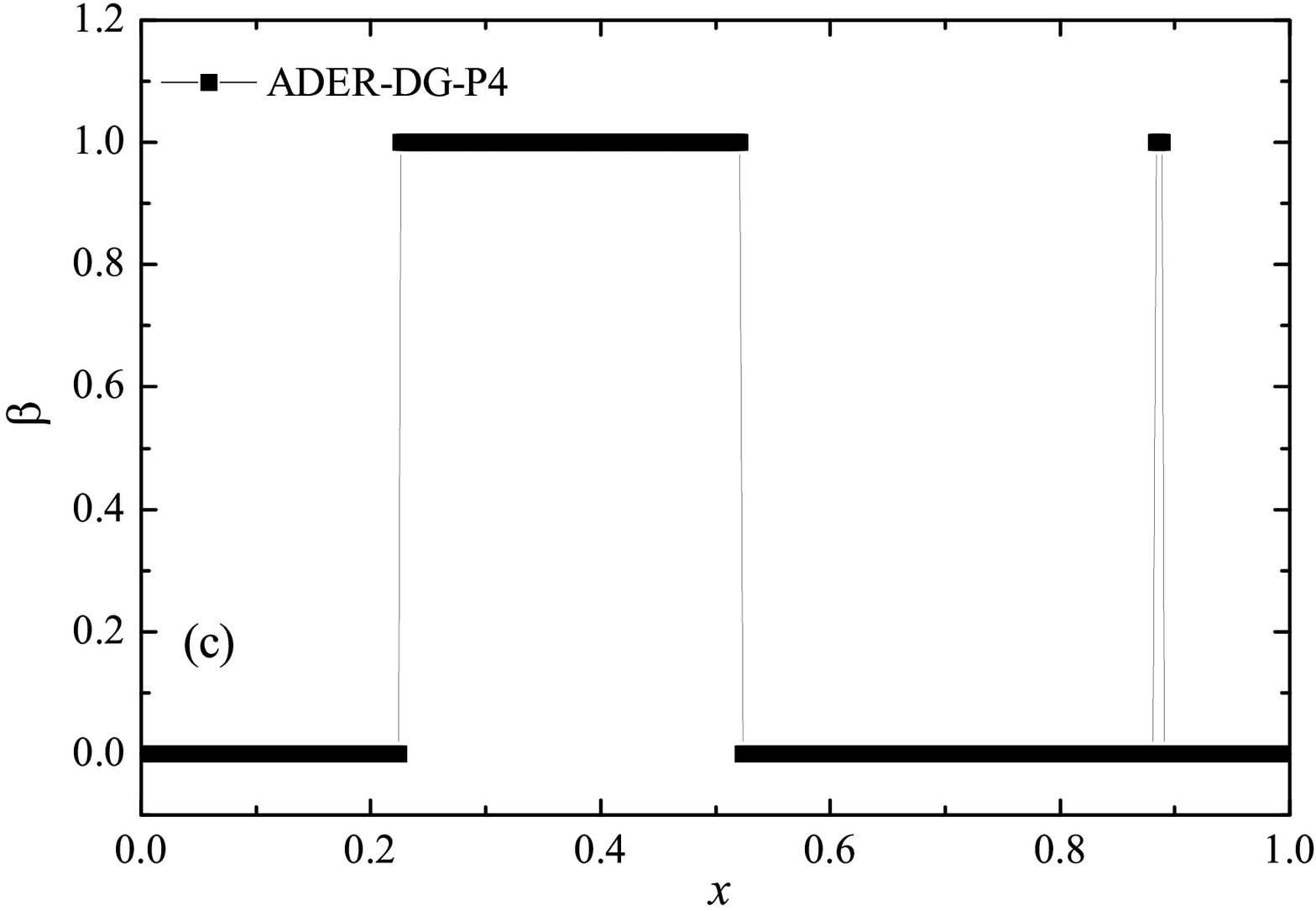}
\includegraphics[width=0.24\textwidth]{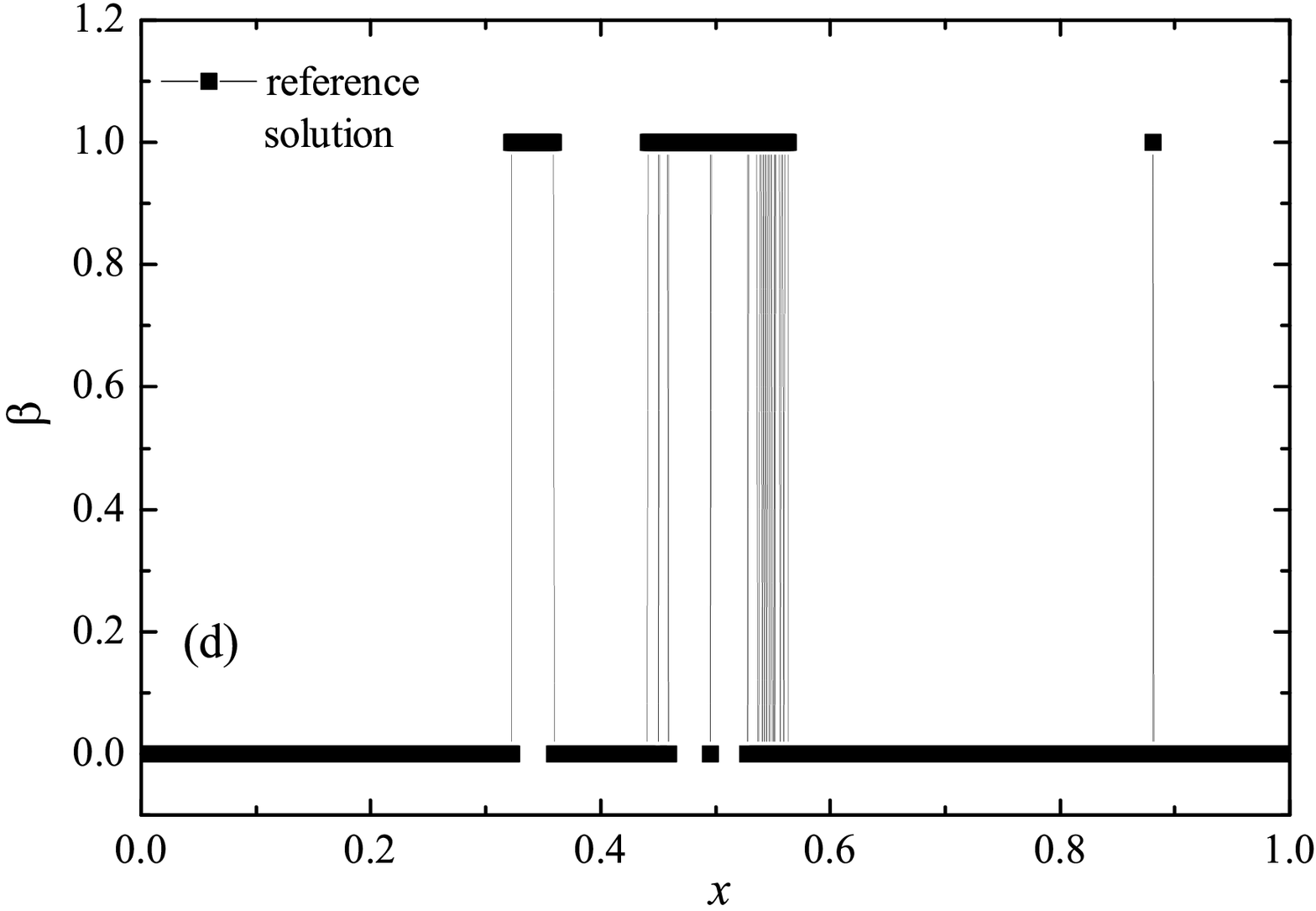}
\caption{%
	Troubled cells indicator $\beta$: $0$ --- normal cell, $1$ --- troubled cell,
	for numerical solution of the problem of the formation of a detonation wave in a 
	two-component medium with a ``fast'' reaction (strong stiff case).
	The graphs show the coordinate dependencies of troubled cells indicator $\beta$
	at the final time $t_{\rm final} = 0.4$ for computational scheme $\mathrm{ADER}$-$\mathrm{DG}$-$\mathbb{P}_2$ (a),
	$\mathrm{ADER}$-$\mathrm{DG}$-$\mathbb{P}_3$ (b), $\mathrm{ADER}$-$\mathrm{DG}$-$\mathbb{P}_4$ (c)
	and for reference solution (c): $\mathrm{ADER}$-$\mathrm{DG}$-$\mathbb{P}_1$ with $3200$ cells.
}
\label{fig:s_dwt_f_limiter}
\end{figure*} 

As a test example of the flow of a multicomponent medium with chemical reactions, a model~\cite{frac_steps_detwave_sim_2000} of the formation and propagation of a detonation wave in media  with discrete ignition temperature kinetics model was chosen. In work~\cite{ader_stiff_2}, this model was used as an example of the study of viscous flow with chemical reactions. The original work~\cite{frac_steps_detwave_sim_2000} describes in detail the problems that one has to face when solving such problems in the framework of the methods of splitting into physical processes and fractional steps. In particular, due to significant differences in the times of the processes occurring in the medium associated with hydrodynamic flow and chemical reactions, it is often impossible to obtain a correct numerical solution in principle. The reagent ``quickly'' burns out, while the correct distribution of the released energy does not occur. As a result, a stationary, in the reference frame accompanying the detonation front, ZND (Zel'dovich, von Neumann, and D\"{o}ring) structure of a CJ (Chapman-Jouguet) detonation wave does not arise in the numerical solution, in particular, it is not possible to obtain the correct propagation velocity of the detonation wave --- the non-stationary two-wave structure is formed, where the shock wave has a higher propagation velocity than is assumed by the conditions of stationary detonation. The work~\cite{frac_steps_detwave_sim_2000} proposes a rather complicated (both from the standpoint of universality, and from the standpoint of modifications to the original numerical method) method for obtaining the correct solution using splitting by physical processes. The work~\cite{chem_kin_hrs_weno} notes the existence of such problems, as a result of which it presents the results of a comparison with the ``standard'' numerical splitting scheme, in which it is not possible to correctly resolve the motion of a detonation wave in the flow.

In this work, the space-time adaptive ADER-DG finite element method with LST-DG predictor and a posteriori sub-cell WENO finite-volume limiting is used for simulation of non-stationary compressible multicomponent reactive flows without using the splitting method --- hydrodynamic processes are considered simultaneously with the kinetics of chemical reactions. The only modification compared to the classic original method~\cite{ader_dg_ideal_flows} is the inclusion of the method of adaptive change in the time step.

The problem statement was chosen similarly to the problem in the works~\cite{frac_steps_detwave_sim_2000, ader_stiff_2}. The spatial domain of the flow was chosen as $x \in [0.0, 1.0]$. The initial discontinuity was located at the coordinate $x_{\rm c} = 0.5$. The initial conditions were chosen in the form: to the left of the discontinuity $\rho_{L} = 1.4$, $u_{L} = 0.0$, $p_{L} = 1.0$; to the right of the discontinuity $\rho_{R} = 0.887565$, $u_{R} = -0.57735$, $p_{R} = 0.191709$. The medium is chosen as a two-component one, with a monomolecular reaction $A_{1} \rightarrow A_{2}$, where the first component $A_{1}$ is the reaction reagent, the second component $A_{2}$ is the reaction product. To the left of the discontinuity, the gas was assumed to be burnt, with mass concentrations $c_{1, L} = 0$, $c_{2, L} = 1$; to the right of the discontinuity, the gas was assumed to be unburnt, with mass concentrations $c_{1, R} = 1$, $c_{2, R} = 0$. The final time of the simulation was chosen as $t_{\rm final} = 0.4$. The discrete ignition temperature kinetics model was represented by the reaction rate constant in the following form:
\begin{eqnarray}\label{rate_constant}
k(T) = \left\{
\begin{array}{cl}
\frac{1}{\tau_{0}},& \mathrm{if}\, T \geqslant T_{\rm ign},\\[2mm]
0,& T < T_{\rm ign},
\end{array}
\right.
\end{eqnarray}
where $T$ is the temperature, $T_{\rm ign}$ is the ignition temperature, and $\tau_{0}$ is the time scale of the chemical reaction. The ignition temperature was chosen $T_{\rm ign} = 0.25$. Dimensional scales and parameters of medium (such as molar mass $\mu$) are chosen such that $T = p/\rho$. The specific energy yield was chosen $q_{0} = 1$. In this work, two tests were considered, the parameters for which were chosen similarly to the work~\cite{ader_stiff_2}: $\tau_{0} = 0.1$ (``slow'' reaction) and $\tau_{0} = 4 \cdot 10^{-3}$ (``fast'' reaction). The first case corresponds to the classical relation between the time scales of the reacting flow; the second is the case of a high stiffness of the solution. The reference solutions are computed on a fine mesh using a second order ADER-DG-$\mathbb{P}_1$ numerical scheme with $\mathrm{ADER}$-$\mathrm{WENO}2$ limiter on $3200$ cells and are shown in Fig.~\ref{fig:figs_s_dwt_s_ref_solution} for the ``slow'' reaction case and in Fig.~\ref{fig:figs_s_dwt_f_ref_solution} for the ``fast'' reaction case. The reference solution corresponds well to the solution presented in the works~\cite{frac_steps_detwave_sim_2000, ader_stiff_2}.

Numerical schemes ADER-DG-$\mathbb{P}_{N}$ with the degrees $N = 1$, $2$, $3$ and $4$ of polynomials were investigated. On the boundaries of the spatial domain of the solution, the boundary conditions of the free boundary were set. The sub-cell limiter for the numerical scheme ADER-DG-$\mathbb{P}_{N}$ was the ADER-WENO ($\mathbb{P}_{N}$) finite-volume scheme. The numerical solution was obtained on a mesh with $N_{\rm cells} = 400$ finite-element cells. The Courant number was chosen $\mathtt{CFL}\_\mathtt{number} = 0.4$.

The numerical solution for the non-stiff problem (``slow'' reaction case) obtained by the ADER-DG-$\mathbb{P}_{4}$ numerical scheme and its comparison with the reference solution is shown in Fig.~\ref{fig:s_dwt_s_numerical}. The numerical solution shows well agreement with the reference solution and shows the main peculiarities of the problem solution --- a CJ detonation wave arises and demonstrates the main features of the wave structure. Non-physical effects and artifacts of the numerical solution, which are typical for schemes with splitting by physical processes, do not appear.

The numerical solution for the stiff problem (``fast'' reaction case) obtained by the ADER-DG-$\mathbb{P}_{4}$ numerical scheme and its comparison with the reference solution is shown in Fig.~\ref{fig:s_dwt_f_numerical}. The numerical solution shows well agreement with the reference solution. The classical nature of the development of stationary detonation is observed. A classical ZND patterns of the formation and evolution of stationary detonation is observed, such as Zel'dovich spike, with complete burnout of the reagent behind the detonation wave --- a spatial region with a high reaction rate appears behind the shock wave front, the energy released as a result of the reaction is correctly redistributed in the flow, forming a stationary detonation structure.

The detonation front practically does not spread, the expansion of the shock wave front occurs by no more than $1$-$2$ finite-element cells. This is characteristic of the solution in both cases. However, it should be noted that in the case of a problem with high stiffness, the magnitude of the peak in numerical solution is smaller than in the case of the reference solution. In this case, the shock wave front itself is somewhat ahead of the shock wave front in the reference solution, by a distance of about $1$-$3$ cells.

In addition to the results obtained in Fig.~\ref{fig:s_dwt_s_limiter} and Fig.~\ref{fig:s_dwt_f_limiter} the coordinate dependencies of the indicator $\beta$ are presented, which determine the ``troubled'' cells ($\beta = 0$ corresponds to the normal cell, $\beta = 1$ corresponds to the ``troubled'' cell) in which the solution obtained by the method is recalculated by the limiter. It can be seen that in both cases the problematic cells occupy about 25-35\% of the nodes of the spatial grid. Problematic cells located on the right arise due to the passage of a shock wave through them and, as a result, a violation of the monotonically of the numerical scheme occurs. The problematic cells located on the left are related to the discontinuity decay perturbations propagating to the left under the initial conditions.

\section*{Conclusion}
\label{sec:conclusion}

In this work, the space-time adaptive ADER finite element DG method with a posteriori correction technique of solutions on subcells by the finite-volume ADER-WENO limiter was used to simulate non-stationary compressible multicomponent reactive flows. The multicomponent composition of the reacting medium and the reactions occurring in it were described by expanding the original system of Euler equations by a system of non-stationary convection-reaction equations. The use of this method to simulate high stiff problems associated with reactions occurring in a multicomponent medium requires the use of the adaptive change in the time step, for which a suitable formulation was proposed in this paper. The solution of classical test problems based on Riemann problems, generalized to the case of multicomponent medium, was carried out. It is shown that the use of the method in the case of multicomponent medium does not lead to the emergence of new non-physical peculiarities and artifacts of the numerical solution. The solution of the classical problem related to the formation and propagation of a ZND detonation wave is carried out in a non-stiff and high-stiff formulation of the problem. It is shown that the space-time adaptive ADER-DG finite element method with LST-DG predictor and a posteriori sub-cell WENO finite-volume limiting allows simulate such problems without using of splitting in directions and fractional step methods.

As a result of the work carried out, it was shown that the space-time adaptive ADER finite element DG method with a posteriori correction technique of solutions on subcells by the finite-volume ADER-WENO limiter, proposed and developed in the works~\cite{ader_dg_ideal_flows, ader_dg_dev_1, ader_dg_dev_2, ader_weno_lstdg, ader_dg_diss_flows, ader_dg_ale, ader_dg_grmhd, ader_dg_gr_prd, ader_dg_PNPM, PNPM_DG, ader_dg_eff_impl, fron_phys, exahype, ader_dg_hpc_impl_1, ader_dg_hpc_impl_2, ader_dg_hpc_impl_3, ader_dg_hpc_impl_4}, can be used to simulate flows without using splitting methods.

\section*{Declarations}

\begin{acknowledgements}
The reported study was support of the Russian Science Foundation grant No. 21-71-00118:\\ \url{https://rscf.ru/en/project/21-71-00118/}.

The simulations were supported in through computational resources provided by the Laboratory of Applied Theoretical Physics and Parallel Computation of Dostoevsky OmSU, by the Shared Facility Center ``Data Center of FEB RAS'' (Khabarovsk) and by Moscow Joint Supercomputer Center of the RAS.

I would like to thank Popova A.P. for help in correcting the English text.
\end{acknowledgements}

\subsection*{Data Availability}
The datasets generated during and/or analysed during the current study are available from the corresponding author on reasonable request.

\subsection*{Conflict of interest}
The author declares that he has no conflict of interest.

\end{document}